\documentclass[useAMS, usenatbib]{mn2e}
\usepackage{graphicx}
\usepackage{dcolumn}
\usepackage{amsmath}
\usepackage{bm}
\usepackage{multirow}
\usepackage{subfig}
\usepackage{tabularx}
\usepackage{float}

\newcommand{\fgas}{f_{\mathrm{gas}\,}}

\title{Testing cosmic acceleration for $w(z)$ parameterizations
using $\fgas$ measurements in galaxy clusters}
\author[Maga\~na, Motta, C\'ardenas, Fo\"ex]{Juan Maga\~na$^{1}$\thanks{juan.magana@uv.cl},
V. Motta$^{1}$\thanks{veronica.motta@uv.cl},
V\'ictor H. C\'ardenas$^{1}$\thanks{victor.cardenas@uv.cl}, and
G. Fo\"ex$^{1,2}$\thanks{gfoex@mpe.mpg.de}
\\$^{1}$Instituto de F\'{\i}sica y Astronom\'ia, Facultad
de Ciencias, Universidad de Valpara\'iso, Avda. Gran Breta\~na 1111,
Valpara\'iso, Chile.\\$^{2}$Max Planck Institute for Extraterrestrial Physics, Giessenbachstrasse, 85748 Garching, Germany
}

\begin{document}

\date{\today}

\maketitle

\begin{abstract}
In this paper we study the cosmic acceleration for five dynamical dark energy models whose equation of state varies 
with redshift. The cosmological parameters of these models are constrained by performing a MCMC 
analysis using mainly gas mass fraction, $\fgas$, measurements in two samples of galaxy clusters:
one reported by Allen et al. (2004), which consists of $42$ points spanning the redshift range $0.05<z<1.1$,
and the other by Hasselfield et al. (2013) from the Atacama Cosmology Telescope survey, 
which consists of $91$ data points in the redshift range $0.118 < \mathrm{z} < 1.36$. 
In addition, we perform a joint analysis with the measurements of the Hubble parameter $H(z)$, 
baryon acoustic oscillations and the cosmic microwave background radiation from WMAP 
and Planck measurements to estimate the equation of state parameters. We obtained that both $\fgas$ samples
provide consistent constraints on the cosmological parameters. 
We found that the $\fgas$ data is consistent at the $2\sigma$ confidence level with a cosmic slowing down 
of the acceleration at late times for most of the parameterizations.
The constraints of the joint analysis using WMAP and Planck measurements show that this trend disappears.
We have confirmed that the $\fgas$ probe provides competitive constraints on the dark energy parameters
when a $w(z)$ is assumed.
 \end{abstract}


\section{Introduction}
The standard cosmological paradigm states that the Universe evolves
from a decelerated to an accelerated phase at late times. The
evidence of this cosmic acceleration (CA) comes not only from type Ia
supernova (SNIa) \citep{Perlmutter:1999,Riess:1998}, but also from several other cosmological
observations \citep{Mortonson:2014,Davis:2014, Li:2013,Albrecht:2006}. 

There are two approaches to explain this feature of the Universe:
one is to modify the right hand side of Einstein's equation by
assuming the existence of an energy source, dubbed dark energy (DE), that produces these
effects, and the other is to modify the left hand side by modifying
the gravity theory \citep{Joyce:2016}. The main property of the DE source that generates the
CA is an equation of state (EoS) parameter $w$, which is the
ratio between its pressure and energy density. When $w=-1$ we obtain
the well-known cosmological constant, $\Lambda$.

The $\Lambda$ cold dark matter ($\Lambda$CDM) model satisfactorily explains
the CA. However it presents several unsolved problems, i.e., the ∼120 orders of magnitude between
the quantum field theory prediction and the cosmological measurements and
why the DE density is similar to that of dark matter (DM) today
\citep{Weinberg:89,cop06}.
Additionally, the increase in number and precision of cosmological observations
has allowed confronting and explore beyond the standard $\Lambda$CDM
\citep{DiValentino:2015ola,Gong:2013,magana:2015,Marra:2013rba,PlanckXVI}. 

One of the methods used to explore departures from the $\Lambda$CDM
model is to assume a redshift dependent EoS
parameter $w(z)$, given by a suitable parameterization. The most used
parameterization for $w(z)$ is the Chevallier-Polarski-Linder (CPL)
ansatz \citep{Chevallier:2000qy, Linder:2003nc} that expands
$w(z)=w_{0}+ w_{1}z/(1+z)$, where $w_{0}$ is the present value of
the EoS and $w_{1}$ is its derivative with respect to redshift.

An intriguing feature emerges once the CPL function for $w(z)$ is used
with SNIa. \citet{shafi2009} found evidence of a low redshift
transition of the reconstructed deceleration parameter $q(z)$, the
so called slowing down of the CA, that shows a turn
around at $z\sim0.3$ \citep[see also][]{Guimaraes:2011}. We have found that such a feature does not only
appear with SNIa but also with gas mass fraction ($\fgas$) data in clusters \citep{victor_fgas}.

Furthermore, another curious behavior emerges when the high
redshift data sets are taken into account (as cosmic microwave
background radiation, CMB, or baryon acoustic oscillations, BAO):
the transition feature at low redshift disappears, making it consistent with the $\Lambda$CDM model 
\citep{shafi2009, Li:2011, victor_rivera,victor_fgas}.
Then, the key point is to separate the low redshift and high redshift data
to study this effect.

\citet{magana:2014} investigated this trend in 
$q(z)$ for five different EoS parameterizations using also
four SNIa data sets; the Constitution \citep{Hicken:2009dk}, Union $2$ \citep{Union2}, 
Union $2.1$ \citep{Susuki:2012} and Lick Observatory Supernova Search (LOSS) \citep{Ganeshalingam:2013mia} SNIa samples. 
They found that, using only SNIa data, the acceleration of the Universe seems to have already peaked and it is
evolving towards lower rates of acceleration. This result is
independent of the EoS parameterization used and it is observed in
the Constitution, Union $2$ and LOSS data sets. However,
this behavior disappears when the Union $2.1$ sample is used. As in
previous studies, if the cosmological observations at high redshift
are included in the analysis, the results change $q(z)$
from a decelerated to an accelerated phase, very similar to
the case of the cosmological constant.

Recently, \citet{Hu:2015} performed a similar analysis for several $w(z)$ functions 
using the latest cosmological data from SNIa, CMB and BAO. As in \citet{magana:2014}, the authors confirm that
the evolution of CA is independent of the EoS parameterization. In addition, they found that 
the Legacy Survey \citep[SNLS$3$,][]{Conley:2011} SNIa data favors the slowing down of the CA 
while the Joint Light-Curve Analysis \citep[JLA,][]{Betoule:2014} sample prefers an eternal CA. 
They also shown that the effects of different BAO data on CA are negligible.
By revisiting the role of the spatial curvature, \citet{Hu:2015} found that a non-flat Universe 
prefers a slowing down of the CA. This result is consistent with those obtained by \citet{victor_rivera}.
An interesting result is that the Planck 2015 data \citep{Planck2015:XIII,Planck2015:XIV} 
favored the slowing down of the CA, while in previous works, 
this trend disappears when the CMB data is taken into account in the analysis.

Here we are interested in testing the transition in the cosmic acceleration as well as the possibility of 
the slowing down in the acceleration for dynamical DE with a $w(z)$ parameterization using galaxy clusters
observations. Several authors have addressed that the gas mass fraction measurements in galaxy clusters can be used
to put constraints on cosmological models \citep{Sasaki:1996, Allen:2004cd, Bonamente:2006, Allen:2007ue,LaRoque:2006, Ettori:2009,Mantz:2014}.
Therefore, $\fgas$ data constitute a complementary geometric method
to constrain the parameter space of the same EoS parameterizations
studied in \citet{magana:2014}.

The paper is organized as follows: in the next section we introduce
the cosmological framework for a flat Universe case. In section \S
\ref{sec:models} we present the $w(z)$ parameterizations. In section
\S \ref{sec:data} we describe the methodology and data used to constrain the parameters
of the models. In section \S \ref{sec:results} we present and discuss the results. Finally, 
we present our conclusions in section \S \ref{sec:conclusions}.

\section{Cosmological framework}
We consider a flat Friedmann-Lema\^itre-Robertson-Walker (FLRW)
Universe with scale factor $a$ and whose DE component has a dynamical EoS $w(z)$. The dimensionless
Hubble parameter $E(z)$ for this Universe is given by
\begin{equation}
 E^{2}(z)\equiv H^{2}(z)/H^{2}_{0}=\Omega_{m}(1+z)^{3} + \Omega_{r}(1+z)^{4} +\Omega_{de}X(z),\quad
 \label{eq:Ez}
\end{equation}
where $H_{0}=H(0)=h\times 100\,\mathrm{km s}^{-1}\mathrm{Mpc}^{-1}$, $\Omega_{m}$ and $\Omega_{r}$ are the density parameter at present day for matter and
radiation respectively. We compute $\Omega_{r}=2.469\times10^{-5}h^{-2}(1+0.2271 N_{eff})$, where $N_{eff}=3.04$ is
the standard number of relativistic species \citep{Komatsu:2011}.
The density parameter for DE is written as
$\Omega_{de}=1-\Omega_{m}-\Omega_{r}$, and the function $X(z)$ reads as
\begin{equation}
X(z)\equiv \frac{\rho_{de}(z)}{\rho_{de}(0)}=
\mathrm{exp}\left(3\int^{z}_{0}\frac{1+w(z)}{1+z}\mathrm{dz}\right),
\label{eq:fz}
\end{equation}
where $\rho_{de}(z)$ is the energy density of DE at redshift $z$, and
$\rho_{de}(0)$ its present value. The comoving distance from the
observer to redshift $z$ is given by
\begin{equation}
r(z)=\frac{c}{H_0}\int_0^z \frac{dz'}{E(z')},
\label{eq:rz}
\end{equation}
and it is related to the angular diameter distance as
\begin{equation}
D_{A} (z) =  \frac{r(z)}{(1+z)}.
\label{eq:dA}
\end{equation}

%
%
Since we are interested in testing the CA for dynamical DE models,
we study the deceleration parameter \textit{q(z)} defined as
\begin{equation}
q(z) = - \frac{\ddot{a}(z)a(z)}{\dot{a}^{2}(z)},
\label{eq:qa}
\end{equation}
where the dot stands for the derivative with respect to the cosmic time. 
Using eq. (\ref{eq:Ez}), this expression can be rewritten as
\begin{equation}
q(z) = \frac{(1+z)}{E(z)} \frac{dE(z)}{dz}-1.
\label{eq:qz}
\end{equation}

\section{\lowercase{$w(z)$} parameterizations} \label{sec:models}
\begin{table*}
\centering
\begin{tabular}{lll}
\hline\hline\\[-0.3cm]
Parameterization & $w(\infty)$ &  $X(z)$\\ \hline \\
JBP & $w_{0}$ & $(1+z)^{3(1+w_{0})}\mathrm{exp}\left[\frac{3}{2}\frac{w_{1}z^{2}}{(1+z)^{2}}\right]$\\[0.3cm]
BA & $w_{0}+w_{1}$ & $(1+z)^{3(1+w_0)}(1+z^2)^{\frac{3}{2}w_{1}}$ \\[0.3cm]
FSLL I & $w_{0}$ & $(1+z)^{3(1+w_{0})}\mathrm{exp}\left[\frac{3w_{1}}{2}\mathrm{arctan(z)}\right]
\left(1+z^{2}\right)^{\frac{3}{4}w_{1}}\left(1+z\right)^{-\frac{3}{2}w_{1}}$\\[0.3cm]
FSLL II& $w_{0}+w_{1}$ & $(1+z)^{3(1+w_{0})}\mathrm{exp}\left[- \frac{3w_{1}}{2}\mathrm{arctan(z)}\right]
\left(1+z^{2}\right)^{\frac{3}{4}w_{1}}\left(1+z\right)^{\frac{3}{2}w_{1}}$\\[0.3cm]
SL  & $\frac{1}{2} (-1-8 w_{0}+9 w_{0.5})$ & $(1+z)^{\frac{3}{2}(1-8w_{0}+9w_{0.5})}
\mathrm{exp}\left[\frac{3z\left\{w_{0}(52z+40)-9w_{0.5}(5z+4)+7z+4\right\}}{8(1+z)^2}\right]$\\[0.3cm]
\hline \hline
\end{tabular}
\caption{Properties for each $w(z)$ parameterization. The second column shows
the amplitude of $w(z)$ when $z\rightarrow\infty$. The third column shows the function
$X(z)$ given by the Eq. (\ref{eq:fz})}
\label{tab:functions}
\end{table*}

One natural extension to the cosmological constant is to explore dynamical DE where
the EoS parameter $w(z)$ varies with redshift through an explicit parameterization. Here
we consider the following five parameterizations: Jassal-Bagla-Padmanabhan
\citep[JBP,][]{Jassal:2005a, Jassal:2005b}, Barbosa-Alcaniz \citep[BA,][]{Barboza:2008rh},
Feng-Shen-Li-Li \citep[FSLL,][]{Feng:2012gf}, and
Sendra-Lazkoz \citep[SL,][]{Sendra:2011pt}

\begin{eqnarray}
w(z)&=&w_{0} + w_{1}\frac{z}{\left(1+z\right)^{2}} \quad \mbox{JBP},\\
w(z)&=&w_0 + w_1 \frac{z(1+z)}{1+z^2} \quad \mbox{BA},\\
w(z)&=&w_{0} + w_{1}\frac{z}{1+z^{2}} \qquad \mbox{FSLL I},\\
w(z)&=&w_{0} + w_{1}\frac{z^{2}}{1+z^{2}} \qquad \mbox{FSLL II},\\
w(z)&=&-1 + c_{1}\left(\frac{1+2z}{1+z}\right) + c_{2}\left(\frac{1+2z}{1+z}\right)^{2} \quad \mbox{SL},
\end{eqnarray}
where $w_{0}=w(0)$ for all parameterizations, $w_{1}=w'(0)$ ($'$ denotes derivative with respect to redshift), 
the constant $c_{1}=(16w_{0}-9w_{0.5}+7)/4$, and $c_{2}=-3w_{0}+ (9w_{0.5}-3)/4$, being
$w_{0.5}$ the value of the EoS at $z=0.5$. Table \ref{tab:functions} shows
the amplitude of the EoS when $z\rightarrow \infty$
and the function $X(z)$ given by Eq. (\ref{eq:fz}) for each parameterization.

\section{Methodology and data} \label{sec:data}
The EoS parameters for each $w(z)$ function are constrained by performing
a Markov Chain Monte Carlo (MCMC) analysis using the following
cosmological observations: the $\fgas$ measurements in galaxy clusters,
$H(z)$ estimations, BAO data, and the CMB information.

\subsection{The gas mass fraction in galaxy clusters}
\subsubsection{$\fgas$ modeling data}
The gas mass fraction is defined as $\fgas\equiv M_{gas}/M_{tot}$,
where $M_{gas}$ is the X-ray gas mass and $M_{tot}$ is the dynamical total mass of the galaxy cluster. 
The $\fgas$ is useful as a cosmological probe when a scaling relation between
the baryonic gas mass fraction in galaxy clusters and the global fraction of baryonic matter
and dark matter of the Universe $\Omega_{b}/\Omega_{0m}$ is assumed.
The $\fgas$ estimations for any cosmological model fitted to the reference $\Lambda$CDM data can be obtained
by writing $M_{gas}$ and $M_{tot}$ in terms of $D_{A}(z)$ \citep{Sasaki:1996, Nesseris:2006er}:

\begin{eqnarray}
\fgas(z) \equiv
\frac{b}{1+\alpha}\frac{\Omega_b}{\Omega_{m}}\left(\frac{D_{A}^{\Lambda
CDM} (z)}{D_A(z)}\right)^{3/2}, \label{eq:one}
\end{eqnarray}

\noindent where ${D_{A}}(z)$ is the angular diameter distance for any cosmological model,
$D_{A}^{\Lambda CDM}(z)$ is the angular diameter distance for the $\Lambda$CDM reference model,
$\Omega_{b}$ is the baryonic density parameter, and $\Omega_{m}$
is the current DM density parameter.
Here $b$ is a bias factor which relates the baryonic fraction in clusters
with the one in the Universe. The constant $\alpha$ relates the
baryonic luminous mass and the baryonic gas mass.
This constant is $\alpha\approx 0.19 \sqrt{h}$ \citep{Allen:2004cd}.
Hereafter, we refer to the fitting function given by the Eq. (\ref{eq:one}) as A04.
From here on, we assume that the gas mass fraction measurements are uncorrelated.
To constrain the parameters of the $w(z)$ functions introduced in \S $3$,
we use the following chi-square in the MCMC analysis
\begin{eqnarray}
\chi_{\fgas}^{2} &=& \left(\sum_{i=1}^{N}\frac{[\fgas(z_i)-\fgas^{obs}(z_i)]^2}{\sigma_{f_{gas,i}}^2}\right)
+\left(\frac{h-h_{,prior}}{\sigma_{h,prior}}\right)^{2}\nonumber\\
&+& \left(\frac{\Omega_b h^{2}-\Omega_b h^{2}_{,prior}}{\sigma_{\Omega_b h^{2},prior}}\right)^{2}
+\left(\frac{b-0.824}{0.089}\right)^{2},
\label{eq:chifgasa04}
\end{eqnarray}
where $\fgas(z)$ is calculated using A04, $\fgas^{obs}$ is the observed gas mass fraction, 
and $\sigma_{\fgas}$ is the error in the measurements.
$N=42$ ($91$) for the Allen (ACT) data set. Table \ref{tab:priors}
gives the adopted priors for $h$ and $\Omega_b h^{2}$ 
from WMAP \citep{Hinshaw:2013} and Planck measurements \citep{PlanckXVI}.

\begin{table}
\centering
\begin{tabular}{|cc|}
\hline
Parameter&Allowance\\
\hline
\multicolumn{2}{|c|}{}\\
$h$& $0.73\pm 0.0175$ (Gaussian)\\
$\Omega_b h^{2}$ & $0.02202\pm 0.00046$ (Gaussian)\\
$\Omega_{m}$ & $[0.2,0.4]$ (Uniform) \\
$w_{0}$ & $[-2,0]$ (Uniform)\\
$w_{1} (w_{0.5})$ & $[-10,2]$ (Uniform)\\
\hline
\multicolumn{2}{|c|}{A04}\\
$b$&$0.824\pm0.089$ (Gaussian)\\
\hline
\multicolumn{2}{|c|}{A08}\\
$\eta$&$0.214\pm0.022$ (Gaussian)\\
$\gamma$&$1.0<\gamma<1.1$ (Uniform)\\
$s_{0}/h_{70}^{0.5}$&$0.16\pm0.048$ (Gaussian)\\
$s_{1}$&$-0.2<s_{1}<0.2$ (Uniform)\\
$b_{0}$&$0.65<b_{0}<1.0$ (Uniform) \\
$b_{1}$&$-0.1<b_{1}<0.1$ (Uniform) \\
$K$ & $1.0\pm0.1$ (Gaussian)\\
\hline

\end{tabular}
\caption{Priors on $h$ \citep{Riess:2016}, $\Omega_b h^{2}$ \citep{Cooke:2014}, and the different parameters 
of the fitting functions A04 \citep[Eq. \ref{eq:one},][]{Allen:2004cd} and A08 \citep[Eq. \ref{eq:fgasA08},][]{Allen:2007ue}}
\label{tab:priors}
\end{table}

\citet{Allen:2007ue} modified the Eq. (\ref{eq:one}) to include more corrections related 
to the cluster physics and the cosmological model as:
\begin{equation}
\fgas(z) = \frac{ K A \gamma b(z)} {1+s(z) }
\left( \frac{\Omega_{\rm b}}{\Omega_{\rm m}} \right)
\left[ \frac{D_{\rm A}^{\rm \Lambda CDM}(z)}{D_{\rm A}(z)} \right]^{1.5}.
\label{eq:fgasA08}
\end{equation}
The factor $A$ in Eq. (\ref{eq:fgasA08}) is given by
\begin{equation}
A= \left( \frac{ \theta_{2500}^{\rm \Lambda CDM}}{\theta_{2500}} \right)^\eta \approx
\left( \frac{ H(z) D_{\rm A}(z)} { \left[ H(z) D_{\rm A}(z)\right] ^{\rm \Lambda CDM}} \right)^\eta,
\end{equation}
which accounts for the change in the angle subtended by $r_{2500}$ 
\footnote{the radius within which the cluster average density is
$2500$ times the critical density.} as the underlying cosmology is varied.
Here, $\eta$ is the slope of the $\fgas$ data in
the region of $r_{2500}$ as measured for the reference $\Lambda$CDM
cosmology. The parameter $\gamma$ takes into account the non-thermal
pressure support in the clusters. The parameter $s(z)=s_{0}(1+s_{1}z$)
models the baryonic mass fraction in stars as function of redshift.
The bias factor also depends on the redshift as $b(z)=b_{0}(1+b_{1}z)$. $K$
parameterizes the residual uncertainty in the accuracy of the instrument
calibration and X-ray modelling. Hereafter, we refer to $\fgas$ function given by the Eq. (\ref{eq:fgasA08}) as A08.
We also consider the fitting function $A08$ to estimate 
(Table \ref{tab:priors} lists the adopted priors for $A08$) the EoS parameters using 
\begin{eqnarray}
\chi_{\fgas}^{2} &=& \left(\sum_{i=1}^{N}\frac{[\fgas(z_i)-\fgas^{obs}(z_i)]^2}{\sigma_{f_{gas,i}}^2}\right)
+\left(\frac{h-h_{,prior}}{\sigma_{h,prior}}\right)^{2}\nonumber\\
&+& \left(\frac{\Omega_b h^{2}-\Omega_b h^{2}_{,prior}}{\sigma_{\Omega_b h^{2},prior}}\right)^{2}
+\left(\frac{\eta-0.214}{0.022}\right)^{2}\nonumber\\
&+&\left(\frac{s_{0}/\sqrt{(100/70)h}-0.16}{0.048} \right)^{2}
+\left(\frac{K-1.0}{0.1} \right)^{2}.\qquad
\label{eq:chifgasa08}
\end{eqnarray}

\subsubsection{$\fgas$ samples}
We use two galaxy cluster gas mass fraction samples: the sample by \citet{Allen:2004cd}, $\fgas^{Allen}$, 
which consists of $42$ points spanning the redshift range $0.05<z<1.1$
and the ACT sample, $\fgas^{ACT}$, which consists of $91$ data points in the redshift range
$0.118 < \mbox{z} < 1.36$. While \citet{Allen:2004cd} provide the $\fgas$ measurements, 
the ACT sample gives the mass \footnote{The mass $M_{500}$ is defined as the one measured within 
the radius $R_{500}$ at which the enclosed mean density is $500$ times the critical density at the cluster redshift.} 
$M_{500}$ calculated using the one-parameter family of universal pressure profiles 
for the cluster gas \citep{Hasselfield:2013}. 
The relation between $M_{500}$ and $\fgas$ is
\citep[see][for more details]{Goncalves,Vikhlinin:2009}.

\begin{equation}
\fgas=(0.0764\pm 0.004)\,h^{-1.5} + (0.037\pm0.006) \log{M_{15}},
\label{eq:fgasVik}
\end{equation}
where $M_{15}$ is the cluster total mass, $M_{500}$, in units of $10^{15}h^{-1}M_{\odot}$.
Notice the computed $\fgas$ data points could be biased due to the Eq. (\ref{eq:fgasVik}), 
since it depends on a fiducial cosmology. 
In addition, the constraints also could be deteriorated due to the intrinsic scatter
in the scaling relation. In Appendix A we study the impact of considering $15\%$ error
($3\sigma$) in each parameter of the Eq. \ref{eq:fgasVik} for the JBP model.
We show that the best fits obtained are within the $1\sigma$ confidence level.

\subsection{$H(z)$ measurements}
Our Bayesian analysis also considers the measurements of the Hubble parameter
which is directly related to the expansion history of the Universe.
We used $34$ points in the redshift range $0.07\le z \le2.3$
compiled by \citet{Sharov:2014} (see also Farooq \& Ratra 2013 and Chen et al. 2013).
The data set is shown in Table \ref{tab:Hz} of the Appendix \ref{Ap:Hz}.
It is worth to notice that some $H(z)$ points might be correlated
to BAO measurements, i.e. BAO points together with the best fit value of the sound horizon 
from CMB data could be used to estimate the $H(z)$ value. Although some of the 
$H(z)$ points in the Table 3 were obtained in this manner, through this work we assume
there is no correlation between $H(z)$ and BAO.
Then, the $\chi^2_{H}$ can be written as
\begin{equation}
\chi_{H}^2 = \sum_{i=1}^{34} \frac{ \left[ H(z_{i}) -H_{obs}(z_{i})\right]^2 }{ \sigma_{H_i}^{2} },
\end{equation}
where $H_{obs}(z_{i})$ is the observational Hubble parameter at $z_{i}$,
$\sigma_{H_i}$ its error, and $H(z_{i})$ is the theoretical value for a model.


\subsection{BAOs measurements}
Another independent test to put constraints on cosmological parameters is to use the
BAO signature. The BAO measurements considered in our analysis are obtained from
the Six-degree-Field Galaxy Survey (6dFGS) BAO data \citep{Beutler2011:6dF},
the WiggleZ experiment \citep{Blake2011:wigglez},
the Sloan Digital Sky Survey (SDSS) Data Release 7 (DR7)
BAO distance measurements \citep{Percival2010:dr7}, the Baryon Oscillation Spectroscopic Survey
(BOSS) SDSS Data Release 9 BAO distance measurements (SDSS DR9) \citep{Anderson2012:dr9} and
the most recent BAO distance estimations from Data release $11$ (DR11) of the BOSS (quasars) of SDSS.
WiggleZ points are obtained using the low redshift data from SDSS and 6dFGS,
thus they might be correlated. Nevertheless, in this work we assume 
that they are not correlated.

The 6dFGS BAO estimated the distance ratio $d_{z}=0.336\pm0.015$ at $z=0.106$ \citep{Beutler2011:6dF},
where
\begin{equation}
d_{z} = \frac{r_{d}}{D_V(z)},
\label{eq:dz}
\end{equation}
where $r_{d}=r_{s}(z_{d})$ and the distance scale $D_V$ is defined as
\begin{equation}
D_V(z)=\frac{1}{H_0}\left[(1+z)^2D_A(z)^2\frac{cz}{E(z)}\right]^{1/3}.
\end{equation}
In the standard scenario $D_A(z)$ relates to the luminosity distance through
$D_A(z)=D_L(z)/(1+z)^2$.
The comoving sound horizon, $r_s(z)$, is defined as
\begin{equation}
 r_s(z) = c \int_z^\infty \frac{c_s(z')}{H(z')}dz',
 \end{equation}
where the sound speed $c_s(z) = 1/\sqrt{3\left(1+\bar{R_b}/\left(1+z\right)\right)}$, with
$\bar{R_b} = 31500\, \Omega_{b}h^2(T_{CMB}/2.7\rm{K})^{-4}$, and
$T_{CMB}$ is the CMB temperature. 

The redshift $z_d$ at the baryon drag epoch is well fitted with the
formula proposed by \citet{Eisenstein98},
\begin{equation}
z_d =\frac{1291(\Omega_{m}h^2)^{0.251}}{1+0.659\,(\Omega_{m}h^2)^{0.828}}[1+b_1(\Omega_b
h^2)^{b_2}],
\end{equation}
where
\begin{eqnarray}
b_1 &=& 0.313\left(\Omega_{m}\,h^2\right)^{-0.419}\left[1+0.607\left(\Omega_{m}\,h^2\right)^{0.674}\right], \\
b_2 &=& 0.238\left(\Omega_{m}\,h^2\right)^{0.223}.
\end{eqnarray}

It is worth to note that these equations were calculated for the standard 
cosmology, here we assume that they are valid for dynamical DE models.
Therefore, the chi-square for the 6dFGS BAO measurement is given by
\begin{equation}
\chi^2_{\mathrm{6dFGS}} =
\left(\frac{d_z-0.336}{0.015}\right)^2.
\end{equation}

From WiggleZ, \citet[][see also Gong et al. 2015]{Kazin:2014} estimated the following three points for $d_{z}= 
0.0870\pm0.0042$, $0.0672\pm0.0031$, and $0.0593\pm0.0020$ at redshifts $z=0.44,0.6$ and $0.73$ respectively.

Thus, the $\chi^2$ for the WiggleZ BAO data is given by
\begin{eqnarray}
\chi^2_{\mathrm{WiggleZ}} &=& \left(\frac{d_z(0.44)-0.0870}{0.0042}\right)^2\nonumber\\
&+&\left(\frac{d_z(0.6)-0.0672}{0.0031}\right)^2\nonumber\\
&+&\left(\frac{d_z(0.73)-0.0593}{0.0020}\right)^2.
\end{eqnarray}

Similarly, using the clustering of galaxies from SDSS DR7,
\citet[][see also Percival et al. 2010]{Ross:2015} obtained a consensus measurement $D_{\mathrm{V}}=(664\pm25)(r_{d}/r_{d,fid})$ Mpc
of the BAO scale at $z=0.15$. By adopting the value of the fiducial cosmology $r_{d,fid}$
and inverting, \citet{Gong:2015} estimate $d_{z}$ and the $\chi^2$ can be expressed as
\begin{equation}
\chi^2_{\mathrm{DR7}}=\left(\frac{d_z(0.15)-0.2239}{0.0084}\right)^2, 
\end{equation}

\citet[][see also Anderson et. al. 2012]{Anderson2014b} measure
the BAO signature from the SDSS-III BOSS DR11 and give
the consensus estimations for the distance $D_{v}=(1264\pm25\mathrm{Mpc})(r_{d}/r_{d,fid})$ at redshift $z=0.32$
and $D_{v}=(2056\pm20\mathrm{Mpc})(r_{d}/r_{d,fid})$ at redshift $z=0.57$. Using the 
$r_{d,fid}$ value given by the authors and inverting to obtain $d_{z}$, the
$\chi^2$ is written as
\begin{equation}
\chi^2_{\mathrm{DR11a}}=\left(\frac{d_z(0.32)-0.1181}{0.0023}\right)^2+\left(\frac{d_z(0.57)-0.0726}{0.0007}\right)^2,
\end{equation}

The most recent measured position of the BAO peak from SDSS DR11 determines
$D_{H}/r_{d}=9.18\pm0.28$ at $z=2.34$ \citep{Delubac2014}, and $D_{H}/r_{d}=9.0\pm0.3$ at $z=2.36$ \citep{Font-Ribera:2014}
where $D_{H}(z)=c/H(z)$. Thus, we compute the $\chi^2$ for these points as

\begin{equation}
\chi^2_{\mathrm{DR11b}}=\left(\frac{\frac{D_H(2.34)}{r_{d}}-9.18}{0.28}\right)^2+
\left(\frac{\frac{D_H(2.36)}{r_{d}}-9.00}{0.3}\right)^2.
\end{equation}

The total $\chi^2$ for all the BAO data points can be written as
\begin{equation}
\chi^2_{\mathrm{BAO}} = \chi^2_{\mathrm{6dFGS}}+\chi^2_{\mathrm{ WiggleZ}} +
\chi^2_{\mathrm{DR7}} + \chi^2_{\mathrm{DR11a}} + \chi^2_{\mathrm{DR11b}}.
\end{equation}

\subsection{CMB}
We also include CMB information by using the following distance posteriors: the acoustic scale, $l_{A}$,
the shift parameter, $R$, and the decoupling redshift, $z_{*}$. These quantities can be used to constrain 
cosmological parameters without considering the full CMB measurements \citep{Komatsu:2009,Komatsu:2011,wang:2006,Wright:2007}.
Although these distance posterior are computed assuming an
underlying cosmology, several authors have proved that these
quantities are almost independent on the input DE models \citep{Li:2008,Mukherjee:2008,wang:2012,Planck2015:XIV}. 

The acoustic scale is defined as
\begin{equation}
l_A = \frac{\pi r(z_*)}{r_s(z_*)},
\label{eq:lA}
\end{equation}
where the redshift of decoupling $z_*$ is given by \citep{Hu:1996},
\begin{equation}
z_* = 1048[1+0.00124(\Omega_b h^2)^{-0.738}]
[1+g_1(\Omega_{m}h^2)^{g_2}],
\end{equation}
and
\begin{eqnarray}
g_1 &=& \frac{0.0783(\Omega_b h^2)^{-0.238}}{1+39.5(\Omega_b h^2)^{0.763}},\nonumber\\
g_2 &=& \frac{0.560}{1+21.1(\Omega_b h^2)^{1.81}}.
\end{eqnarray}
It is worth to note that we assume that these fitting
formulae are valid in dynamical DE models.
The shift parameter is defined as \citep{Bond:1997}
\begin{equation}
R = \frac{\sqrt{\Omega_{m}H_{0}^2}}{c} r(z_{*}).
\end{equation}

Thus, the $\chi^2$ for the CMB data is constructed as
\begin{equation}\label{cmbchi}
 \chi^2_{\mathrm{CMB}} = X^T\,\mathrm{Cov}_{\mathrm{CMB}}^{-1}\,X,
\end{equation}
where $\mathrm{Cov}_{\mathrm{CMB}}^{-1}$ is the inverse covariance matrix and
\begin{equation}
 X =\left(
 \begin{array}{c}
 l_A^{th} - l_A^{obs} \\
 R^{th} -  R^{obs}\\
 z_*^{th} - z_{*}^{obs}
\end{array}\right),
\end{equation}
the superscripts $th$ and $obs$ refer to the theoretical and observational estimations respectively.

\citet{Hinshaw:2013} obtained for a flat $\Lambda$CDM Universe, using the WMAP 9-yr temperature power spectrum,
the following distance posteriors $l_{A}^{obs}=302.40$, $R^{obs}=1.7246$, $z_{*}^{obs}=1090.88$, and
the inverse covariance matrix

\begin{equation}
\mathrm{Cov}^{-1}_{\mathrm{WMAP9}} = \left(
\begin{array}{ccc}
3.182 & 18.253 & -1.419\\
18.253 & 11887.879 & -193.808\\
-1.429 & -193.808 & 4.556
\end{array}\right).
\label{eq:Cwmap9}
\end{equation}

On the other hand, for a flat $w$CDM, \citet{Neveu:2016} estimated from Planck measurements: 
$l_{A}^{obs}=301.787\pm0.089$, $R^{obs}=1.7492\pm0.0049$, 
$z_{*}^{obs}=1089.99\pm0.29$. 
They also provide the following the inverse covariance matrix, $\mathrm{Cov}^{-1}_{Pl}$, of these quantities
\begin{equation}
 \mathrm{Cov}^{-1}_{\mathrm{Pl}}= \left(
\begin{array}{ccc}
162.48 & -1529.4 & 2.0688 \\
 -1529.4 & 207232 & -2866.8 \\
2.0688 & -2866.8 & 53.572 \\
\end{array}
\right).
\end{equation}

\section{Results and discussion}
\label{sec:results}
In all our estimations, we use Gaussian priors on $h$ \citep{Riess:2016}, 
and $\Omega_{b}h^{2}$ \citep{Cooke:2014}, which are derived from measurements
independent from the CMB and useful in dynamical DE models, 
thus leaving $\Omega_{m}$, $w_{0}$, 
and $w_{1}$ ($w_{0.5}$ for the SL model) as the only free parameters of the analysis (see Table \ref{tab:priors}).
To calculate these constraints we assume a Gaussian likelihood $\mathcal{L}\propto \exp(-\chi^{2}/2)$. First, 
we obtain the constraints using the measurements of $\fgas$ alone. As mentioned before, we consider 
the $\fgas^{Allen}$ or the $\fgas^{ACT}$ sample. We also estimated
the parameters from the combination of $H(z)+$BAO$+$CMB (WMAP or Planck)
and the joint analysis of $\fgas+H(z)+$BAO$+$CMB.
Thus, with the $\chi^{2}$ functions defined in section \S \ref{sec:data} for each data, 
we construct the total $\chi^{2}$-function for these three cases:
$\chi^{2}=\chi^{2}_{\fgas}$, $\chi^{2}=\chi^{2}_{H}+\chi^{2}_{\mathrm{BAO}}+\chi^{2}_{\mathrm{CMB}}$, 
and $\chi^{2}=\chi^{2}_{\fgas}+\chi^{2}_{H}+\chi^{2}_{\mathrm{BAO}}+\chi^{2}_{\mathrm{CMB}}$.
We perform our Bayesian analysis using the \emph{emcee} Python module \citep{emcee}.
In all our computations, we consider a total of $6500$ steps with $500$ walkers 
and $1000$ steps were removed which correspond to the burn-in phase to stabilize the estimations. 
We start the walkers in a small ball around the expected points of maximum probability estimated with a differential evolution
method. To judge the convergence of the
sampler we ask that the acceptance fraction is in the $[0.2-0.5]$ range and check the
autocorrelation time which is found to be $\mathcal{O}(60)$ and $\mathcal{O}(80)$ for the A04 and 
A08 fitting respectively.

\subsection{Impact of $\fgas$ fitting function}
\begin{figure}
\centering
\includegraphics[width=8cm,height=8cm]{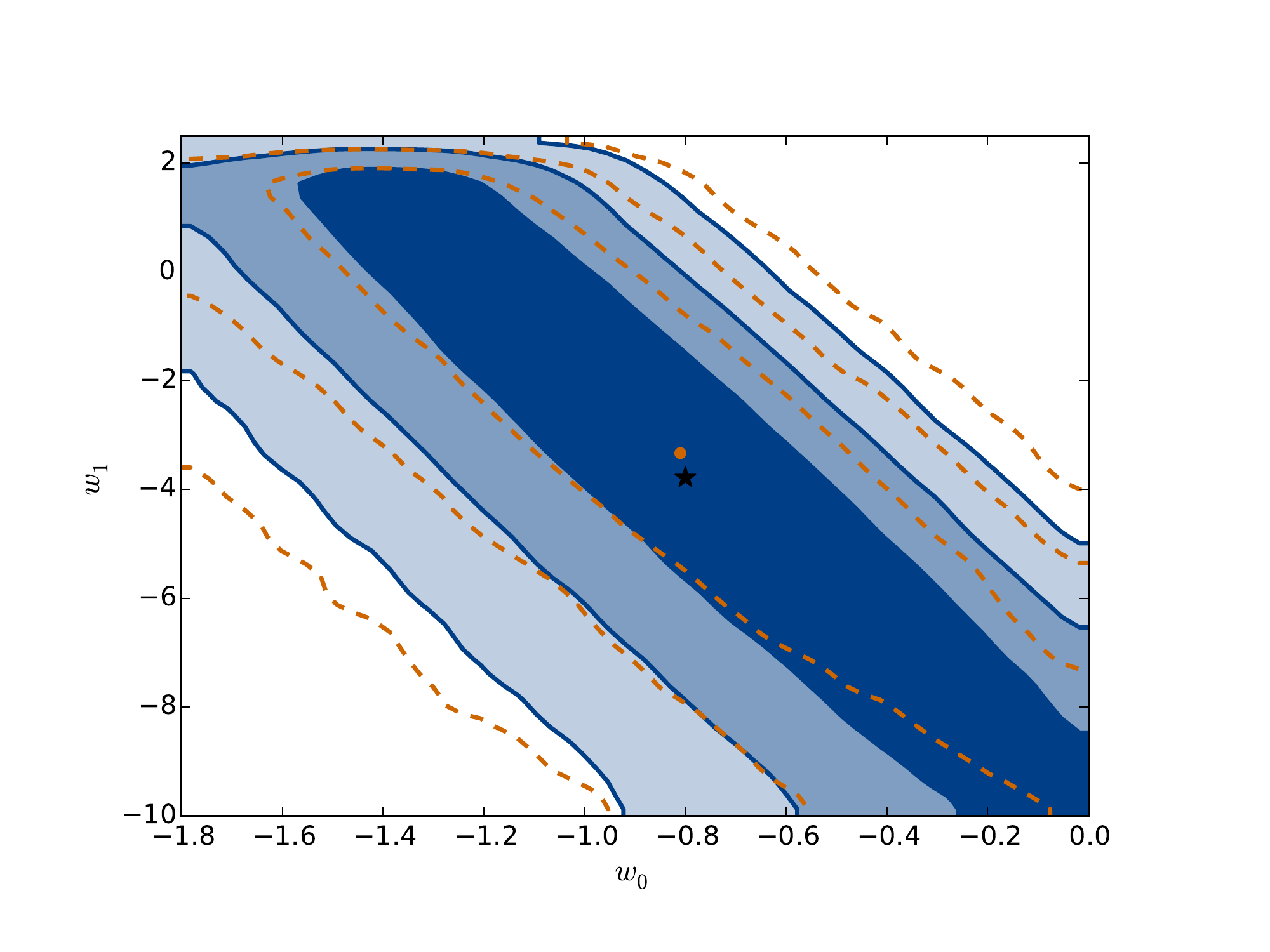}
\caption{Comparison of the $w_{0}-w_{1}$ confidence contours for the JBP parameterization 
using the $\fgas^{\mathrm{Allen}}$ data alone with the A$04$ and A$08$ equations. 
The blue filled contours show the $68\%$, $95\%$ and $99\%$ confidence levels
obtained by considering Eq. (\ref{eq:one}), while the orange dashed contours show the same by considering
Eq. (\ref{eq:fgasA08}). The star and the dot marks represent the best fit obtained using 
each one of these $\fgas$ fitting function respectively.
Observe that there is no significative statistically difference in the $w_{0}$ and $w_{1}$ bounds.
We found the same result for the other $w(z)$ functions}
\label{fig:comparison}
\end{figure}

For all $w(z)$ parameterizations we estimated the cosmological parameters using the chi-squares functions for
both A$04$ and A$08$ $\fgas$ fitting equations. The first one (Eq. \ref{eq:one}) assumes 
that the baryonic fraction from clusters does not depend on redshift, while the second one (Eq. \ref{eq:fgasA08}) assumes a linear dependence 
for $s$ and $b$ parameters with redshift. In addition, the A$08$ equation considers more correction
factors due to the underlying cosmological model and the cluster physics. 
Fig. \ref{fig:comparison} shows that there is no significative difference in the $w_{0}$ and $w_{1}$
constraints for the JBP parameterization derived from $\fgas^{\mathrm{Allen}}$ sample when A$04$ or A$08$ function are used.
When the $\fgas^{\mathrm{ACT}}$ sample is used a similar result is found, 
and the $q(z)$ parameter trend is the same for both A$04$ or A$08$ functions (see Appendix B for more details). 
In addition, we found the same result for the other $w(z)$ functions.
Thus, in what follows, we only present the results obtained using the A$04$ equation. 
For completeness, Tables \ref{tab:jbp}-C5 (Tables C4-C5 are available online only)
give the best fits for $\Omega_{m}$, $w_{0}$ and $w_{1}$, by considering the A$08$ fitting function 
(Eq. \ref{eq:fgasA08}) in the parameter estimation. 

\subsection{Impact of $\fgas$ measurements on the parameter estimation}
\begin{figure}
\centering
\includegraphics[width=8cm,height=8cm]{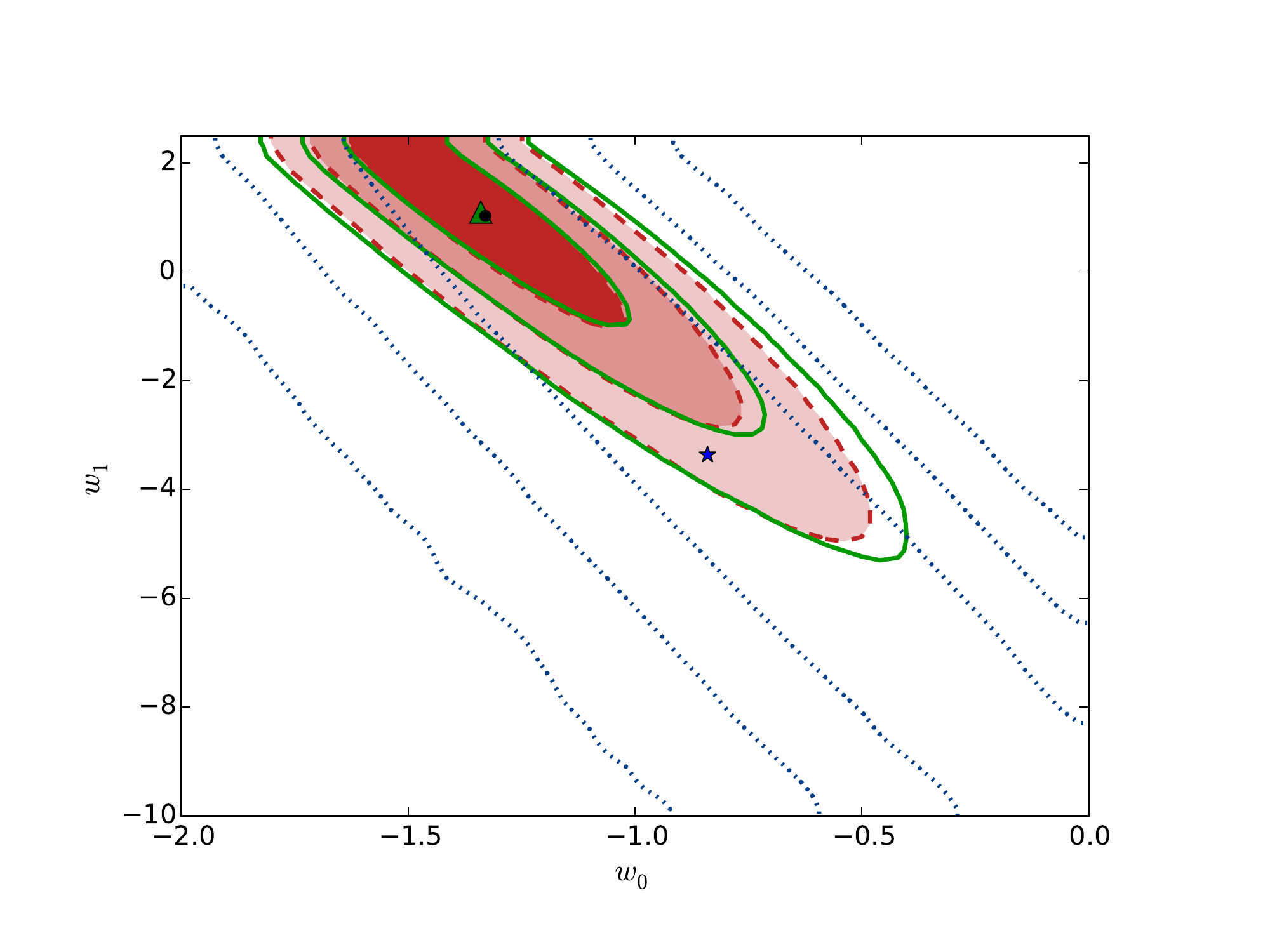}
\caption{Comparison of the $w_{0}-w_{1}$ $68\%$, $95\%$ and $99\%$ confidence levels, for the JBP model using 
$\fgas^{\mathrm{Allen}}$ (no-filled dotted blue contours), $H(z)+$BAO$+$WMAP (no-filled green contours), and 
the joint analysis $\fgas^{Allen}+H(z)+$BAO$+$WMAP (filled dashed red contours) data. 
The star, triangle, and dot represent the best fit obtained using each one 
of these data set respectively. The best fit obtained using $f_{gas}$ is consistent
at $3\sigma$ level with those obtained from $H(z)+$BAO$+$WMAP and the joint analysis.}
\label{fig:comparison_data}
\end{figure}

Tables and Figures \ref{tab:jbp}-C5 (Tables and Figures C4-C5 are available online only)
summarize our results for $\Omega_{m}$ and the 
EoS parameters for each $w(z)$ parameterization estimated with the different tests. 
Notice that, in general, the estimations on $\Omega_{m}$ derived from the $\fgas$ data
are in very good agreement to those obtained from the combination 
of $H(z)+$BAO$+$WMAP and of the joint analysis of the all data sets 
(see priors in Table \ref{tab:priors}). 
Thus we confirm that $\fgas$ measurements in galaxy clusters is a suitable probe
to put bounds on the content of dark matter in the Universe. 
For the EoS parameters, $w_{0}$ and $w_{1}$, the best fits 
show different estimations when using $\fgas$, $H(z)+$BAO$+$WMAP, and the joint analysis of all data sets. 
However, these bounds are consistent at the $3\sigma$ confidence level 
(see Fig. \ref{fig:comparison_data} for the comparison of the JBP estimations).
On the other hand, the bounds estimated with the joint analysis of the data are consistent, at $68\%$ confidence level,
with the cosmological constant, i.e., $w_{0}\approx -1$ and $w_{1}\approx 0$. 
Since our interest is to focus on the gas mass fraction probe, in what follows, we only present the results obtained using the $\fgas$ data and the 
joint analysis $\fgas+H(z)+$BAO$+$CMB. 

\subsection{Cosmic acceleration for the $w(z)$ parameterizations}

\begin{figure*}
\begin{tabular}{cc}
\includegraphics[width=0.4\textwidth]{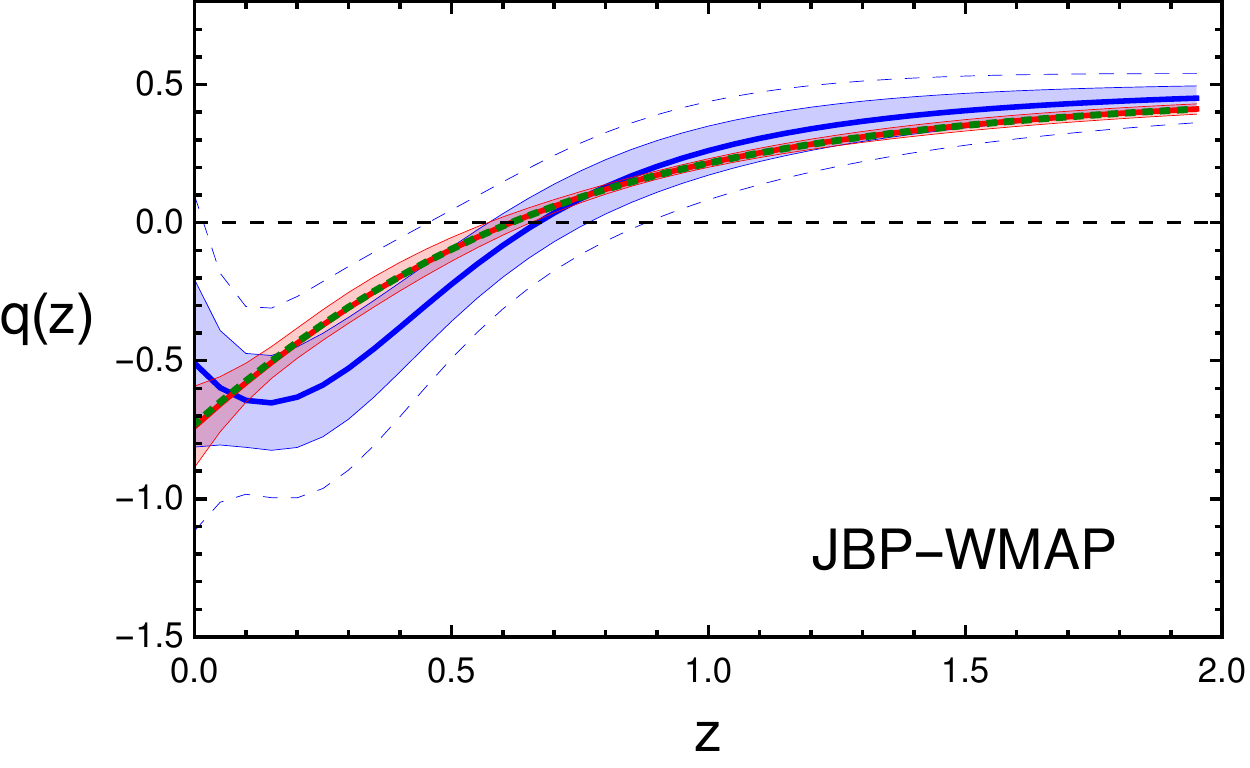} & \includegraphics[width=0.4\textwidth]{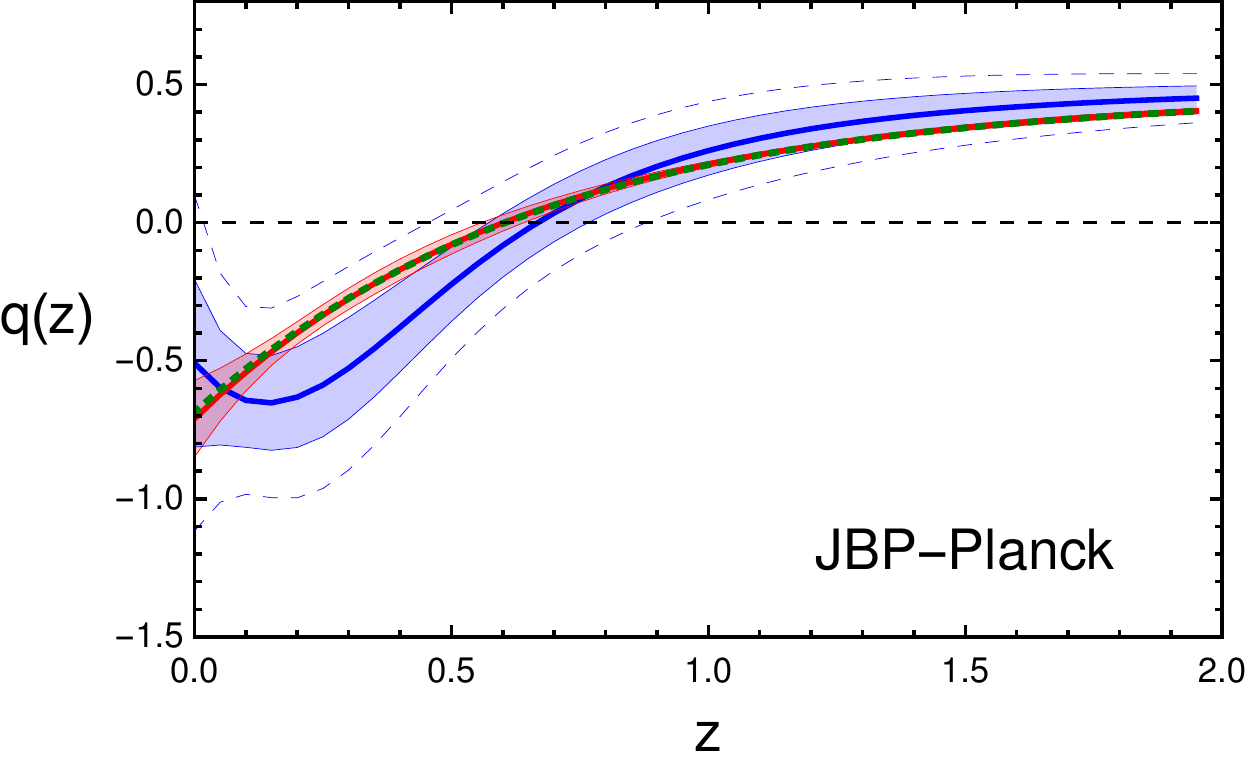}\\
\includegraphics[width=0.4\textwidth]{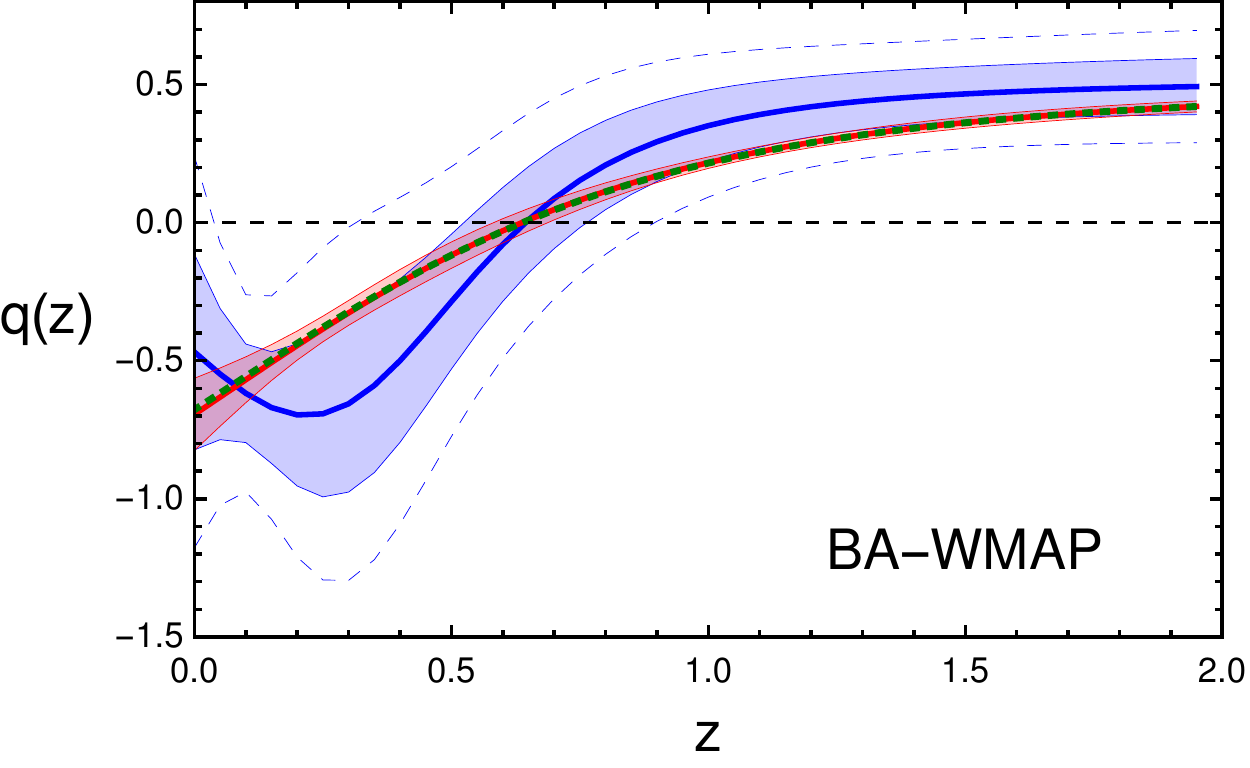} & \includegraphics[width=0.4\textwidth]{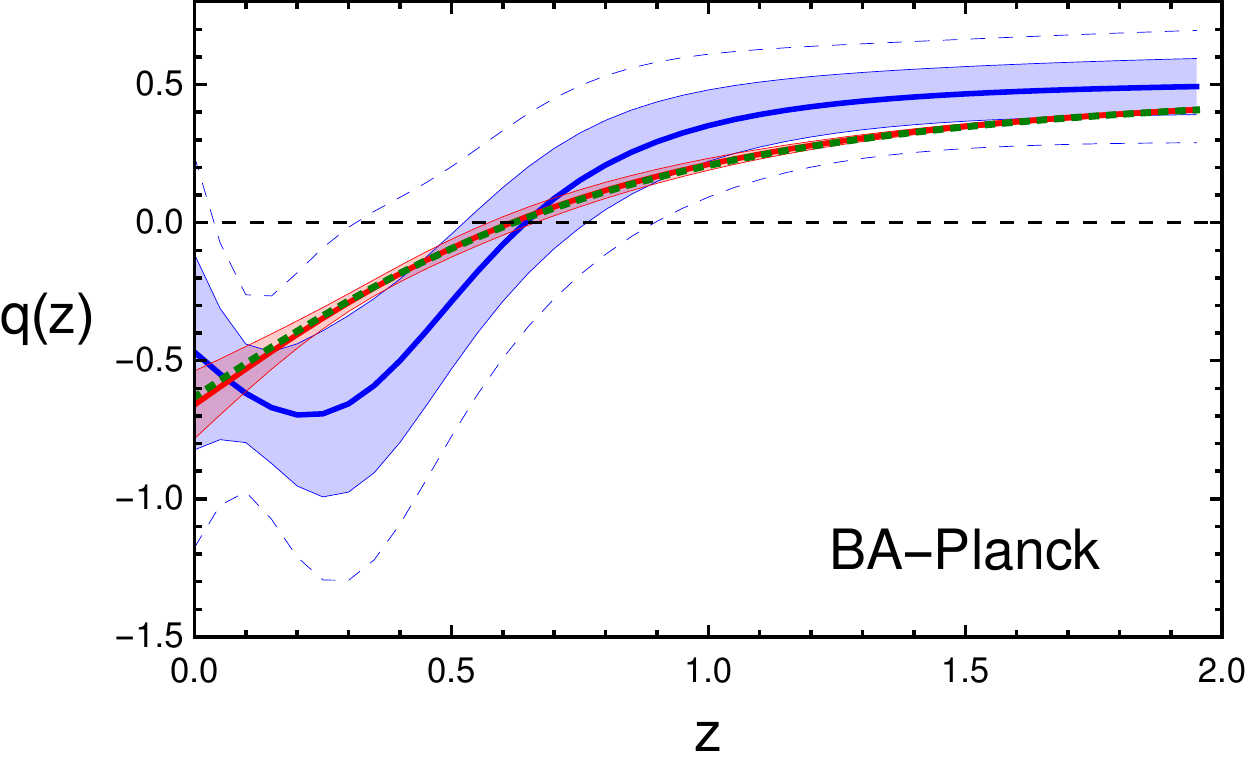}\\
\includegraphics[width=0.4\textwidth]{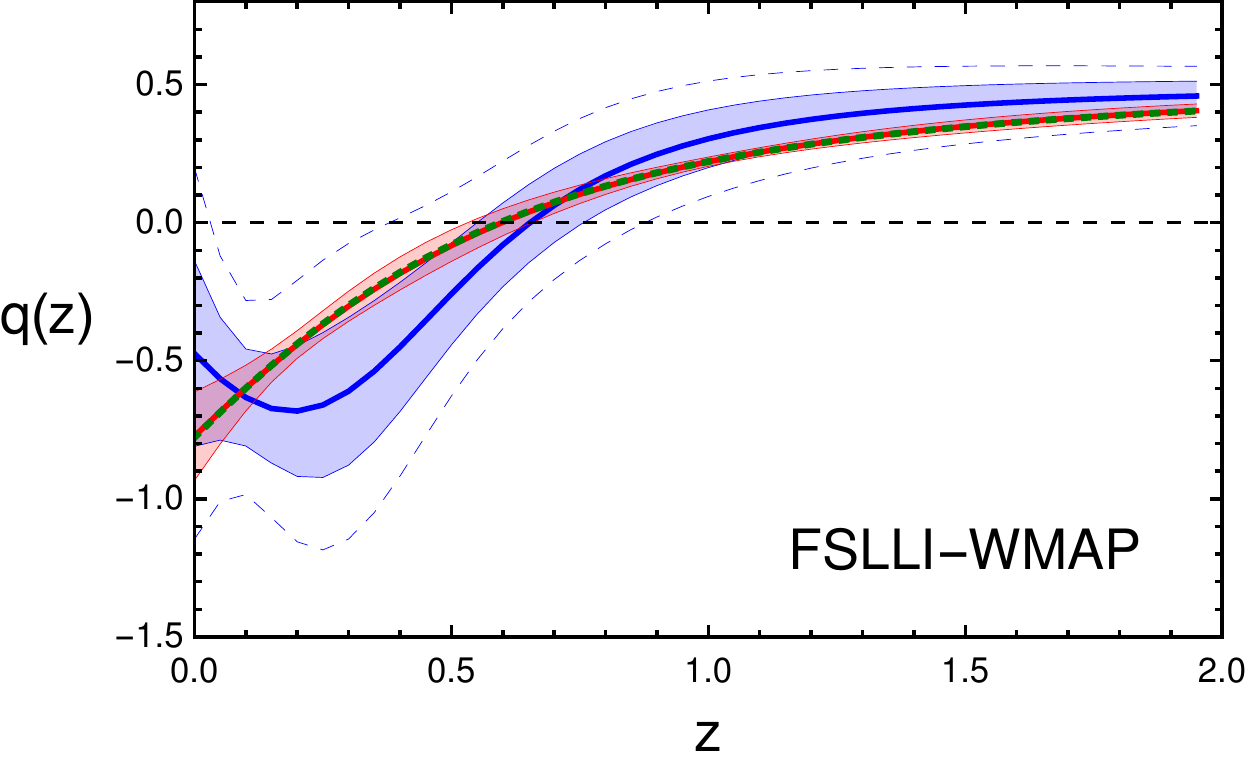} & \includegraphics[width=0.4\textwidth]{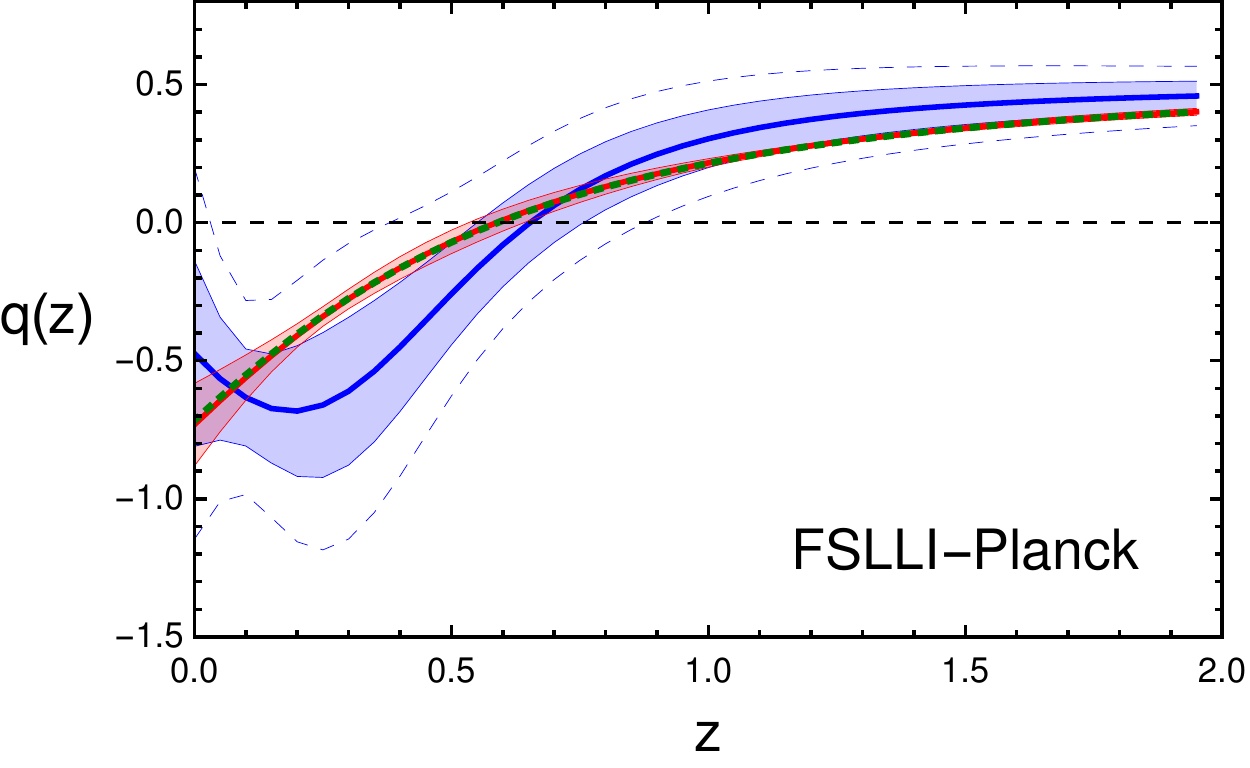}\\
\includegraphics[width=0.4\textwidth]{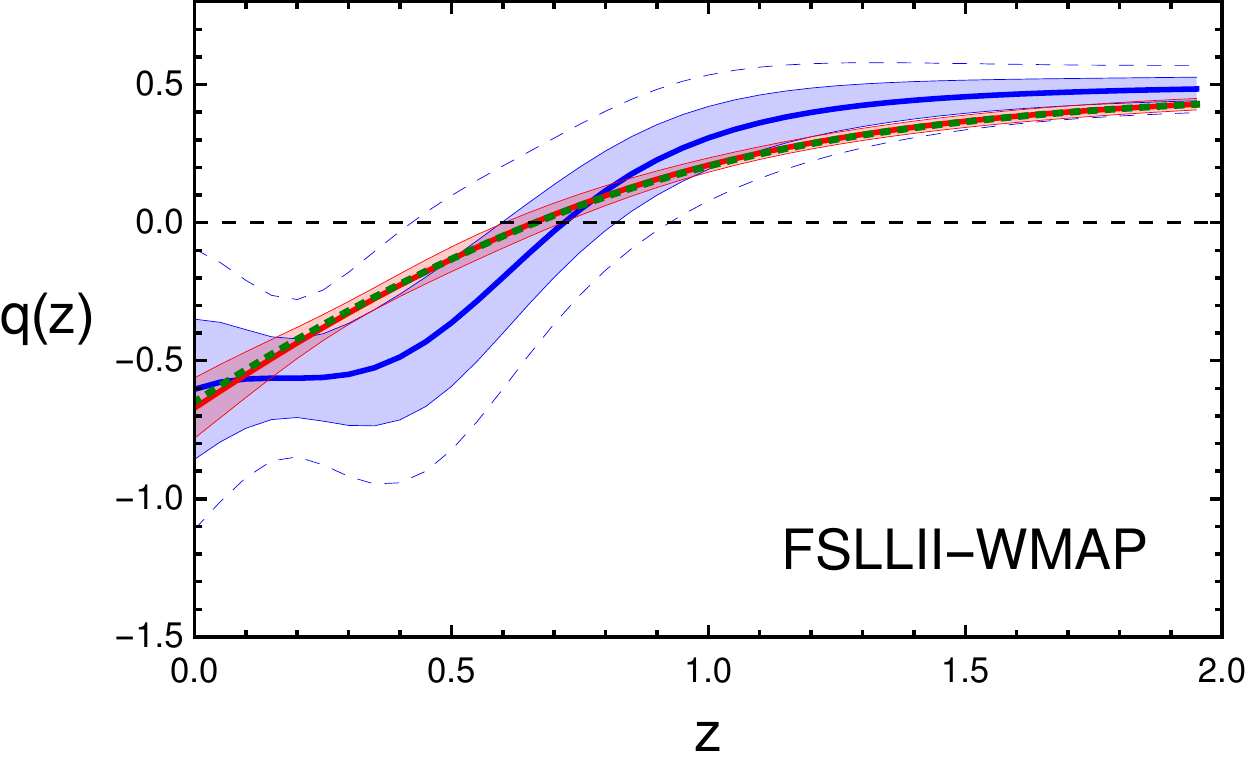} & \includegraphics[width=0.4\textwidth]{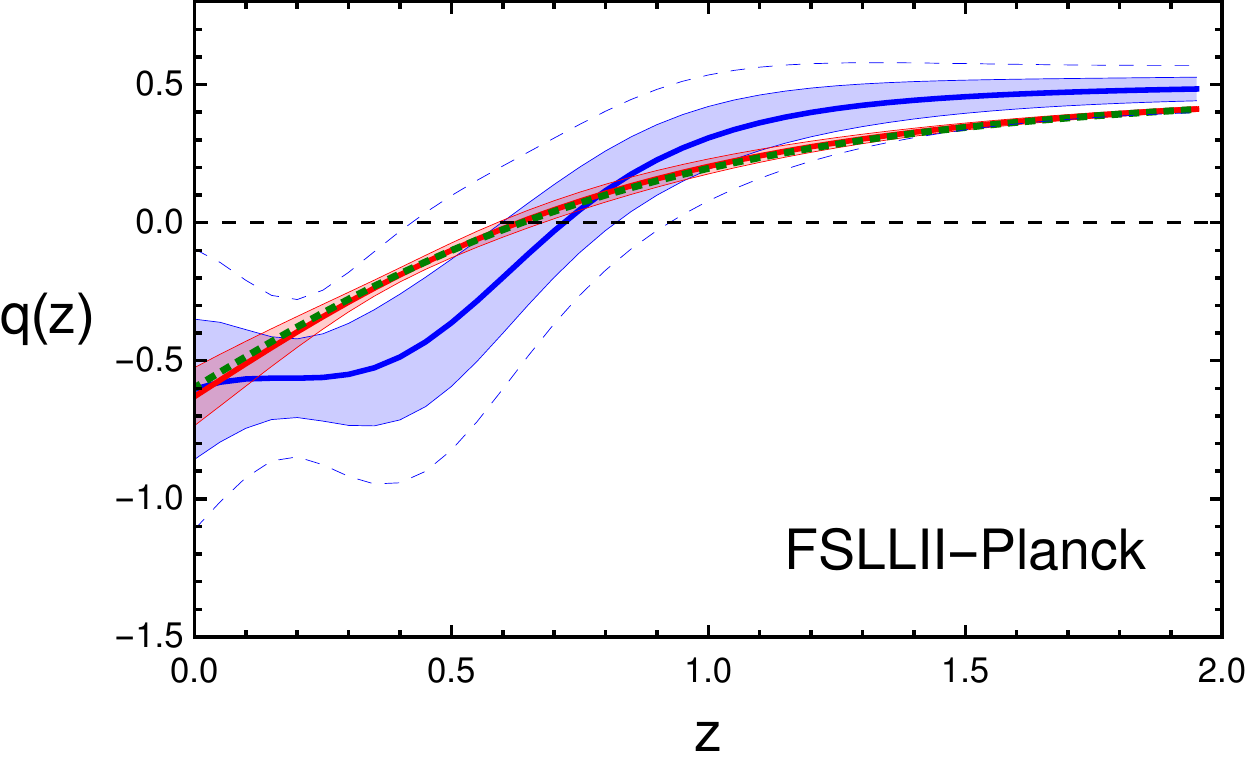}\\
\includegraphics[width=0.4\textwidth]{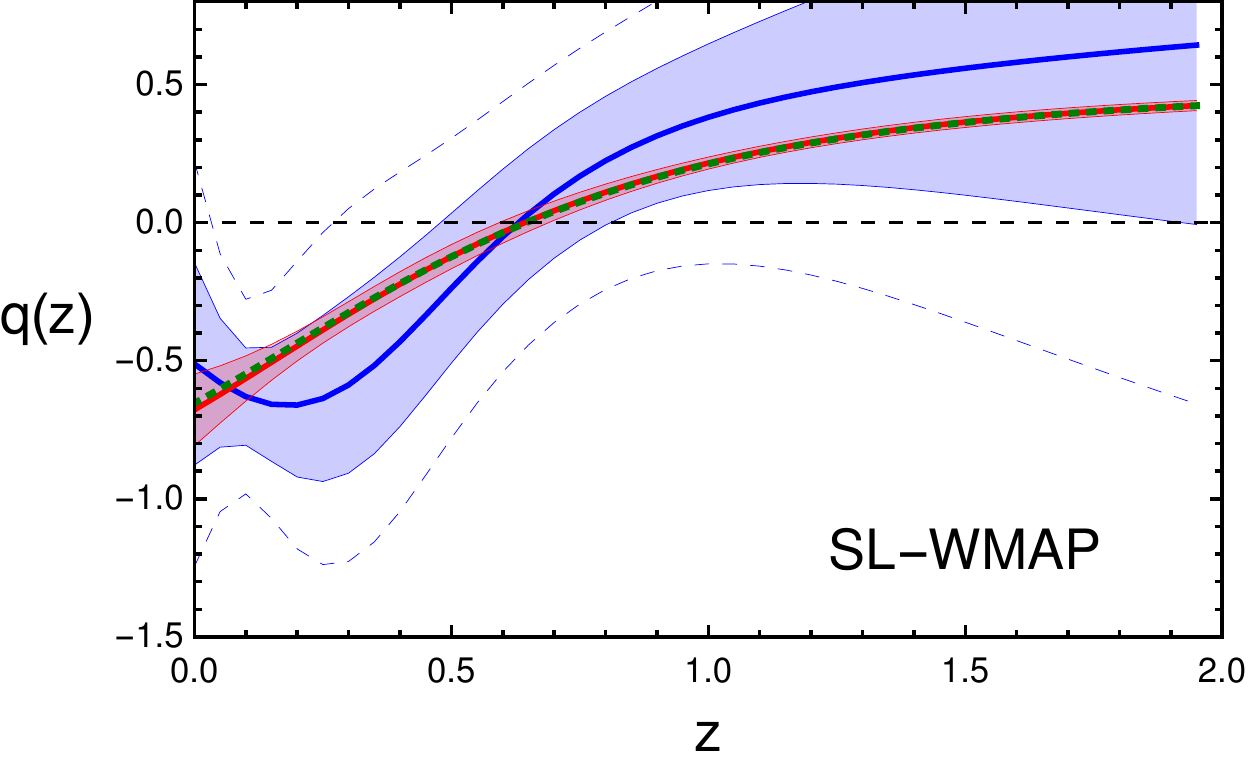} & \includegraphics[width=0.4\textwidth]{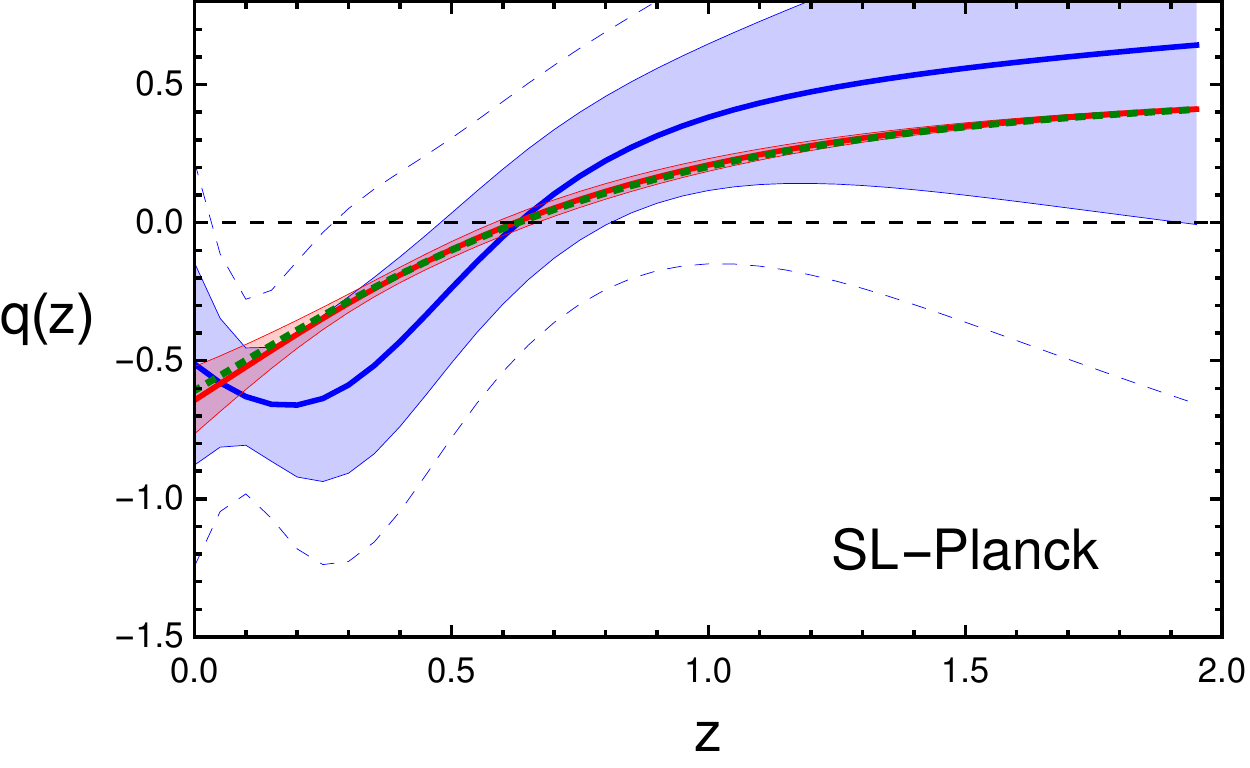}\\
\end{tabular}
\caption{Reconstructed $q(z)$ deceleration parameter for each $w(z)$ parameterization: from the top to bottom panels, 
JBP, BA, FSLLI, FSLLII, and SL parameterizations respectively. The left and right show
the results obtained using WMAP and Planck data respectively. 
The $q(z)$ computed from the $\fgas^{\mathrm{Allen}}$ constraints is shown in blue solid line.
The result obtained from the $\fgas^{\mathrm{Allen}}+H(z)+$BAO$+$CMB limits is shown in red solid line. 
The shadow (dashed lines) regions show the $1\sigma$ ($2\sigma$) region calculated with the error propagation for each data set. 
For completeness, the dotted green line shows the reconstructed $q(z)$ using the $H(z)+$BAO$+$CMB limits.}
\label{fig:qz}
\end{figure*}

Tables \ref{tab:jbp}-C5 give the estimated redshift of the transition, $z_{t}$, 
from a decelerated to an accelerated phase. If the slowing down of the CA emerges, we also give the redshift when it occurs.
Both redshifts were calculated using the following Monte Carlo approach.
We randomly selected a 1:20 subsample of the posterior sample and 
calculated for each point the $q(z)$ on a grid in $z$. Then,
we obtained the $z_{t}$ and $z_{sd}$ distributions which are well-approximated by a Gaussian
funtion. Therefore, by producing a histrogram, we estimated the central values and its errors.
We found that the $z_{t}$ value estimated using only the $\fgas^{Allen}$ data is lower than those
obtained with the joint analysis of $\fgas^{Allen}+H(z)+$BAO$+$CMB data. 
We reconstructed the cosmological evolution of the $q(z)$ parameter 
and propagated its error as follows. By considering the subsample,
we produced a well-approximated Gaussian distribution of $q(z_{i})$ for each $z_{i}$ on the grid in $z$
and determined the central value, $1$ and $2\sigma$ levels.
The behaviour of the $q(z)$ parameter for each parameterization using the $\fgas^{Allen}$,
$H(z)+$BAO$+$WMAP, and the joint analysis of $\fgas^{Allen}+H(z)+$BAO$+$WMAP data are shown
in the left panel of Figure \ref{fig:qz}. We found that for all parameterizations 
(except for FSLLII, which presents a behavior slightly oscillatory) the CA has two transitions 
when the $\fgas$ constraints are used. At redshifts $0.67<z_{t}<0.72$, the Universe begins its
accelerated expansion, however, it reaches a peak of acceleration at $0.19<z_{sd}<0.24$ and then the 
acceleration slows down. Moreover, this trend is supported at $2\sigma$ confidence level
(dashed lines in Fig. \ref{fig:qz}). Our results are in good agreement
with those obtained by \citet{Hu:2015}. They found that in the JBP, BA and FSLL parameterizations
the cosmic acceleration presents a slowing down at $z\sim0.24$, $0.26$, and $0.28$ respectively within the $2\sigma$ confidence level.
Nevertheless, we calculated the Akaike information criteria \citep[AIC,][]{Akaike:1974} and Bayesian information criteria \citep[BIC,][]{Schwarz:1978}
for the $\Lambda$CDM and the five dynamical DE models. By comparing with the $\Lambda$CDM, 
we obtained that the dynamical models lead larger AICs ($\sim3$) and BICs ($\sim12$) values,
indicating that a dynamical DE model is not necessary to explain the $\fgas$ data. 
\citep[this result is similar to that obtained by][]{Hu:2015}.
Since that the slowing down of cosmic acceleration
depend on the $w_{0}-w_{1}$ values, which are completely degenerated, 
it is crucial to provide narrow EoS constraints to prevent misleading results.
A future increase in $\fgas$ points (i.e. decrease in the 
systematic uncertainties) will produce a better parameter estimation 
for $w(z)$ parameterizations. For instance, \citet{Mantz:2014}
simulated the $\fgas$ measurements for $400$ galaxy clusters and obtained 
for an evolving DE model an increase in the Figure-of-Merit (FoM) by a factor $\sim15$ 
in an optimistic scenario of future surveys (this implies a 
reduction of the area enclosed by the $w_{0}-w_{1}$ $95\%$ confidence contour
and the EoS constraints are very close to the cosmological constant). 
Therefore the slowing down of cosmic acceleration obtained from $\fgas$ constraints could be only a statistical fluctuation.

On the other hand, if the $H(z)+$BAO$+$WMAP constraints are used
to reconstruct $q(z)$, we found that it evolves similarly to that of the cosmological constant. 
When we used the estimations obtained with the joint analysis of the all data, 
the transition from a decelerated to an accelerated phase occurs at $0.64<z_{t}<0.70$ and
$q(z)\sim -0.7$ at $z\rightarrow0$, i.e., the slowing down of the CA disappears, 
and it evolves as the cosmological constant. Furthermore,  we found that the evolution of the CA is independent 
of the EoS parameterization (see the different panels of Fig. \ref{fig:qz}). This same scenario was found by 
\citet{magana:2014} when several SNIa data sets are considered \citep[see also][]{Hu:2015,Shi:2011}.
Although to discern which model is the preferred one by observations is not the scope of this paper,
it can be determined by comparing the $\chi_{min}$, given in Tables \ref{tab:jbp}-C5, among data sets. 
Any $w(z)$ parametrization could be plausible to model a dynamical dark energy.

\begin{figure*}
\begin{tabular}{cc}
\includegraphics[width=0.4\textwidth]{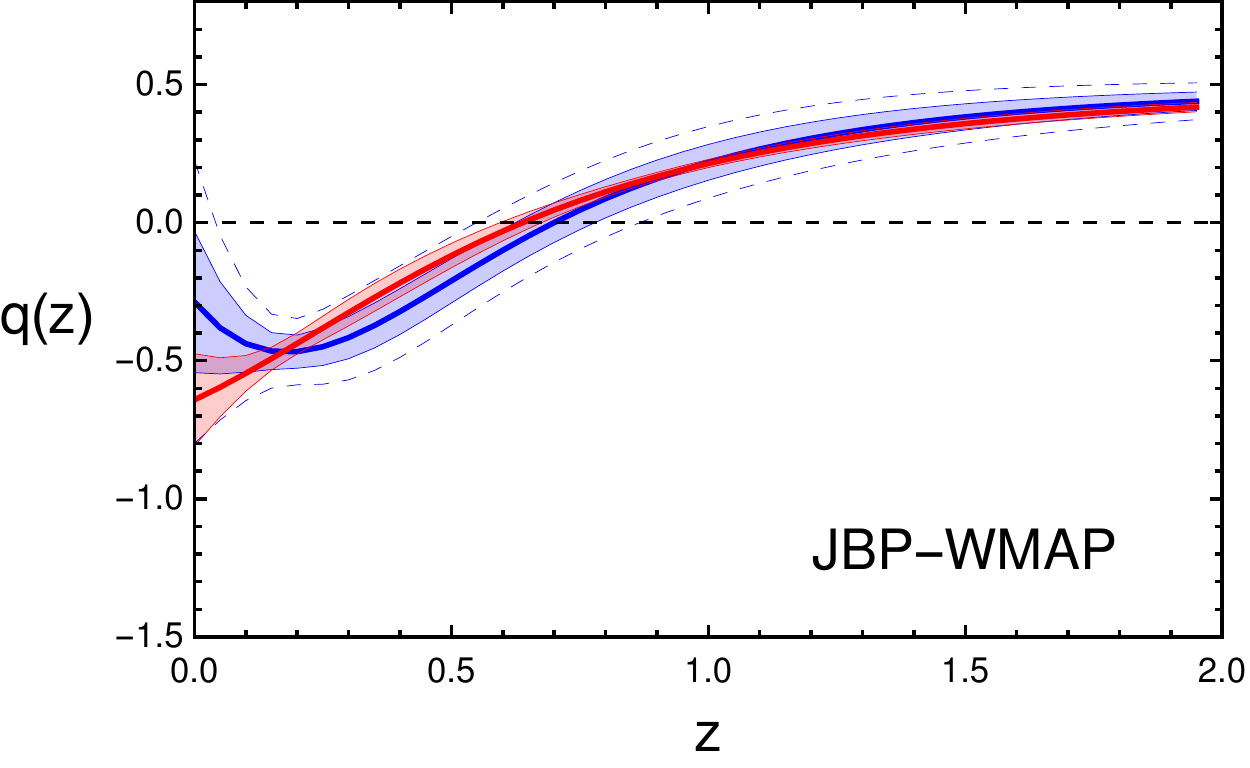} & \includegraphics[width=0.4\textwidth]{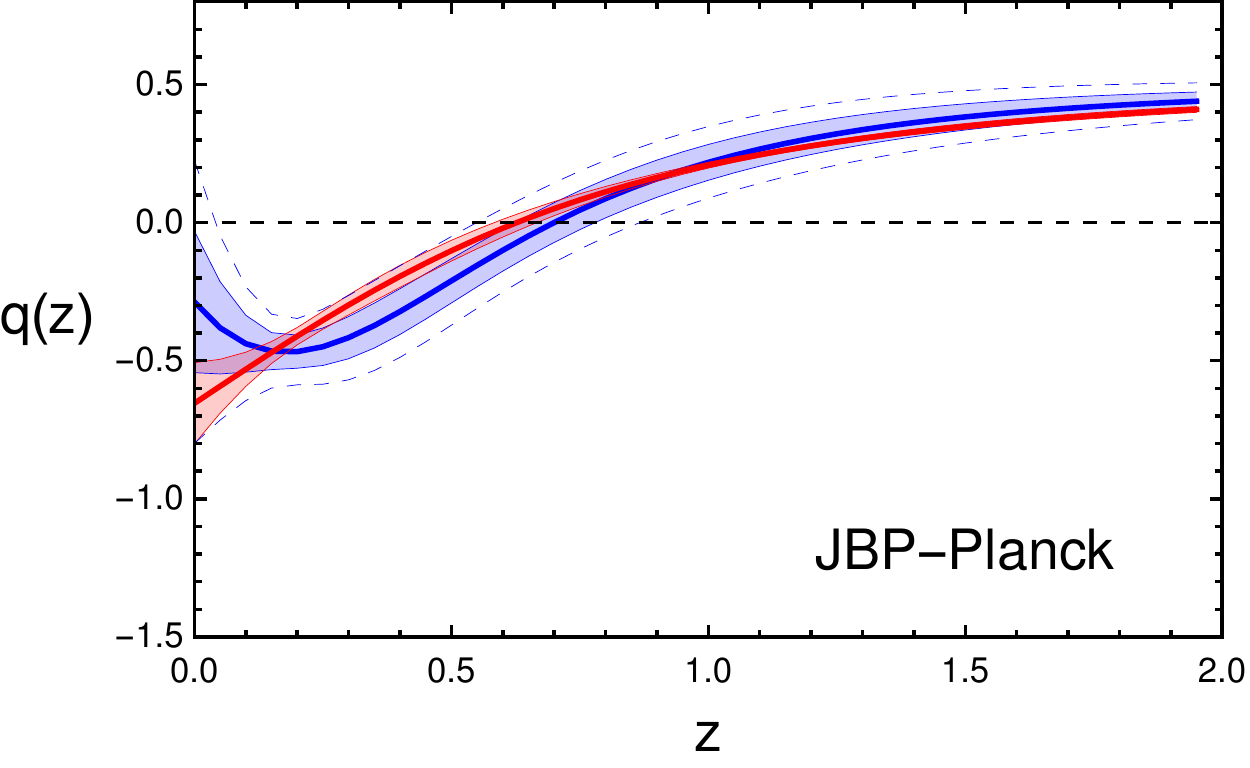}\\
\includegraphics[width=0.4\textwidth]{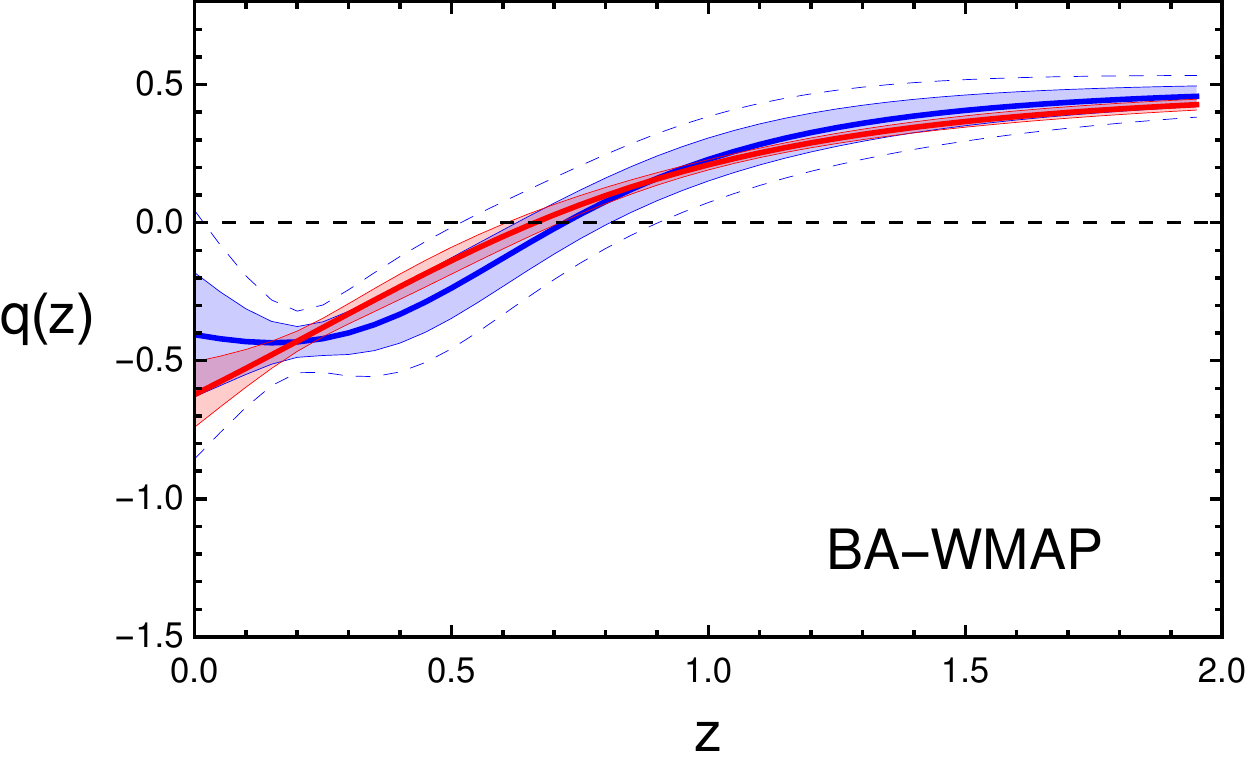} & \includegraphics[width=0.4\textwidth]{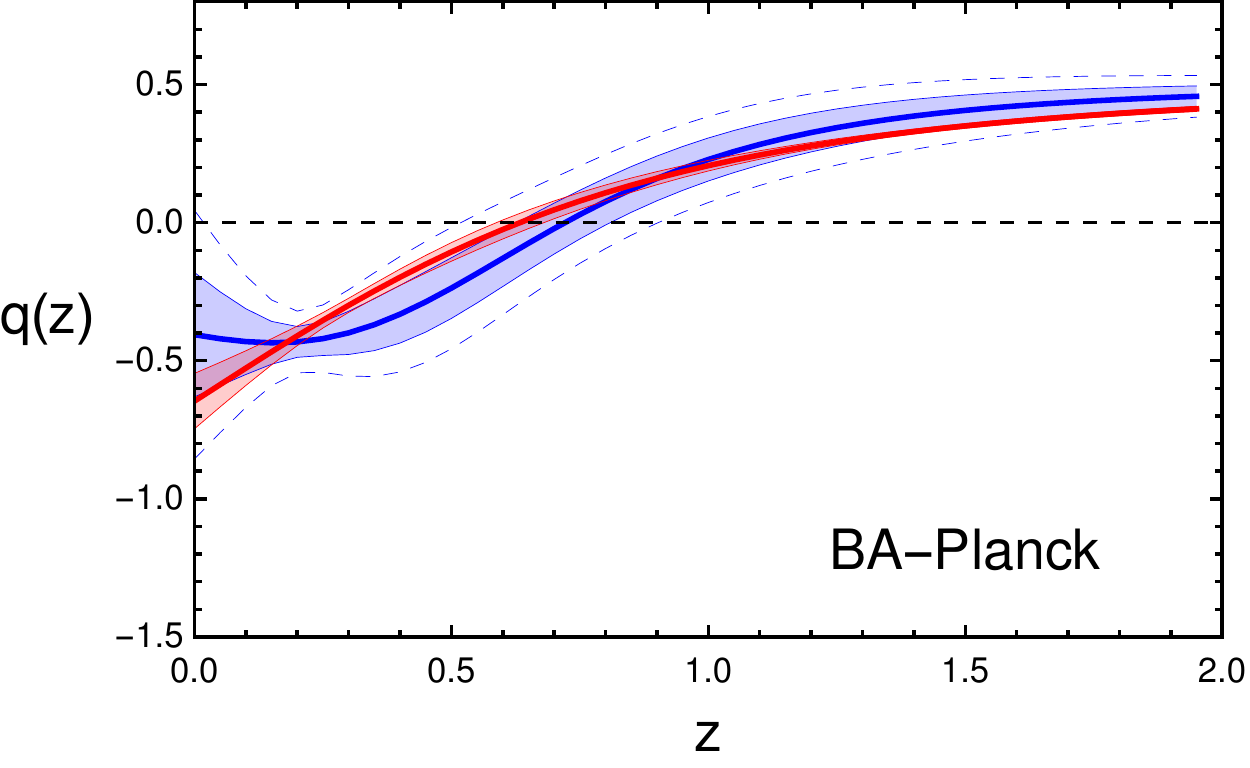}\\
\includegraphics[width=0.4\textwidth]{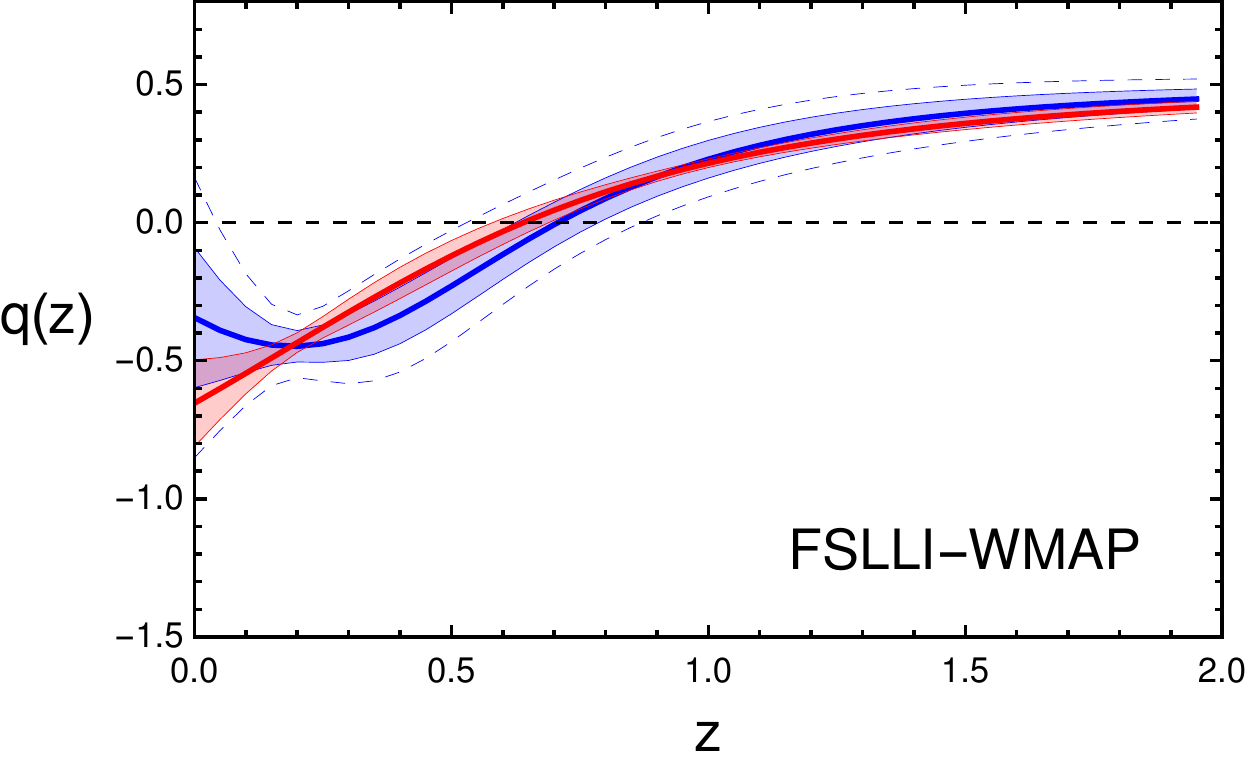} & \includegraphics[width=0.4\textwidth]{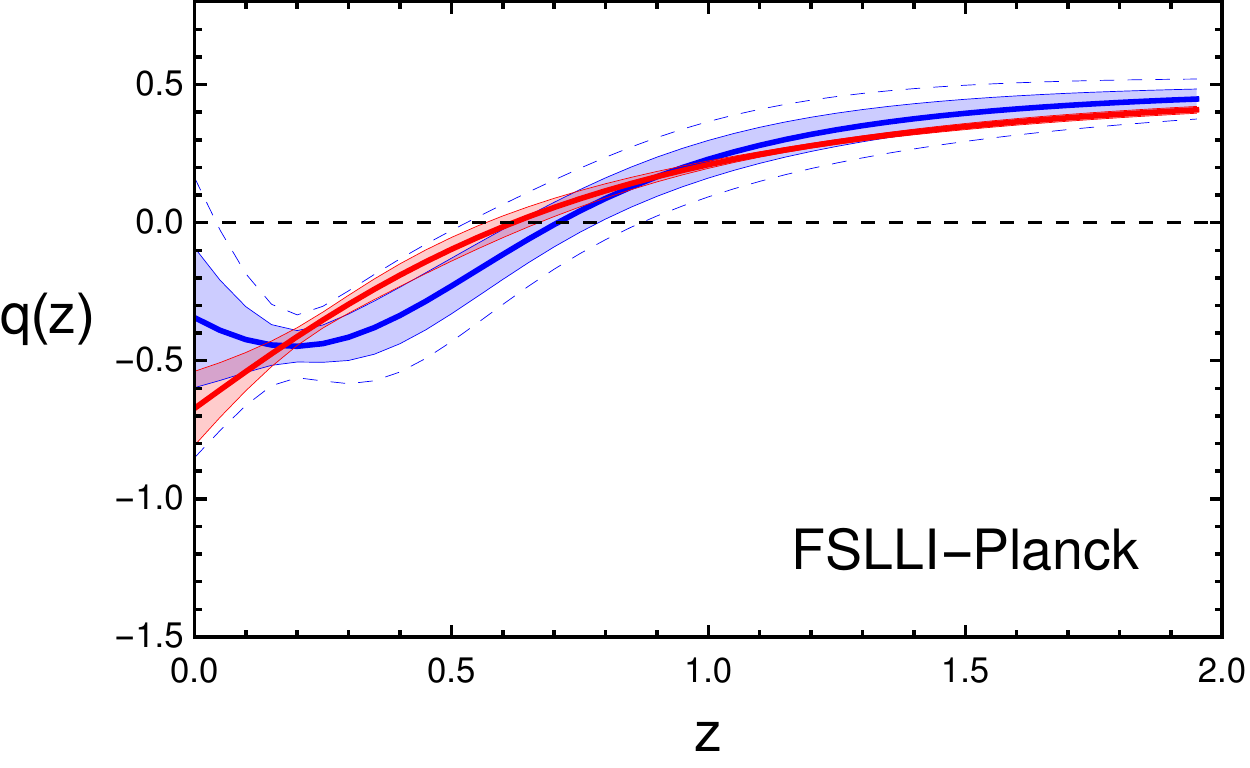}\\
\includegraphics[width=0.4\textwidth]{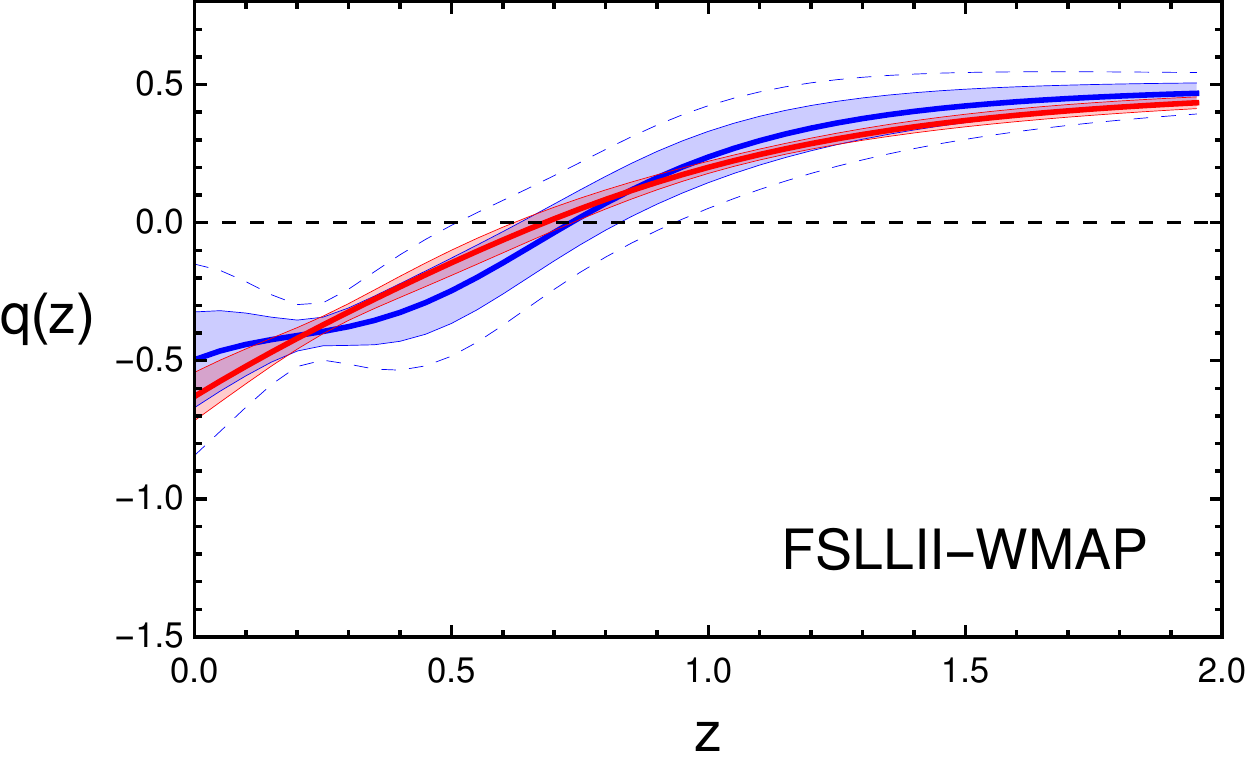} & \includegraphics[width=0.4\textwidth]{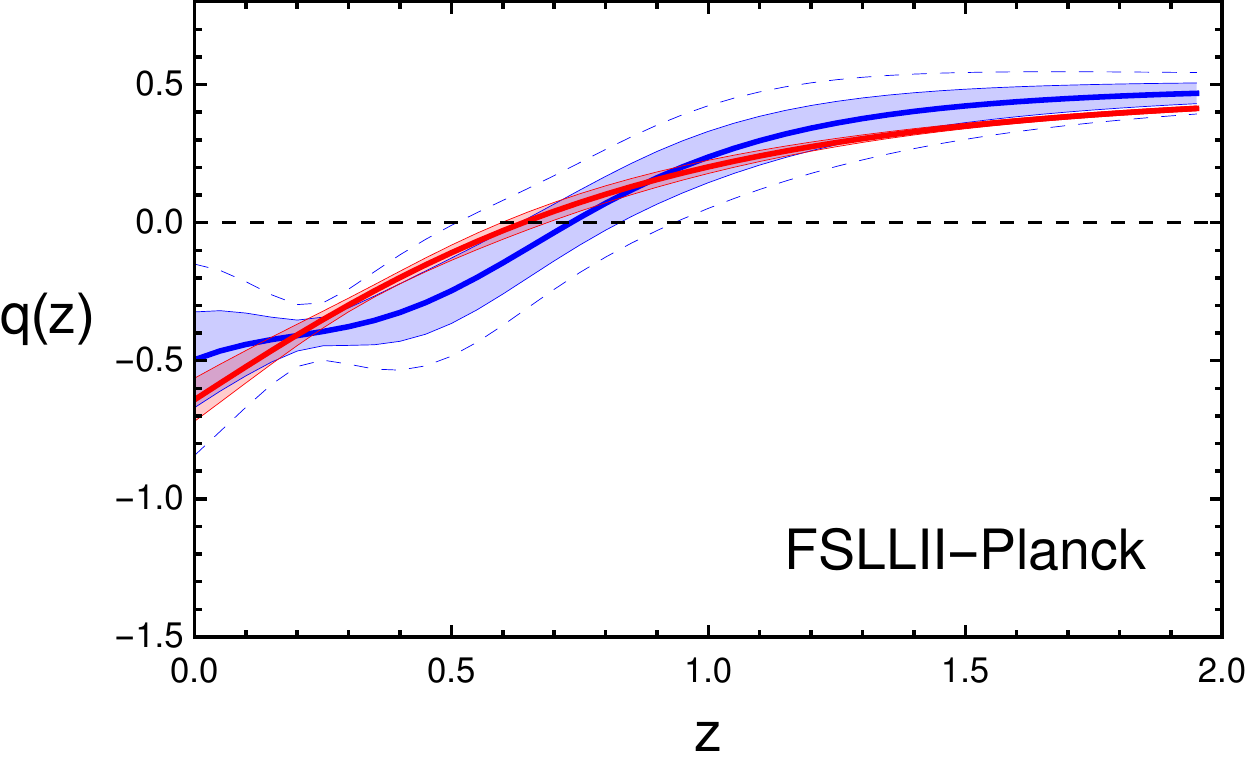}\\
\includegraphics[width=0.4\textwidth]{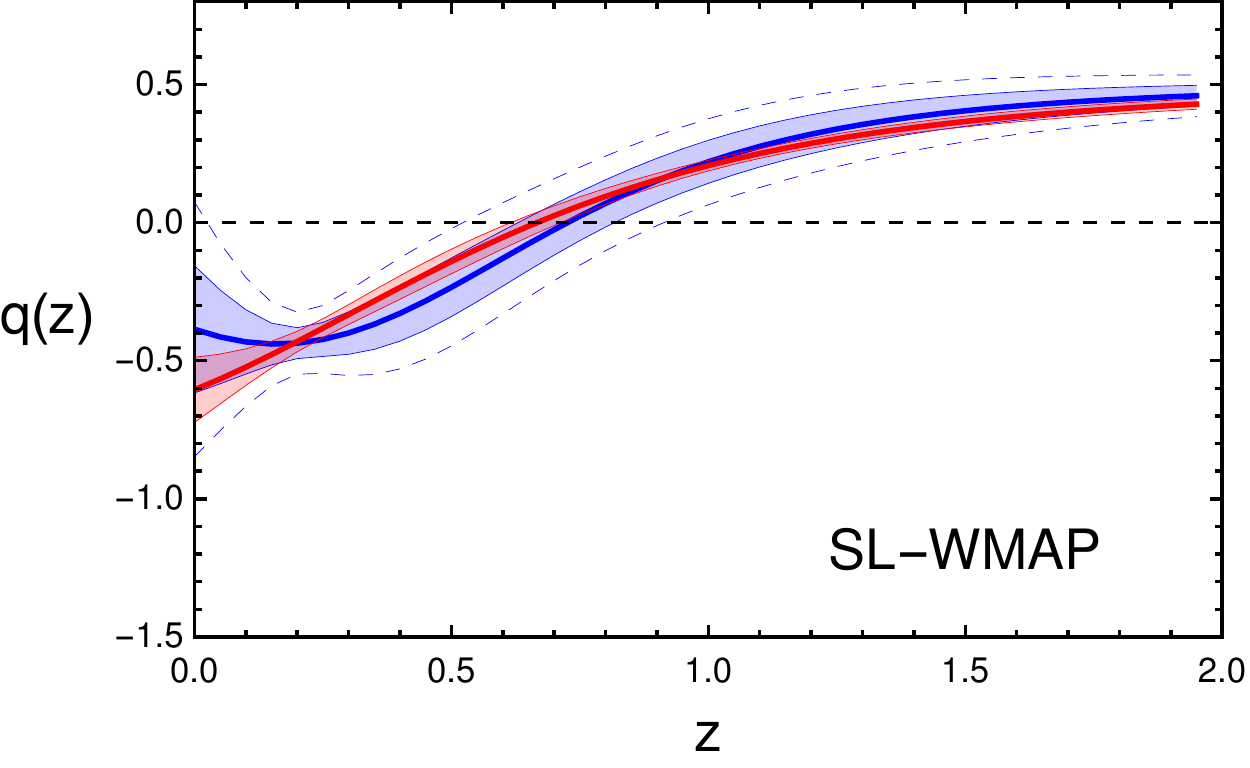} & \includegraphics[width=0.4\textwidth]{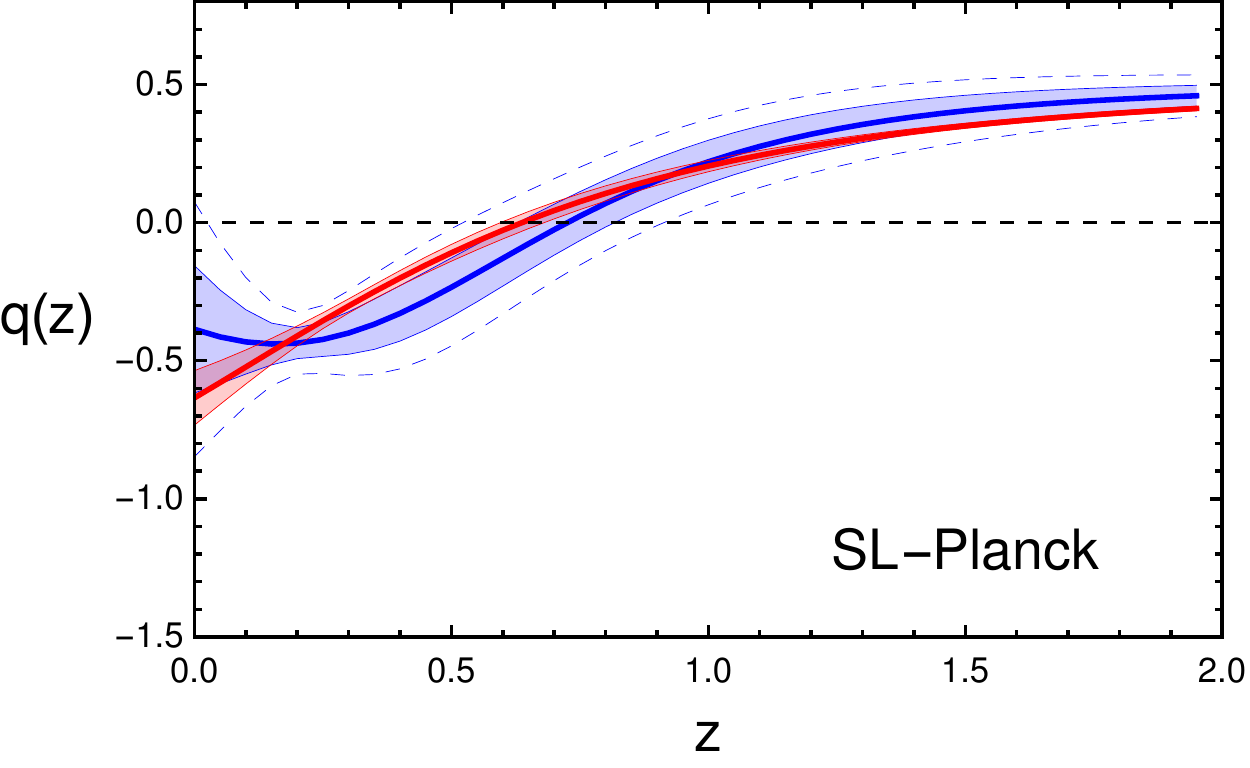}\\
\end{tabular}
\caption{Reconstructed $q(z)$ deceleration parameter for each $w(z)$ parameterization: from the top to bottom panels, 
JBP, BA, FSLLI, FSLLII, and SL parameterizations respectively. The left and right show
the results obtained using WMAP and Planck data respectively. 
The $q(z)$ computed from the $\fgas^{\mathrm{ACT}}$ constraints is shown in blue solid line.
The result obtained from the $\fgas^{\mathrm{ACT}}+H(z)+$BAO$+$CMB limits is shown in red solid line. 
The shadow (dashed lines) regions show the $1\sigma$ ($2\sigma$) region calculated with the error propagation.}
\label{fig:qz_ACT}
\end{figure*}

\subsection{Impact of CMB data on $q(z)$}
As shown in the previous section, when the CMB data from WMAP measurements are included, 
the slowing down of the CA, which emerges when only low-redshift data are used, disappears 
\citep[see also][]{shafi2009, Li:2011, victor_rivera,victor_fgas,magana:2014}. Nevertheless,
in a recent paper, \citet{Hu:2015} found that for the CPL parameterization, when the Planck data
\citep[from the $2013$ and $2015$ measurements,][]{PlanckXVI,Planck2015:XIII, Planck2015:XIV} 
in combination with the BAO and SNIa (JLA sample) data are considered in the analysis, the slowing down of the CA is 
still present at the $1\sigma$ and $2\sigma$ levels. To prove this result, we also consider the 
Planck $2015$ data in the $\fgas$ and $\fgas+H(z)+$BAO$+$CMB analysis. 
We found that there are no significative differences between the best fits on $\Omega_{m}$ obtained using Planck 
with those from WMAP (see tables \ref{tab:jbp}-C5 and Figures \ref{fig:JBP}-C5). 
Additionally, the $w_{0}$ and $w_{1}$ constraints obtained using Planck 
data in the joint analysis are consistent at $1\sigma$ confidence level to those estimated including WMAP measurements.
When $q(z)$ is reconstructed for each parameterization using the constraints derived of the combination of all data,
we obtain that the universe evolves from a decelerated phase to one accelerated phase
at $0.63<z_{t}<0.68$, and at $z\rightarrow0$ the cosmic acceleration is consistent with that
of the standard model (see right panels of Fig. \ref{fig:qz}).
This result is in agreement with those obtained using WMAP data where the slowing down of CA 
obtained using the gas mass fraction disappears by adding the CMB (high-redshift) information.
Nevertheless, our result is in tension with that found by \citet{Hu:2015} and we check
that the Planck 2015 measurements do not favor a slowing down of the CA.

\subsection{Impact of the $\fgas^{\mathrm{ACT}}$ sample on the constraints}
In the previous sections, we have presented the results using the $\fgas^{\mathrm{Allen}}$ in the
$w(z)$ fitting. Here we investigate whether the slowing down of the CA appears 
when the $\fgas^{\mathrm{ACT}}$ sample, with about twice the data points by $\fgas^{\mathrm{Allen}}$ set,
is used in the parameter estimation. As before, we estimate the $\Omega_{m}$, $w_{0}$, and $w_{1}$ constraints
using only the $\fgas$ data and the joint analysis $\fgas+H(z)+$BAO$+$CMB (WMAP or Planck). 
We give the best fits for each parameterization in Tables \ref{tab:jbp}-C5. 
Notice that the $\Omega_{m}$ constraints from the different data sets are very similar to those obtained
when the $\fgas^{\mathrm{Allen}}$ sample was used. Although there is a slightly tension in the 
$w_{0}$ and $w_{1}$ best fits obtained from both gas mass fraction samples, they are consistent
within their $68\%$ confidence regions. When the $q(z)$ deceleration parameter for each $w(z)$ 
is reconstructed using the $\fgas^{\mathrm{ACT}}$ constraints, 
the transition from a decelerated phase to an accelerated phase occurs at $\simeq0.75$.
We also found that the $\fgas^{\mathrm{ACT}}$ is consistent at the $2\sigma$ confidence level 
with the slowing down of the CA at $0.20<z_{sd}<0.25$ for the JBP, BA, FSLLI, 
and SL parameterizations (see left panels of Fig. \ref{fig:qz_ACT}). 
Notice that this trend is softened compared with those obtained using the $\fgas^{\mathrm{Allen}}$ sample.
As in the case of SNIa data, we confirm that this trend occurs for several $\fgas$ samples. 
By reconstructing $q(z)$ using the $\fgas^{ACT}+H(z)+$BAO$+$WMAP (Planck) constraints, 
the CA evolves as the cosmological constant and the Universe passes 
from a decelerated phase to an accelerated phase at $0.68<z_{t}<0.72$ ($0.66<z_{t}<0.69$).
It is worth to note that the $\fgas^{\mathrm{ACT}}$ sample could yield biased constraints
due to the cluster gas physics and underlying cosmology assumed in
the $\fgas-M_{500}$ relation (Eq. \ref{eq:fgasVik}) which was used to compute the data point.
To take into account its bias we performed a test using large $\fgas$ uncertainties (see Appendix A)
and we found no significative differences on the $\Omega_{m}$, $w_{0}$, and $w_{1}$ constraints,
leading to similar evolutions of $q(z)$ as those obtained with normal errors.

\section{Conclusions} \label{sec:conclusions}
We investigated the evolution of the deceleration parameter $q(z)$ 
in the following models with the EoS as a function of redshift
using gas mass fraction measurements in galaxy clusters: JBP, BA, FSLLI, FSLLII, and SL.
We used two $\fgas$ data sets: the sample by \citet{Allen:2004cd} which consists of $42$ points spanning 
the redshift range $0.05<z<1.1$ and the ACT sample which consists of $91$ 
data points in the redshift range $0.118 < \mathrm{z} < 1.36$ \citep{Hasselfield:2013}. 
These $\fgas^{\mathrm{ACT}}$ data points were computed using a $\fgas-M_{500}$ relation \citep{Vikhlinin:2009}.
In spite of the difference between the samples, they both provide consistent estimations for the cosmological parameters.
We validated that the gas mass fraction test is a qualified tool to constrain the cosmological parameters
$\Omega_{m}, w_{0}, w_{1}$, and then the CA in these models.

We also considered different fitting function for the $\fgas$ data points: 
A$04$ (Eq. \ref{eq:one}) and A$08$ (Eq. \ref{eq:fgasA08}). The main difference between both functions is that 
the second one considers several corrections due to the underlying cosmology and the cluster physics. 
We found that the A$08$ fitting function does not offer significantly better
cosmological constraints than the simplest A$04$ function (see Fig.\ref{fig:comparison}). 

A MCMC analysis was performed using the $\fgas$ data and the joint
$\fgas+H(z)+$BAO$+$CMB (WMAP and Planck) measurements.
We used these data sets to reconstruct the $q(z)$ parameter. For all parameterizations (except for FSLLII),
a slowing down of CA emerges at late times (at the $2\sigma$ confidence level) when using the $\fgas$ data alone, i.e. 
different $w(z)$ functions does not influence the $q(z)$ cosmic evolution. 
This is in agreement with results obtained by \citet[][see also Shi et al. 2011]{magana:2014} using several SNIa samples.
Our results are also in agreement with those obtained by \citet{Hu:2015}, who performed a comprehensive analysis 
with several cosmological data. Furthermore, our results suggest that this CA 
trend is not a systematic effect produced by 
SNIa or $\fgas$ samples (as suggested previously by \citet{shafi2009, victor_fgas, magana:2014,Hu:2015}). 
Nevertheless, the emergence of the slowing down of cosmic acceleration
depend on the $w_{0}-w_{1}$ values. Therefore, to prevent misleading results, 
it is crucial to provide narrow EoS constraints. 
Here, we used two fgas samples, by Allen and ACT, which have poor statistics. 
A future increase in $\fgas$ points (i.e. decrease in the 
systematic uncertainties) will produce a better parameter estimation 
for $w(z)$ parameterizations. For instance, \citet{Mantz:2014}
simulated the $\fgas$ measurements for $400$ galaxy clusters and obtained 
for an evolving DE model an increase in the FoM by a factor $\sim15$
in an optimistic scenario of future surveys (this implies a reduction
of the area enclosed by the $w_{0}-w_{1}$ $95\%$ confidence contour and the EoS constraints are very close
to the cosmological constant). Therefore the slowing down of cosmic acceleration 
obtained from $\fgas$ constraints could be only a statistical fluctuation.

On the other hand, the CA behavior changes when the $H(z)+$BAO$+$CMB constraints are used.
We obtained for each parameterization that these data favor
an accelerated expansion at late times. 
The $q(z)$ reconstructed from the joint analysis $\fgas+H(z)+$BAO$+$CMB (WMAP and Planck)
shows that the CA behaves as the cosmological constant.
Our result is in tension with those obtained by \citet{Hu:2015}, 
who found a CA slowing down for the CPL parameterization using the Planck data.

We investigated the impact to consider the $\fgas^{\mathrm{ACT}}$ sample and we confirmed 
that, in general, it is consistent at the $2\sigma$ confidence level with
the slowing down CA (except for the FSLLII parameterization).
As in the $\fgas^{\mathrm{Allen}}$ case, this trend disappears by adding the $H(z)$, BAO and WMAP measurements 
in the parameter estimation. 

Finally, our results suggest that several low-redshift cosmological 
data are consistent with a CA slowing down when $w(z)$ is parameterized. 
Although this trend does not emerge in other approaches to study 
the evolution of $q(z)$ \citep[see for example][]{Rani:2015,Zhang2016}
we encourage to the community to further investigate this phenomena in DE models. 
To improve our constraints, a large high precision $\fgas$ sample is needed.

\section{Acknowledgments}
We thank the anonymous referee for thoughtful remarks and suggestions.
J.M. acknowledges support from Gemini $32130024$ and FONDECYT 3160674.
V.M. acknowledges support from ECOS-CONICYT C12U02 and Centro de Astrof\'{\i}sica de Valpara\'{\i}so.
V.C. acknowledges support from DIUV 50/2013,

\appendix 
\section{Impact of the uncertainties on the $\fgas-M_{500}$ relation}
\begin{figure*}
\begin{tabular}{cc}
{\includegraphics[width=3in]{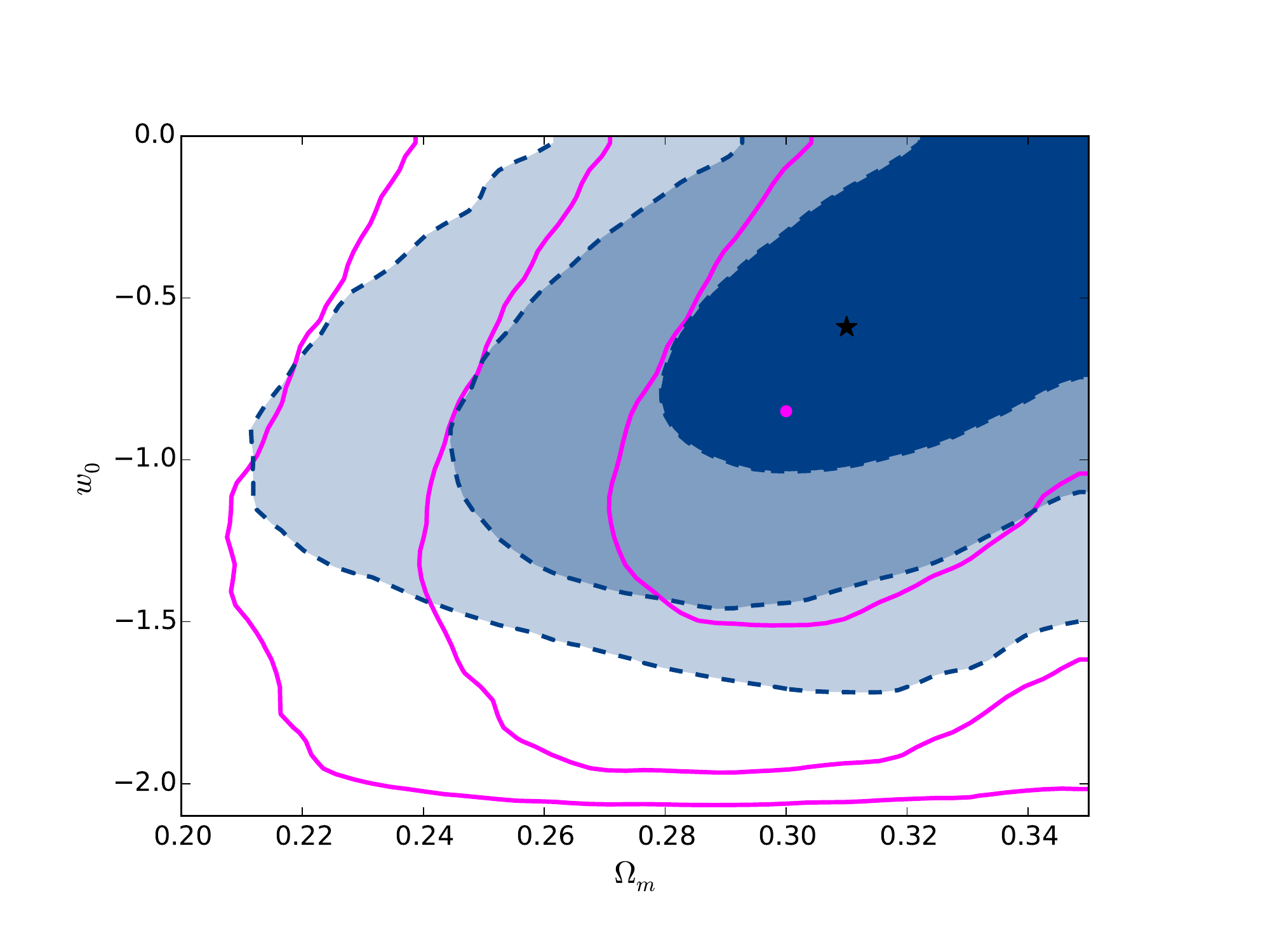}} &
{\includegraphics[width=3in]{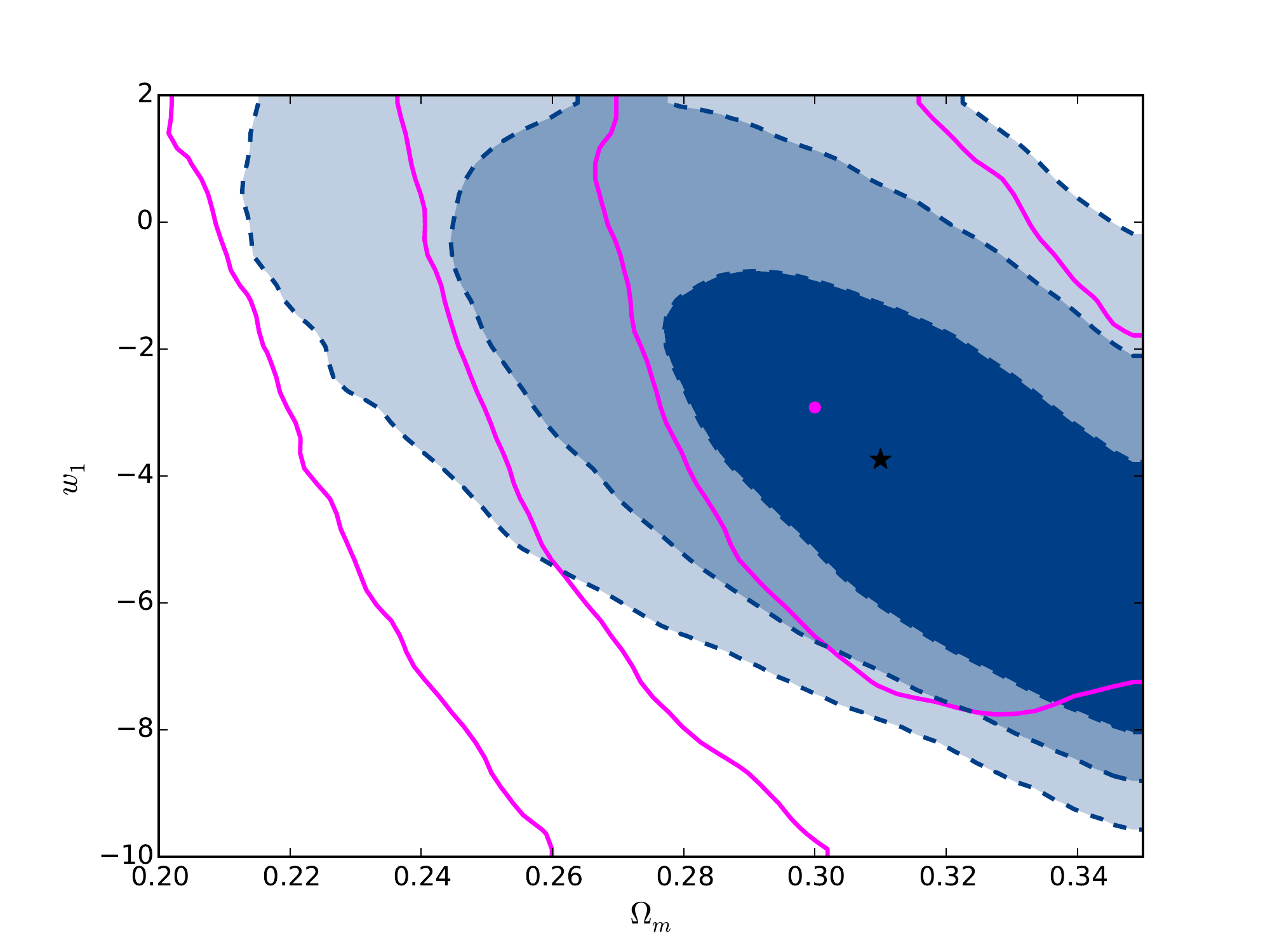}}\\
{\includegraphics[width=3in]{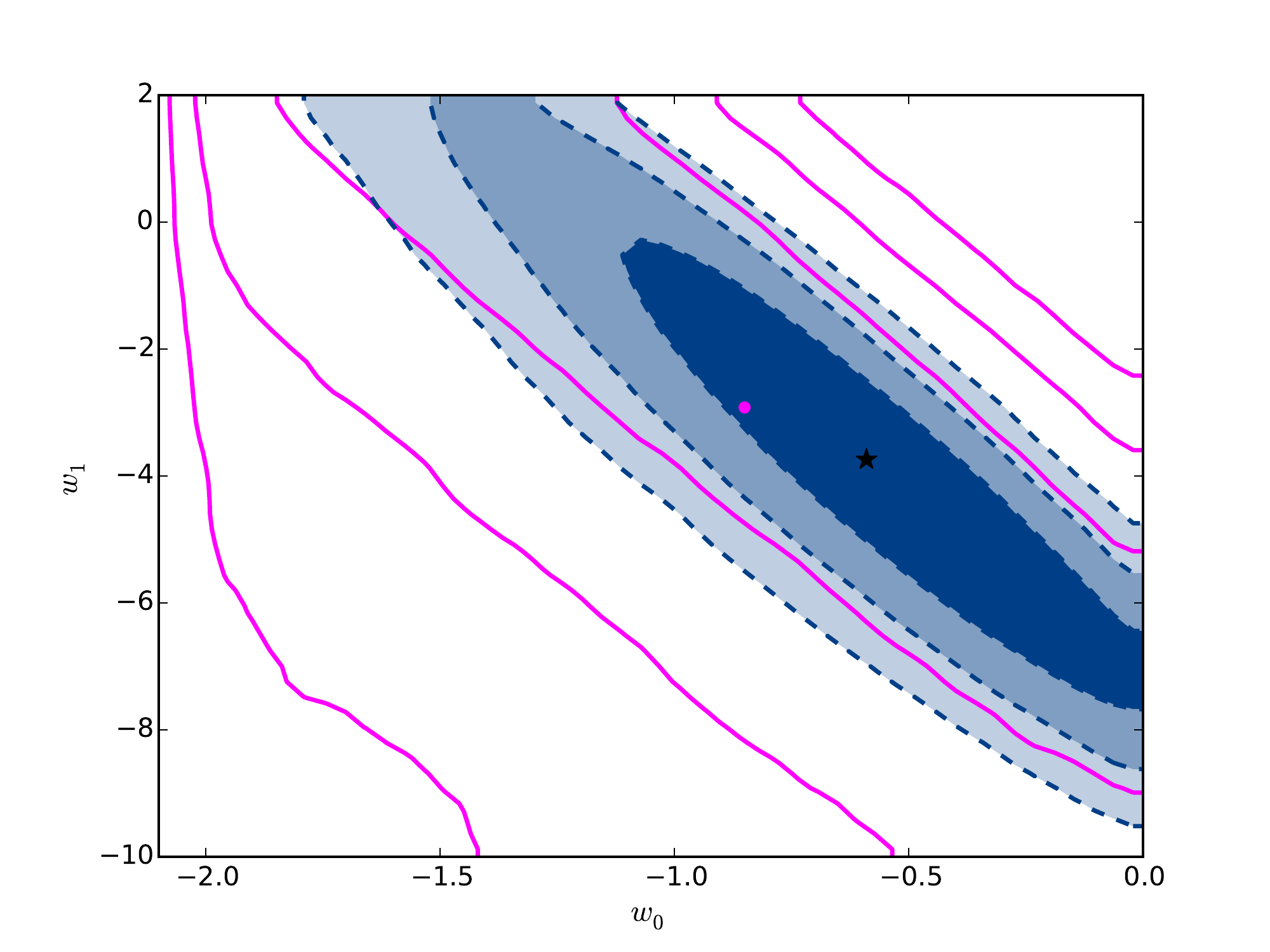}} &
{\includegraphics[width=3in]{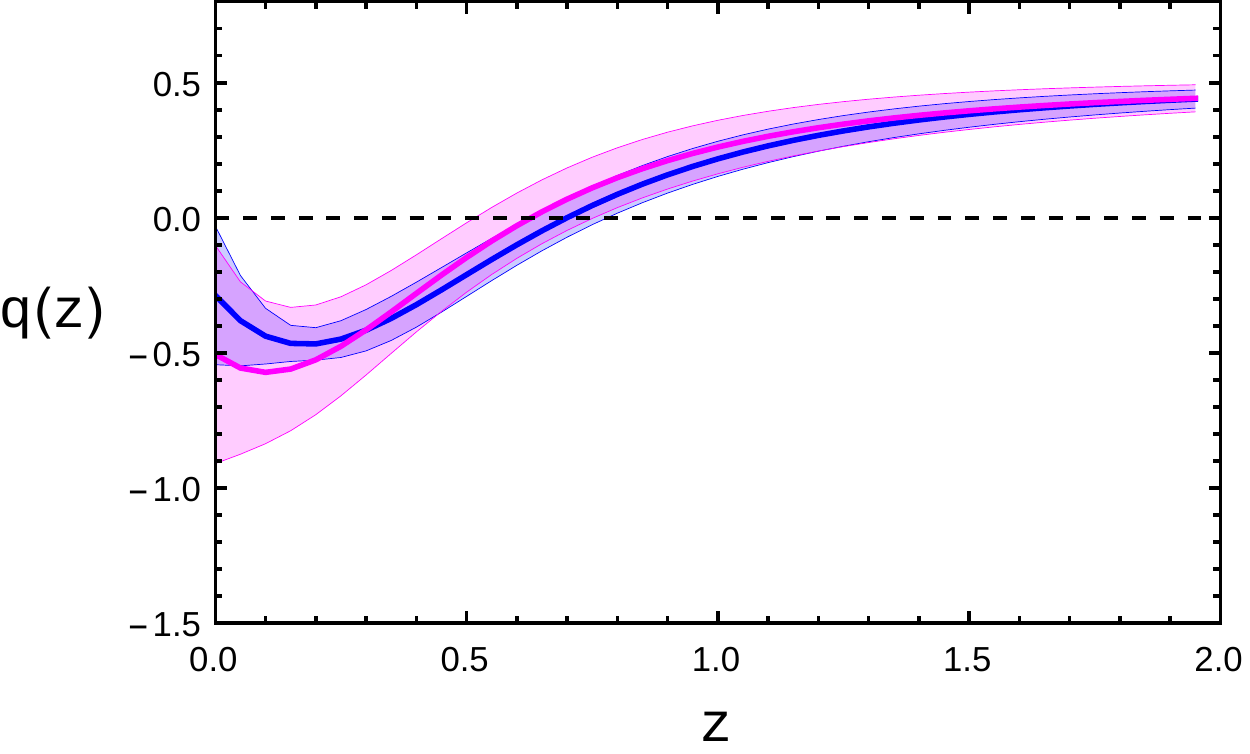}}\\
\end{tabular}
\caption{The top panels and the bottom left panel show the 2D
at $68\%$, $95\%$, $99\%$ confidence levels for the $\Omega_{m}$, $w_{0}$, and $w_{1}$ parameters of the 
JBP parameterization obtained from the $\fgas^{\mathrm{ACT}}$ data alone for two cases:
using the best fit values of the $f_{gas}-M_{500}$ scaling relation (blue filled contours)
and using $15\%$ error in each parameter of the equation (magenta no-filled contours). 
The star and dot marks represent the best value obtained using each one of these cases respectively.
The bottom right panel shows the reconstructed $q(z)$ parameter for these two cases. We found
that a $15\%$ scatter in the $f_{gas}-M_{500}$ scaling relation has little or no significant impact in 
the cosmological parameter estimation up to $1\sigma$ confidence level.}
\label{fig:sampling}
\end{figure*}

The cosmological constraints in the models could be biased due to the uncertainties
on the scaling relation parameters in the Eq. \ref{eq:fgasVik}. To test
the impact of these uncertainties, we constrain, for instance, the JBP model using 
$15\%$ error in each parameter of the equation.
Figure \ref{fig:sampling} shows that the best fit using $15\%$ errors is within the $1\sigma$ confidence 
level of the previous best fit (see Figure C1). Figure \ref{fig:sampling} also shows that the $q(z)$ 
parameter exhibits the slowing down feature at $1\sigma$ level as presented before (Figure 3).
The scatter in the $f_{gas}-M_{500}$ scaling relation has little or no significant impact in 
the cosmological parameter estimation up to $1\sigma$ confidence level.

\section{Impact of the A08 fitting function when the $\fgas^{\mathrm{ACT}}$ is used}
\begin{figure*}
\begin{tabular}{cc}
\includegraphics[width=3in]{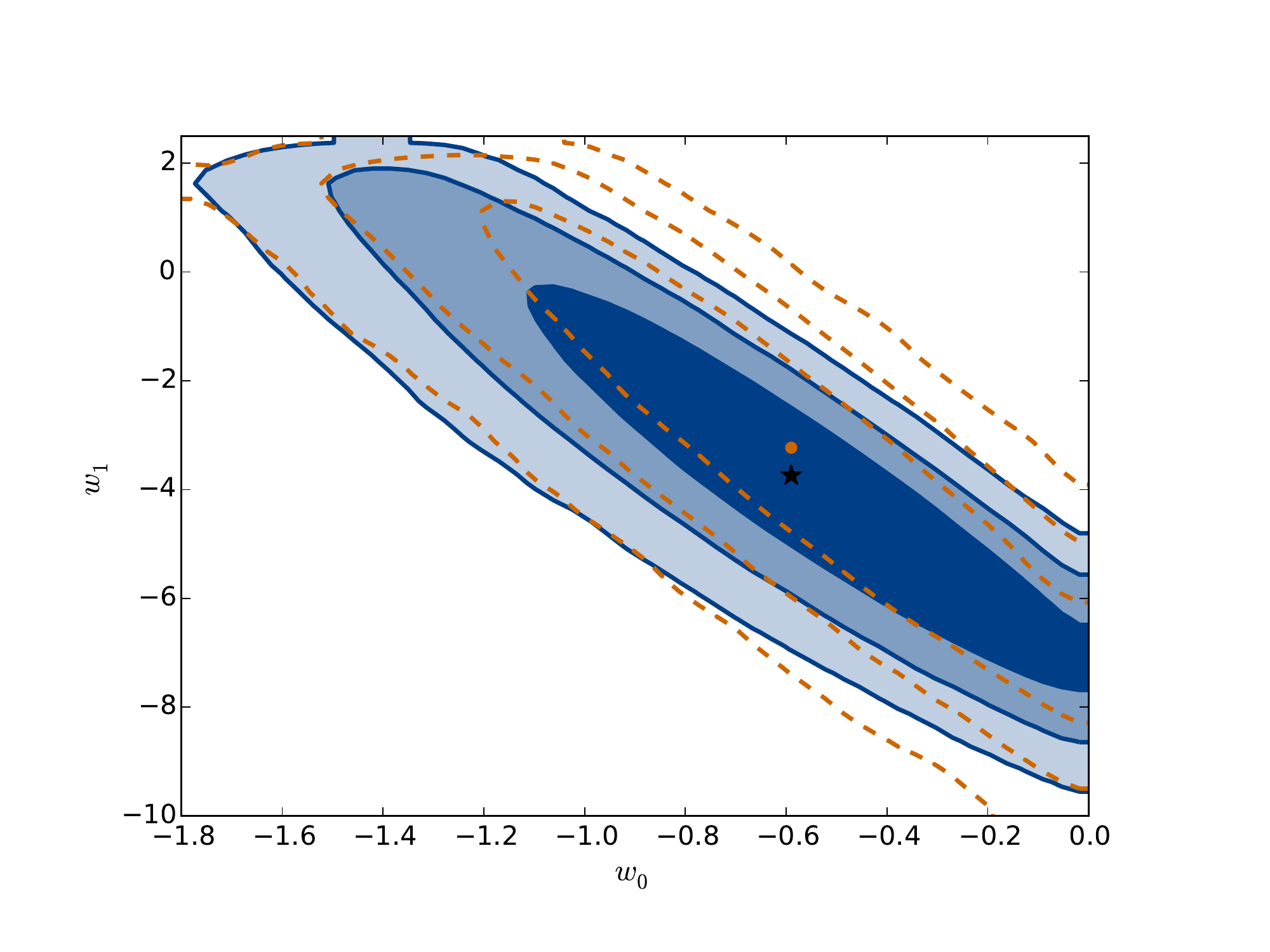}&
\includegraphics[width=3in]{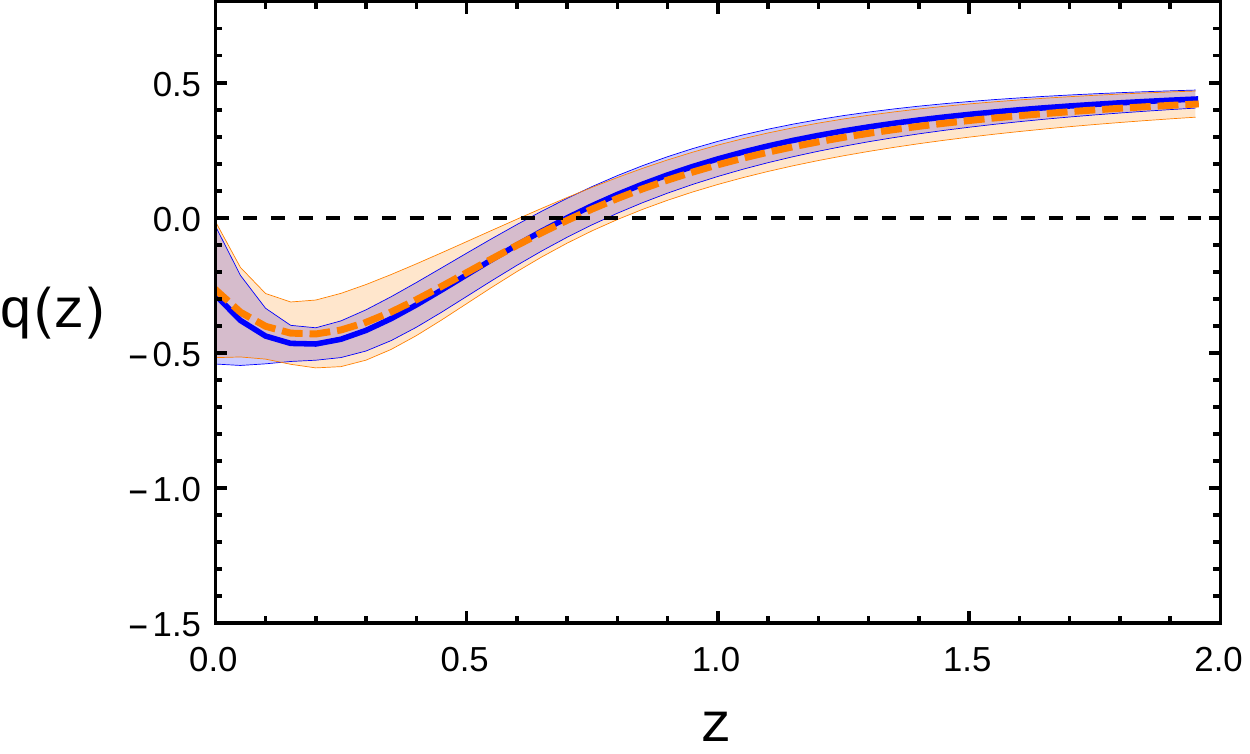}
\end{tabular}
\caption{The left panel shows the $w_{0}-w_{1}$
$68\%$, $95\%$, $99\%$ confidence levels for the 
JBP parameterization obtained from the $\fgas^{\mathrm{ACT}}$ data alone 
using the $A04$ (blue filled contours) and $A08$ (orange no-filled contours) fitting functions. 
The star and dot marks represent the best value obtained using each one of these functions respectively.
The right panel shows the reconstructed $q(z)$ parameter using these two cases: $A04$ (blue solid line )
and $A08$ (orange dashed line). The shadow regions show the $1\sigma$ region calculated with the error
propagation of the cosmological parameters. We found that the $q(z)$ parameter behaviour 
is essentially the same when the best fits derived from both functions are used.}
\label{fig:comparison2}
\end{figure*}

To test whether the constraints derived from the $\fgas^{\mathrm{ACT}}$ sample
are improved using the A$08$ fitting function instead of the A$04$ we constrain the JBP model using both fitting functions. 
The left panel of the Fig. \ref{fig:comparison2} shows that there is no significative difference 
in the $w_{0}$ and $w_{1}$ constraints when A$04$ or A$08$ functions are used.
In addition, the right panel of the Fig. \ref{fig:comparison2} shows 
that the $q(z)$ parameter behaviour is essentially the same when the best fits derived from both functions are used.
We found the same result for the other $w(z)$ functions.

\section{Tables and confidence contours for each model using \lowercase{$\fgas$} and \lowercase{$\fgas$}$+H(z)+$BAO$+$CMB}
\label{Ap:tables_contours}
\begin{table*}
\centering
\begin{tabular}{|lcccccc|}
\multicolumn{7}{c}{JBP parameterization}\\
\hline
Data set& $\chi^{2}_{min}$&$\Omega_{m}$& $w_{0}$&$w_{1}$&$z_{t}$&$z_{sd}$\\
\hline
\multicolumn{7}{|c|}{}\\
H(z)+BAO+W9 & $49.32$ & $0.28^{+0.01}_{-0.01}$ & $-1.25^{+0.22}_{-0.16}$ & $ 0.51^{+0.99}_{-1.47}$&$0.66^{+0.04}_{-0.04}$& No \\
H(z)+BAO+Pl & $67.22$ & $0.29^{+0.01}_{-0.01}$ & $-1.22^{+0.21}_{-0.16}$ & $0.55^{+0.91}_{-1.18}$&$0.65^{+0.04}_{-0.04}$& No \\
\hline
\multicolumn{7}{|c|}{A$04$}\\
\hline
\multicolumn{7}{|c|}{}\\
$\fgas^{Allen}$ & $41.80$ & $0.28^{+0.03}_{-0.03}$& $-0.80^{+0.45}_{-0.45}$ & $-3.78^{+3.73}_{-3.76}$&$0.71^{+0.09}_{-0.09}$& $0.20^{+0.06}_{-0.09}\,(2.6)$\\ 
$\fgas^{ACT}$ & $160.71$ & $0.31^{+0.02}_{-0.03}$& $-0.59^{+0.34}_{-0.39}$ & $-3.74^{+2.50}_{-2.19}$&$0.75^{+0.08}_{-0.07}$& $0.23^{+0.05}_{-0.09}\, (3.2)$\\ 
$\fgas^{Allen}$+H(z)+BAO+W9 & $91.40$ & $0.28^{+0.01}_{-0.01}$ & $-1.26^{+0.21}_{-0.15}$ & $0.53^{+0.97}_{-1.41}$&$0.66^{+0.04}_{-0.04}$& No \\
$\fgas^{Allen}$+H(z)+BAO+Pl & $109.29$ & $0.29^{+0.01}_{-0.01}$ & $-1.24^{+0.19}_{-0.15}$ & $0.65^{+0.85}_{-1.08}$&$0.64^{+0.04}_{-0.03}$& No \\
$\fgas^{ACT}$+H(z)+BAO+W9 & $212.64$ & $0.29^{+0.01}_{-0.01}$ & $-1.12^{+0.23}_{-0.20}$ & $-0.31^{+1.30}_{-1.53}$&$0.68^{+0.03}_{-0.04}$& No \\
$\fgas^{ACT}$+H(z)+BAO+Pl & $231.30$ & $0.29^{+0.01}_{-0.01}$ & $-1.15^{+0.20}_{-0.18}$ & $0.05^{+1.05}_{-1.17}$&$0.67^{+0.04}_{-0.04}$& No \\
\hline
\multicolumn{7}{|c|}{A$08$}\\
\hline
\multicolumn{7}{|c|}{}\\
$\fgas^{Allen}$ & $41.93$ & $0.29^{+0.03}_{-0.04}$& $-0.81^{+0.47}_{-0.48}$ & $-3.33^{+3.44}_{-3.69}$&$0.71^{+0.13}_{-0.11}$& $0.20^{+0.07}_{-0.10}\,(2.3)$\\ 
$\fgas^{ACT}$ & $174.41$ & $0.31^{+0.02}_{-0.03}$& $-0.59^{+0.36}_{-0.40}$ & $-3.23^{+2.93}_{-2.76}$&$0.75^{+0.10}_{-0.10}$& $0.24^{+0.05}_{-0.10}\,(3.2)$\\ 
$\fgas^{Allen}$+H(z)+BAO+W9 & $91.32$ & $0.28^{+0.01}_{-0.01}$ & $-1.24^{+0.21}_{-0.16}$ & $0.44^{+1.03}_{-1.42}$&$0.66^{+0.04}_{-0.04}$& No \\
$\fgas^{Allen}$+H(z)+BAO+Pl & $109.01$ & $0.29^{+0.01}_{-0.01}$ & $-1.22^{+0.19}_{-0.16}$ & $0.54^{+0.90}_{-1.12}$&$0.65^{+0.04}_{-0.04}$& No \\
$\fgas^{ACT}$+H(z)+BAO+W9 & $212.44$ & $0.29^{+0.01}_{-0.01}$ & $-1.12^{+0.23}_{-0.19}$ & $-0.11^{+1.25}_{-1.58}$&$0.67^{+0.04}_{-0.04}$& No \\
$\fgas^{ACT}$+H(z)+BAO+Pl & $231.08$ & $0.29^{+0.01}_{-0.01}$ & $-1.11^{+0.20}_{-0.18}$ & $0.00^{+1.04}_{-1.17}$&$0.66^{+0.04}_{-0.04}$& No \\
\hline
\end{tabular}
\caption{Best fits for the $\Omega_{m}$, $w_{0}$, and $w_{1}$ parameters using 
two $\fgas$ fitting functions (A$04$ and A$08$) and several data sets for the JBP parameterization. We also present
the estimated transition redshift, $z_{t}$, when the Universe changes from a decelerated to an accelerated phase. 
If the slowing down of the CA emerges, we also give the redshift, $z_{sd}$, when it occurs. 
In parenthesis we also give the distance between zero and $z_{sd}$ in units of standard deviation. It is worth to note
that we report asymmetric errors in our estimations and then we consider the D'Agostini criteria \citep{DAgostini:2004} to 
obtain $\sigma_{z_{sd}}=(\sigma_{z_{sd}}^{+}+\sigma_{z_{sd}}^{-})/2$.}
\label{tab:jbp}
\end{table*}

\begin{table*}
\centering
\begin{tabular}{|lcccccc|}
\multicolumn{7}{c}{BA parameterization}\\
\hline
Data set& $\chi^{2}_{min}$&$\Omega_{m}$& $w_{0}$&$w_{1}$&$z_{t}$&$z_{sd}$\\
\hline
\multicolumn{7}{|c|}{}\\
H(z)+BAO+W9 & $49.85$ & $0.28^{+0.01}_{-0.01}$ & $-1.16^{+0.14}_{-0.14}$ & $-0.04^{+0.30}_{-0.37}$&$0.68^{+0.05}_{-0.05}$& No \\
H(z)+BAO+Pl & $67.46$ & $0.29^{+0.01}_{-0.01}$ & $-1.12^{+0.13}_{-0.13}$ & $0.007^{+0.22}_{-0.24}$&$0.67^{+0.05}_{-0.04}$& No \\
\hline
\multicolumn{7}{|c|}{A$04$}\\
\hline
\multicolumn{7}{|c|}{}\\
$\fgas^{Allen}$ & $41.71$ & $0.28^{+0.03}_{-0.03}$& $-0.80^{+0.48}_{-0.50}$ & $-2.18^{+2.32}_{-2.54}$&$0.68^{+0.11}_{-0.12}$& $0.19^{+0.06}_{-0.09}\,(2.5)$\\ 
$\fgas^{ACT}$ & $161.52$ & $0.31^{+0.02}_{-0.03}$& $-0.85^{+0.25}_{-0.25}$ & $-1.07^{+0.79}_{-0.83}$&$0.75^{+0.08}_{-0.09}$& $0.20^{-0.06}_{-0.10}\,(2.5)$\\ 
$\fgas^{Allen}$+H(z)+BAO+W9 & $91.92$ & $0.28^{+0.01}_{-0.01}$ & $-1.17^{+0.13}_{-0.12}$ & $-0.03^{+0.28}_{-0.35}$&$0.68^{+0.05}_{-0.05}$& No \\
$\fgas^{Allen}$+H(z)+BAO+Pl & $109.78$ & $0.29^{+0.01}_{-0.01}$ & $-1.15^{+0.12}_{-0.11}$ & $0.04^{+0.20}_{-0.22}$&$0.66^{+0.04}_{-0.04}$& No \\
$\fgas^{ACT}$+H(z)+BAO+W9 & $212.49$ & $0.29^{+0.01}_{-0.01}$ & $-1.10^{+0.12}_{-0.11}$ & $-0.20^{+0.30}_{-0.37}$&$0.70^{+0.05}_{-0.05}$& No \\
$\fgas^{ACT}$+H(z)+BAO+Pl & $231.30$ & $0.29^{+0.008}_{-0.008}$ & $-1.13^{+0.10}_{-0.10}$ & $-0.02^{+0.20}_{-0.22}$&$0.68^{+0.04}_{-0.04}$& No \\
\hline
\multicolumn{7}{|c|}{A$08$}\\
\hline
\multicolumn{7}{|c|}{}\\
$\fgas^{Allen}$ & $41.85$ & $0.29^{+0.03}_{-0.04}$& $-0.86^{+0.48}_{-0.49}$ & $-1.72^{+1.98}_{-2.31}$&$0.67^{+0.14}_{-0.16}$& $0.18^{+0.07}_{-0.09}\,(2.2)$\\ 
$\fgas^{ACT}$ & $174.33$ & $0.31^{+0.02}_{-0.03}$& $-0.83^{+0.29}_{-0.26}$ & $-0.92^{+1.01}_{-1.28}$&$0.73^{+0.10}_{-0.15}$& $0.19^{+0.06}_{-0.12}\,(2.1)$\\ 
$\fgas^{Allen}$+H(z)+BAO+W9 & $91.75$ & $0.28^{+0.01}_{-0.01}$ & $-1.16^{+0.13}_{-0.13}$ & $-0.06^{+0.29}_{-0.36}$&$0.68^{+0.05}_{-0.05}$& No \\
$\fgas^{Allen}$+H(z)+BAO+Pl & $109.39$ & $0.29^{+0.01}_{-0.01}$ & $-1.13^{+0.12}_{-0.12}$ & $0.01^{+0.21}_{-0.23}$&$0.67^{+0.04}_{-0.04}$& No \\
$\fgas^{ACT}$+H(z)+BAO+W9 & $212.35$ & $0.29^{+0.01}_{-0.01}$ & $-1.07^{+0.12}_{-0.12}$ & $-0.20^{+0.29}_{-0.36}$&$0.70^{+0.05}_{-0.05}$& No \\
$\fgas^{ACT}$+H(z)+BAO+Pl & $229.97$ & $0.30^{+0.01}_{-0.01}$ & $-1.06^{+0.11}_{-0.11}$ & $0.08^{+0.20}_{-0.22}$&$0.68^{+0.05}_{-0.05}$& No \\
\hline
\end{tabular}
\caption{The same as Table C1 for the BA parameterization.}
\label{tab:ba}
\end{table*}

\begin{table*}
\centering
\begin{tabular}{|lcccccc|}
\multicolumn{7}{c}{FSLLI parameterization}\\
\hline
Data set& $\chi^{2}_{min}$&$\Omega_{m}$& $w_{0}$&$w_{1}$&$z_{t}$&$z_{sd}$\\
\hline
\multicolumn{7}{|c|}{}\\
H(z)+BAO+W9 & $49.24$ & $0.28^{+0.01}_{-0.01}$ & $-1.27^{+0.20}_{-0.19}$ & $0.40^{+0.68}_{-0.78}$&$0.64^{+0.06}_{-0.06}$& No \\
H(z)+BAO+Pl & $67.01$ & $0.29^{+0.01}_{-0.01}$ & $-1.22^{+0.18}_{-0.17}$ & $0.32^{+0.54}_{-0.57}$&$0.63^{+0.05}_{-0.05}$& No \\
\hline
\multicolumn{7}{|c|}{A$04$}\\
\hline
\multicolumn{7}{|c|}{}\\
$\fgas^{Allen}$ & $41.75$ & $0.28^{+0.03}_{-0.03}$& $-0.79^{+0.48}_{-0.49}$ & $-2.77^{+2.90}_{-3.07}$&$0.69^{+0.10}_{-0.10}$& $0.21^{+0.07}_{-0.09}\,(2.6)$\\ 
$\fgas^{ACT}$ & $161.59$ & $0.31^{+0.02}_{-0.03}$& $-0.74^{+0.30}_{-0.31}$ & $-1.83^{+1.32}_{-1.30}$&$0.75^{+0.08}_{-0.08}$& $0.24^{+0.08}_{-0.11}\,(2.5)$\\ 
$\fgas^{Allen}$+H(z)+BAO+W9 & $91.35$ & $0.28^{+0.01}_{-0.01}$ & $-1.27^{+0.18}_{-0.17}$ & $0.36^{+0.64}_{-0.73}$&$0.64^{+0.06}_{-0.06}$& No \\
$\fgas^{Allen}$+H(z)+BAO+Pl & $109.22$ & $0.28^{+0.01}_{-0.01}$ & $-1.23^{+0.16}_{-0.15}$ & $0.36^{+0.50}_{-0.52}$&$0.63^{+0.05}_{-0.05}$& No \\
$\fgas^{ACT}$+H(z)+BAO+W9 & $212.60$ & $0.29^{+0.01}_{-0.01}$ & $-1.13^{+0.18}_{-0.17}$ & $-0.15^{+0.67}_{-0.74}$&$0.68^{+0.04}_{-0.06}$& No \\
$\fgas^{ACT}$+H(z)+BAO+Pl & $231.23$ & $0.29^{+0.009}_{-0.009}$ & $-1.16^{+0.15}_{-0.14}$ & $0.07^{+0.49}_{-0.52}$&$0.66^{+0.05}_{-0.05}$& No \\
\hline
\multicolumn{7}{|c|}{A$08$}\\
\hline
\multicolumn{7}{|c|}{}\\
$\fgas^{Allen}$ & $41.86$ & $0.29^{+0.03}_{-0.04}$& $-0.83^{+0.48}_{-0.50}$ & $-2.29^{+2.56}_{-2.87}$&$0.69^{+0.14}_{-0.13}$& $0.23^{+0.12}_{-0.10}\,(2.0)$\\ 
$\fgas^{ACT}$ & $174.33$ & $0.31^{+0.02}_{-0.03}$& $-0.83^{+0.29}_{-0.26}$ & $-0.92^{+1.01}_{-1.28}$&$0.73^{+0.10}_{-0.13}$& $0.26^{+0.08}_{-0.11}\,(2.7)$\\ 
$\fgas^{Allen}$+H(z)+BAO+W9 & $91.33$ & $0.28^{+0.01}_{-0.01}$ & $-1.25^{+0.18}_{-0.17}$ & $0.31^{+0.66}_{-0.74}$&$0.64^{+0.06}_{-0.06}$& No \\
$\fgas^{Allen}$+H(z)+BAO+Pl & $109.01$ & $0.28^{+0.01}_{-0.01}$ & $-1.22^{+0.16}_{-0.16}$ & $0.31^{+0.52}_{-0.54}$&$0.64^{+0.05}_{-0.05}$& No \\
$\fgas^{ACT}$+H(z)+BAO+W9 & $212.35$ & $0.29^{+0.01}_{-0.01}$ & $-1.07^{+0.12}_{-0.12}$ & $-0.20^{+0.29}_{-0.36}$&$0.67^{+0.05}_{-0.06}$& No \\
$\fgas^{ACT}$+H(z)+BAO+Pl & $229.97$ & $0.30^{+0.01}_{-0.01}$ & $-1.06^{+0.11}_{-0.11}$ & $-0.08^{+0.20}_{-0.22}$&$0.66^{+0.05}_{-0.05}$& No \\
\hline
\end{tabular}
\caption{The same as Table C1 for the FSLLI parameterization.}
\label{tab:fslli}
\end{table*}

\begin{table*}
\centering
\begin{tabular}{|lcccccc|}
\multicolumn{7}{c}{FSLLII parameterization}\\
\hline
Data set& $\chi^{2}_{min}$&$\Omega_{m}$& $w_{0}$&$w_{1}$&$z_{t}$&$z_{sd}$\\
\hline
\multicolumn{7}{|c|}{}\\
H(z)+BAO+W9 & $49.82$ & $0.29^{+0.01}_{-0.01}$ & $-1.13^{+0.10}_{-0.10}$ & $-0.36^{+0.58}_{-0.74}$&$0.71^{+0.06}_{-0.05}$& No \\
H(z)+BAO+Pl & $67.61$ & $0.29^{+0.01}_{-0.01}$ & $-1.09^{+0.10}_{-0.10}$ & $-0.13^{+0.39}_{-0.44}$&$0.69^{+0.05}_{-0.04}$& No \\
\hline
\multicolumn{7}{|c|}{A$04$}\\
\hline
\multicolumn{7}{|c|}{}\\
$\fgas^{Allen}$ & $41.50$ & $0.28^{+0.02}_{-0.02}$& $-1.05^{+0.25}_{-0.27}$ & $-4.58^{+4.07}_{-3.64}$&$0.72^{+0.11}_{-0.13}$& No\\ 
$\fgas^{ACT}$ & $161.64$ & $0.31^{+0.02}_{-0.03}$& $-0.99^{+0.17}_{-0.18}$ & $-2.54^{+1.95}_{-2.41}$&$0.75^{+0.10}_{-0.10}$& No\\ 
$\fgas^{Allen}$+H(z)+BAO+W9 & $91.95$ & $0.29^{+0.01}_{-0.01}$ & $-1.15^{+0.09}_{-0.09}$ & $-0.30^{+0.56}_{-0.72}$&$0.70^{+0.05}_{-0.05}$& No \\
$\fgas^{Allen}$+H(z)+BAO+Pl & $109.98$ & $0.29^{+0.01}_{-0.01}$ & $-1.12^{+0.09}_{-0.09}$ & $-0.05^{+0.36}_{-0.40}$&$0.68^{+0.04}_{-0.03}$& No \\
$\fgas^{ACT}$+H(z)+BAO+W9 & $212.28$ & $0.29^{+0.01}_{-0.01}$ & $-1.12^{+0.08}_{-0.07}$ & $-0.55^{+0.58}_{-0.76}$&$0.72^{+0.05}_{-0.05}$& No \\
$\fgas^{ACT}$+H(z)+BAO+Pl & $231.34$ & $0.29^{+0.007}_{-0.007}$ & $-1.13^{+0.07}_{-0.07}$ & $-0.09^{+0.34}_{-0.39}$&$0.69^{+0.04}_{+0.03}$& No \\
\hline
\multicolumn{7}{|c|}{A$08$}\\
\hline
\multicolumn{7}{|c|}{}\\
$\fgas^{Allen}$ & $41.60$ & $0.29^{+0.03}_{-0.04}$& $-1.04^{+0.29}_{-0.32}$ & $-4.30^{+3.87}_{-3.72}$&$0.69^{+0.14}_{-0.16}$& No\\ 
$\fgas^{ACT}$ & $170.02$ & $0.31^{+0.02}_{-0.03}$& $-0.95^{+0.16}_{-0.17}$ & $-1.94^{+2.25}_{-3.38}$&$0.73^{+0.12}_{-0.17}$& No\\ 
$\fgas^{Allen}$+H(z)+BAO+W9 & $91.64$ & $0.29^{+0.01}_{-0.01}$ & $-1.14^{+0.10}_{-0.09}$ & $-0.36^{+0.57}_{-0.76}$&$0.71^{+0.06}_{-0.05}$& No \\
$\fgas^{Allen}$+H(z)+BAO+Pl & $109.53$ & $0.29^{+0.01}_{-0.01}$ & $-1.11^{+0.09}_{-0.09}$ & $-0.08^{+0.36}_{-0.40}$&$0.68^{+0.05}_{-0.03}$& No \\
$\fgas^{ACT}$+H(z)+BAO+W9 & $211.99$ & $0.29^{+0.01}_{-0.01}$ & $-1.08^{+0.08}_{-0.08}$ & $-0.57^{+0.55}_{-0.73}$&$0.73^{+0.05}_{-0.05}$& No \\
$\fgas^{ACT}$+H(z)+BAO+Pl & $229.97$ & $0.30^{+0.01}_{-0.01}$ & $-1.06^{+0.08}_{-0.08}$ & $-0.25^{+0.35}_{-0.38}$&$0.69^{+0.05}_{-0.04}$& No \\
\hline
\end{tabular}
\caption{The same as Table C1 for the FSLLII parameterization.}
\label{tab:fsllii}
\end{table*}

\begin{table*}
\centering
\begin{tabular}{|lcccccc|}
\multicolumn{7}{c}{SL parameterization}\\
\hline
Data set& $\chi^{2}_{min}$&$\Omega_{m}$& $w_{0}$&$w_{1}$&$z_{t}$&$z_{sd}$\\
\hline
\multicolumn{7}{|c|}{}\\
H(z)+BAO+W9 & $50.01$ & $0.29^{+0.01}_{-0.01}$ & $-1.13^{+0.15}_{-0.14}$ & $ -1.20^{+0.09}_{-0.12}$&$0.69^{+0.05}_{-0.04}$& No \\
H(z)+BAO+Pl & $68.52$ & $0.29^{+0.01}_{-0.01}$ & $-1.10^{+0.13}_{-0.13}$ & $-1.13^{+0.05}_{-0.05}$&$0.68^{+0.04}_{-0.03}$& No \\
\hline
\multicolumn{7}{|c|}{A$04$}\\
\hline
\multicolumn{7}{|c|}{}\\
$\fgas^{Allen}$ & $41.71$ & $0.28^{+0.03}_{-0.03}$& $-0.83^{+0.50}_{-0.56}$ & $-1.92^{+0.93}_{-1.02}$&$0.67^{+0.12}_{-0.15}$& $0.24^{+0.12}_{-0.11}\,(2.0)$\\ 
$\fgas^{ACT}$ & $161.64$ & $0.31^{+0.02}_{-0.03}$& $-0.81^{+0.27}_{-0.27}$ & $-1.48^{+0.28}_{-0.29}$&$0.76^{+0.09}_{-0.09}$& $0.25^{+0.09}_{+0.13}\,(2.2)$\\ 
$\fgas^{Allen}$+H(z)+BAO+W9 & $92.11$ & $0.28^{+0.01}_{-0.01}$ & $-1.15^{+0.13}_{-0.12}$ & $-1.21^{+0.09}_{-0.12}$&$0.69^{+0.05}_{-0.04}$& No \\
$\fgas^{Allen}$+H(z)+BAO+Pl & $110.97$ & $0.29^{+0.01}_{-0.01}$ & $-1.13^{+0.12}_{-0.12}$ & $-1.13^{+0.04}_{-0.50}$&$0.67^{+0.04}_{-0.03}$& No \\
$\fgas^{ACT}$+H(z)+BAO+W9 & $212.46$ & $0.29^{+0.01}_{-0.01}$ & $-1.08^{+0.13}_{-0.11}$ & $-1.23^{+0.09}_{-0.11}$&$0.71^{+0.04}_{-0.04}$& No \\
$\fgas^{ACT}$+H(z)+BAO+Pl & $231.91$ & $0.29^{+0.008}_{-0.008}$ & $-1.12^{+0.10}_{-0.10}$ & $-1.15^{+0.04}_{-0.05}$&$0.68^{+0.04}_{-0.03}$& No \\
\hline
\multicolumn{7}{|c|}{A$08$}\\
\hline
\multicolumn{7}{|c|}{}\\
$\fgas^{Allen}$ & $41.76$ & $0.29^{+0.03}_{-0.04}$& $-0.87^{+0.50}_{-0.54}$ & $-1.76^{+0.77}_{-0.93}$&$0.66^{+0.15}_{-0.18}$& $0.25^{+0.13}_{-0.12}\,(2.0)$\\ 
$\fgas^{ACT}$ & $179.20$ & $0.31^{+0.02}_{-0.04}$& $-0.79^{+0.32}_{-0.29}$ & $-1.39^{+0.42}_{-0.49}$&$0.74^{+0.11}_{-0.15}$& $0.27^{+0.09}_{-0.13}\,(2.4)$\\ 
$\fgas^{Allen}$+H(z)+BAO+W9 & $91.87$ & $0.29^{+0.01}_{-0.01}$ & $-1.13^{+0.13}_{-0.13}$ & $-1.21^{+0.09}_{-0.12}$&$0.69^{+0.05}_{-0.04}$& No \\
$\fgas^{Allen}$+H(z)+BAO+Pl & $110.27$ & $0.29^{+0.01}_{-0.01}$ & $-1.11^{+0.12}_{-0.12}$ & $-1.13^{+0.05}_{-0.05}$&$0.67^{+0.04}_{-0.03}$& No \\
$\fgas^{ACT}$+H(z)+BAO+W9 & $212.29$ & $0.29^{+0.01}_{-0.01}$ & $-1.05^{+0.13}_{-0.12}$ & $-1.20^{+0.10}_{-0.12}$&$0.71^{+0.05}_{-0.05}$& No \\
$\fgas^{ACT}$+H(z)+BAO+Pl & $230.83$ & $0.30^{+0.01}_{-0.01}$ & $-1.04^{+0.11}_{-0.11}$ & $-1.12^{+0.05}_{-0.05}$&$0.69^{+0.04}_{-0.04}$& No \\
\hline
\end{tabular}
\caption{The same as Table C1 for the SL parameterization.}
\label{tab:sl}
\end{table*}

\begin{figure*}
\centering
JBP parameterization\par\smallskip
\begin{tabular}{ccc}
\subfloat[$\fgas^{\mathrm{Allen}}$]{\includegraphics[width=0.35\textwidth]{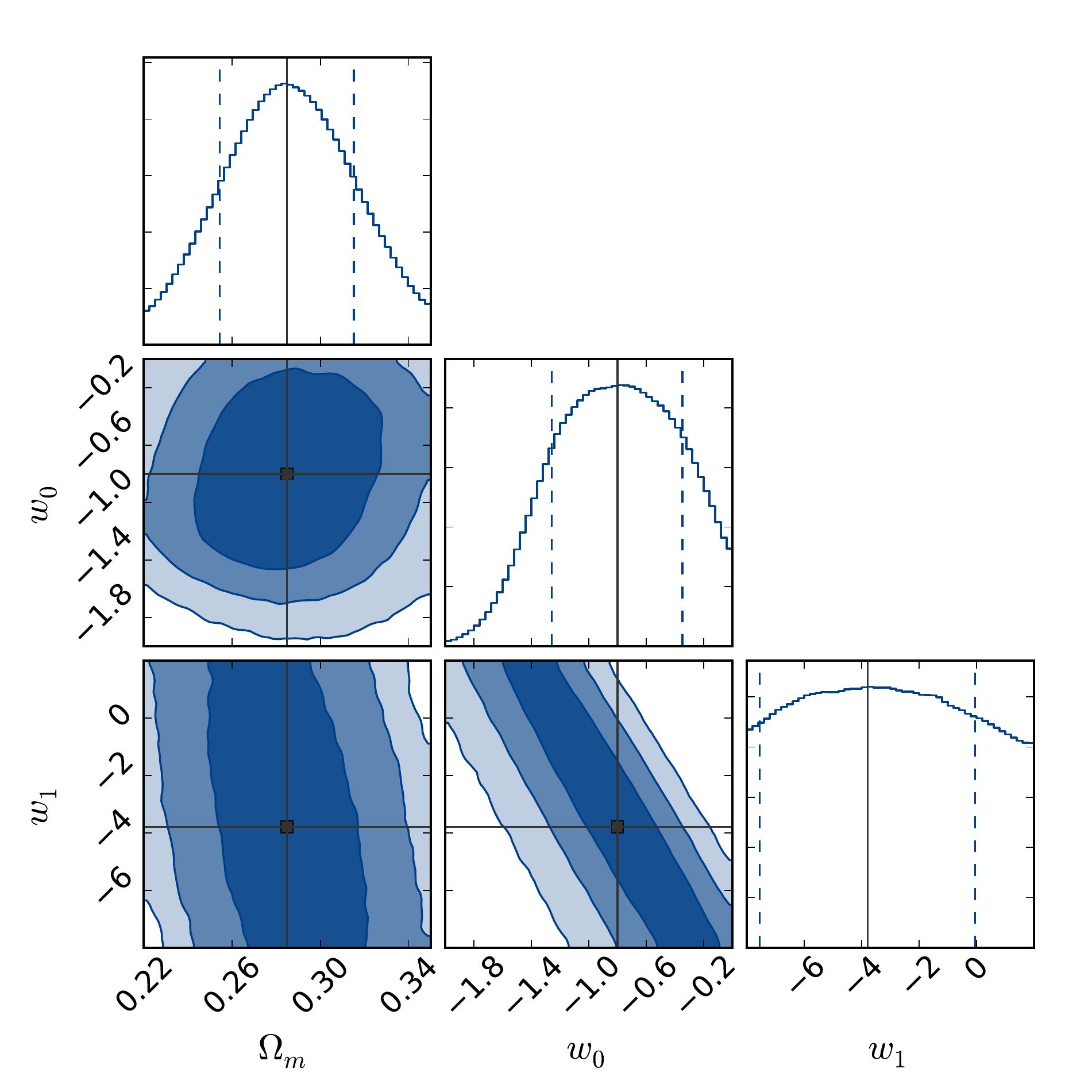}} & 
\subfloat[$\fgas^{\mathrm{Allen}}+H(z)$+BAO+WMAP]{\includegraphics[width=0.35\textwidth]{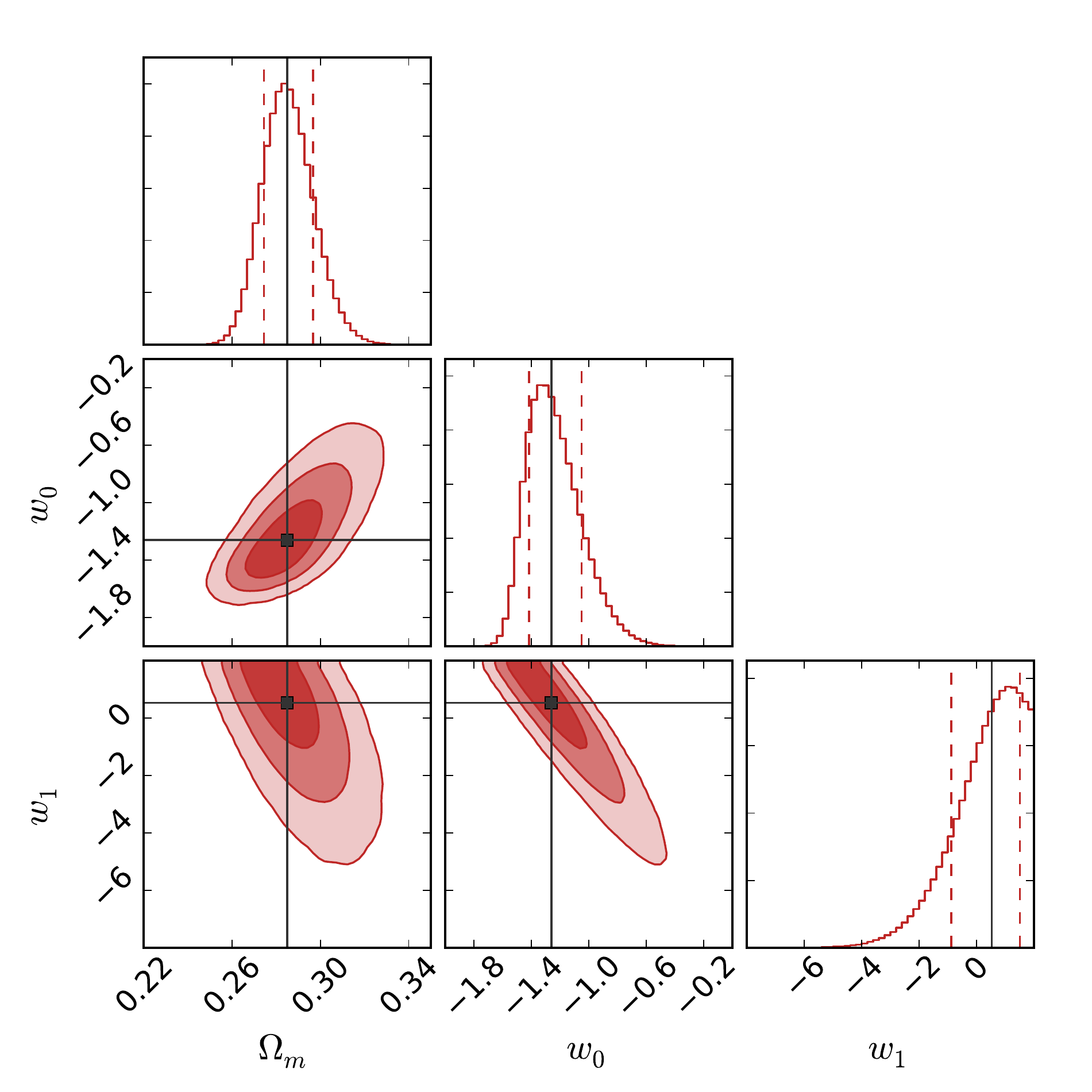}} &
\subfloat[$\fgas^{\mathrm{Allen}}+H(z)$+BAO+Planck]{\includegraphics[width=0.35\textwidth]{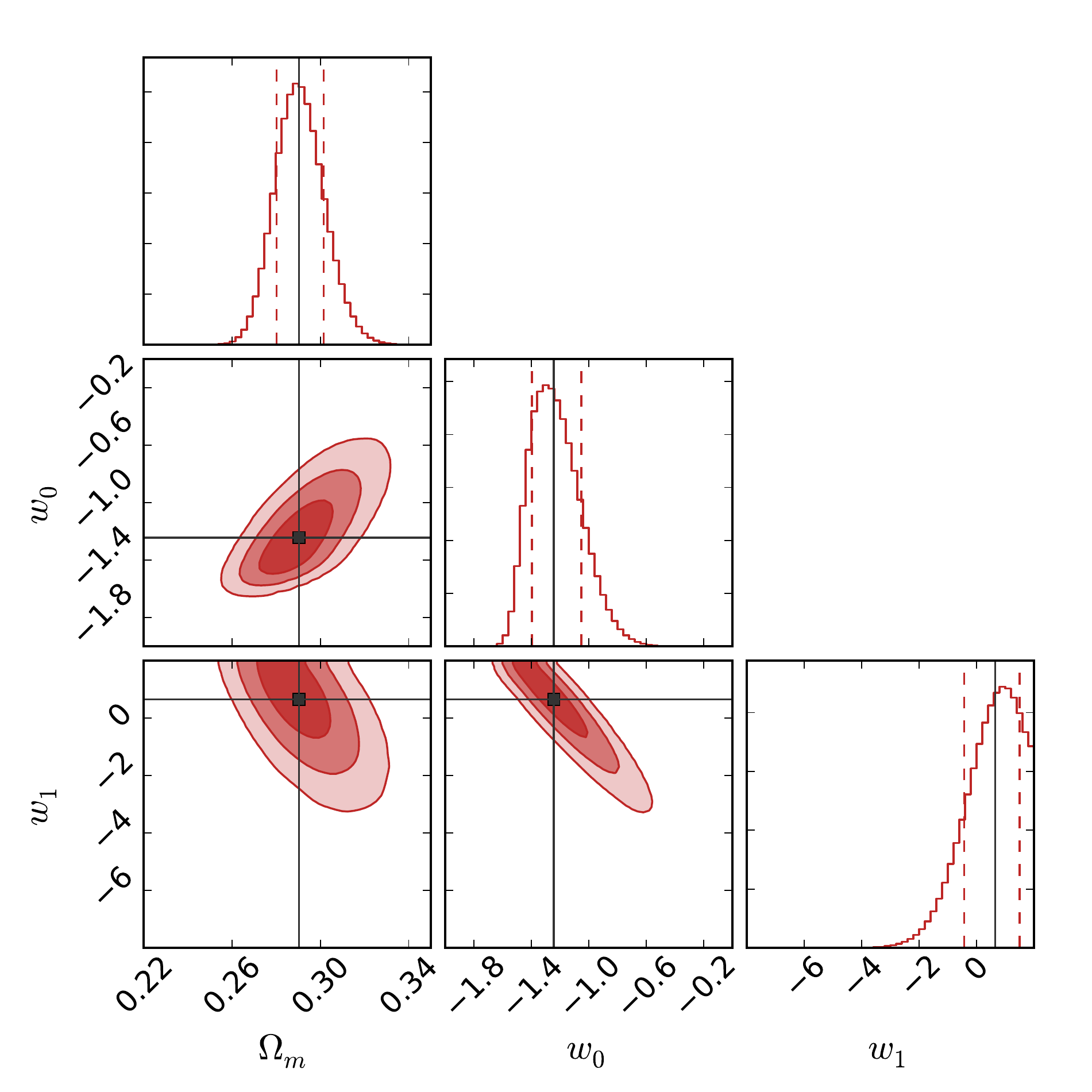}}\\
\subfloat[$\fgas^{\mathrm{ACT}}$]{\includegraphics[width=0.35\textwidth]{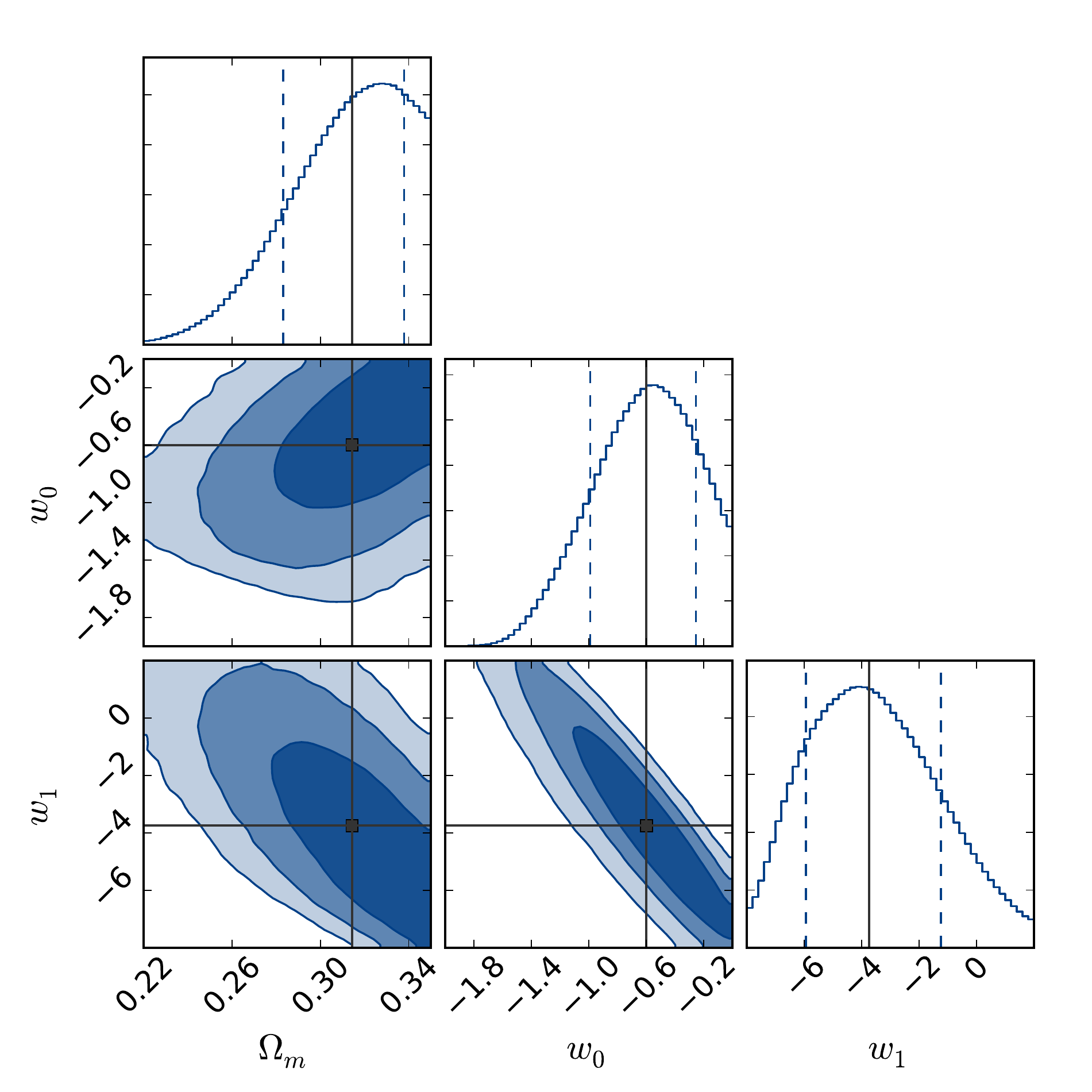}} & 
\subfloat[$\fgas^{\mathrm{ACT}}+H(z)$+BAO+WMAP]{\includegraphics[width=0.35\textwidth]{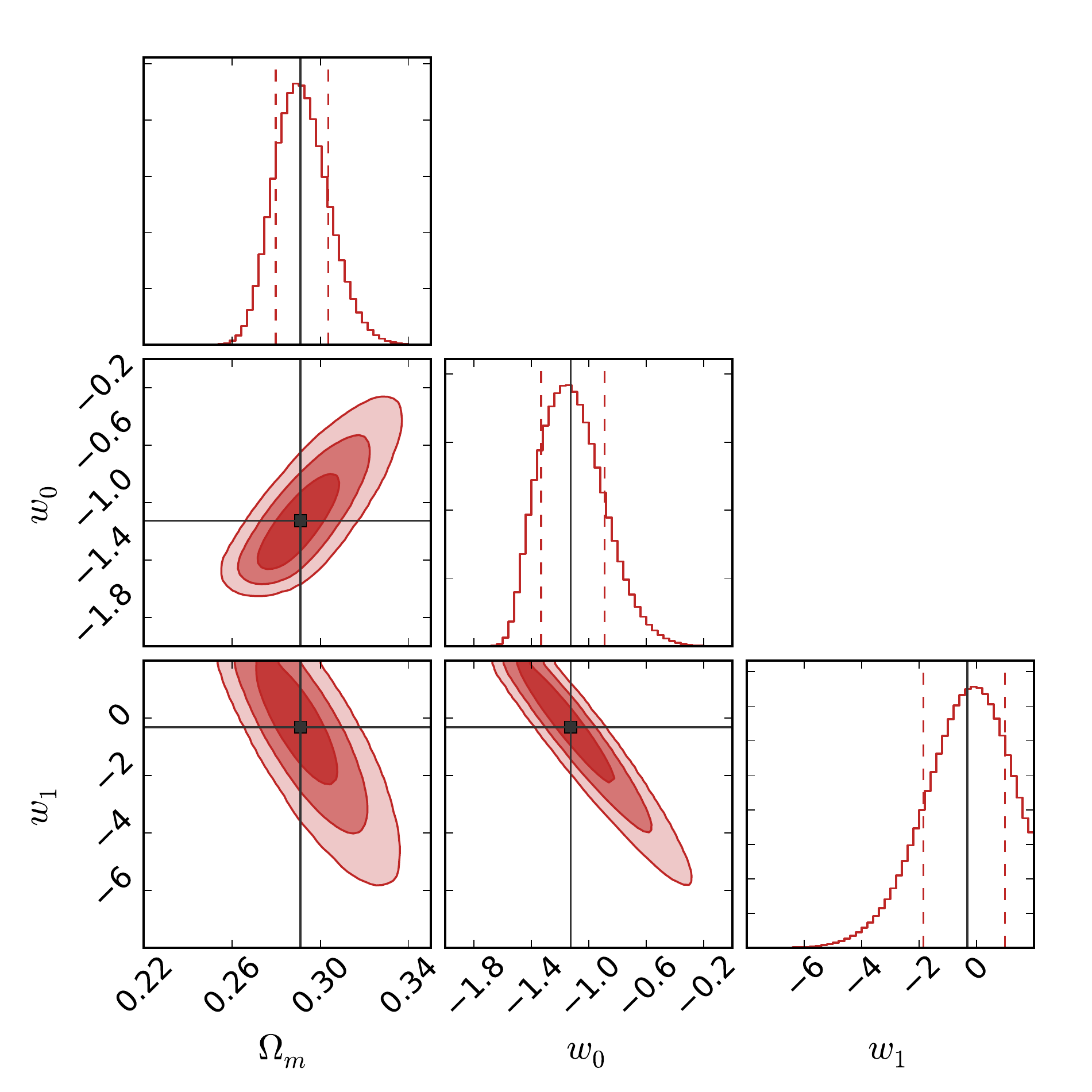}} &
\subfloat[$\fgas^{\mathrm{ACT}}+H(z)$+BAO+Planck]{\includegraphics[width=0.35\textwidth]{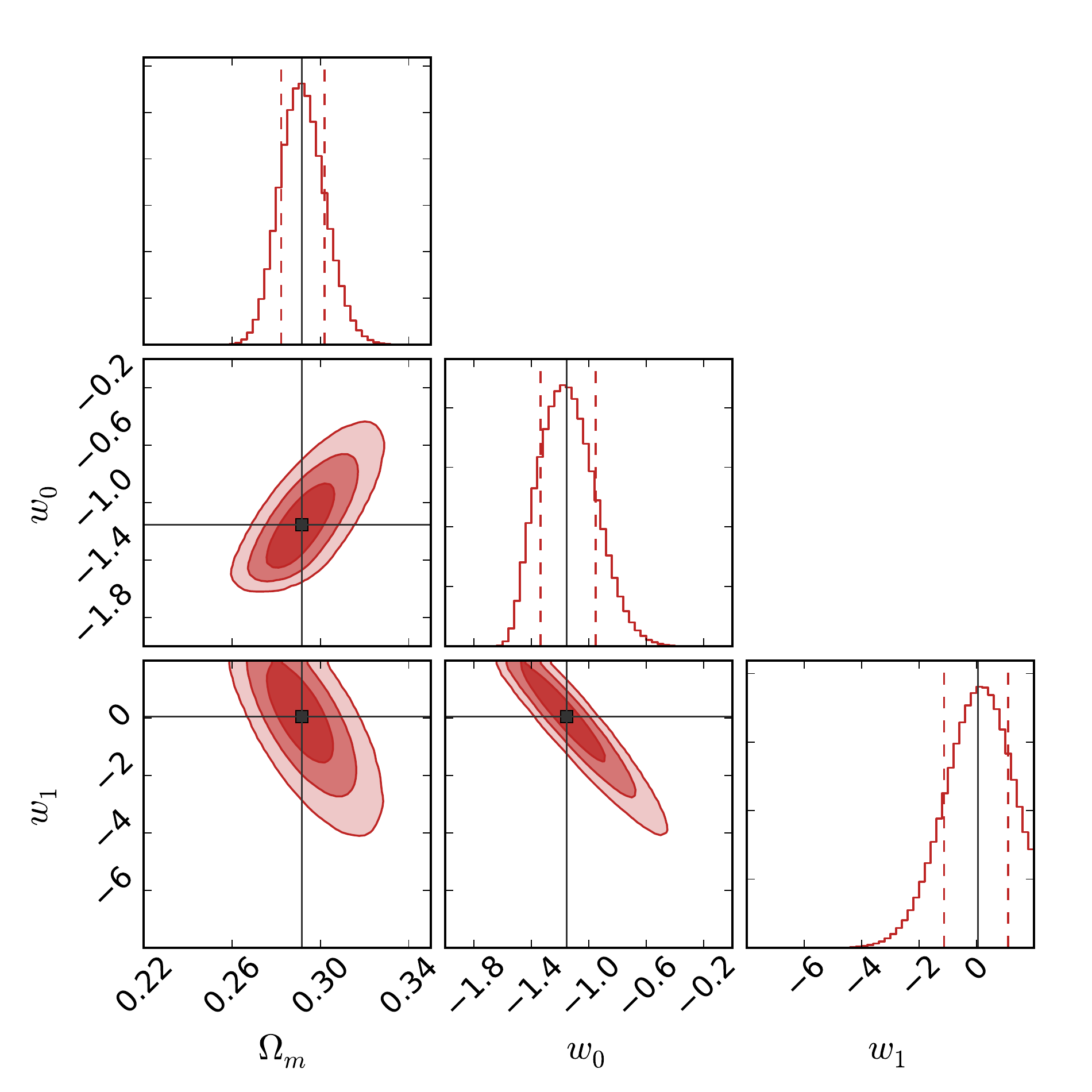}}\\
\end{tabular}
\caption{1D marginalized posterior distributions and 2D $68\%$, $95\%$, $99\%$ confidence levels for the $\Omega_{m}$, $w_{0}$, and $w_{1}$ parameters of the 
JBP parameterization. The top-left panel shows the confidence regions obtained using the 
$\fgas^{\mathrm{Allen}}$ data alone. The top-middle and top-right panels show the same contours computed from
the $\fgas^{\mathrm{Allen}}+H(z)+$BAO$+$CMB joint analysis using WMAP and Planck data respectively.
The bottom panels show the same as the top panels obtained using
the $\fgas^{\mathrm{ACT}}$ data. The squares show the best fit values.}
\label{fig:JBP}
\end{figure*}

\begin{figure*}
\centering
BA parameterization\par\smallskip
\begin{tabular}{ccc}
\subfloat[$\fgas^{\mathrm{Allen}}$]{\includegraphics[width=0.35\textwidth]{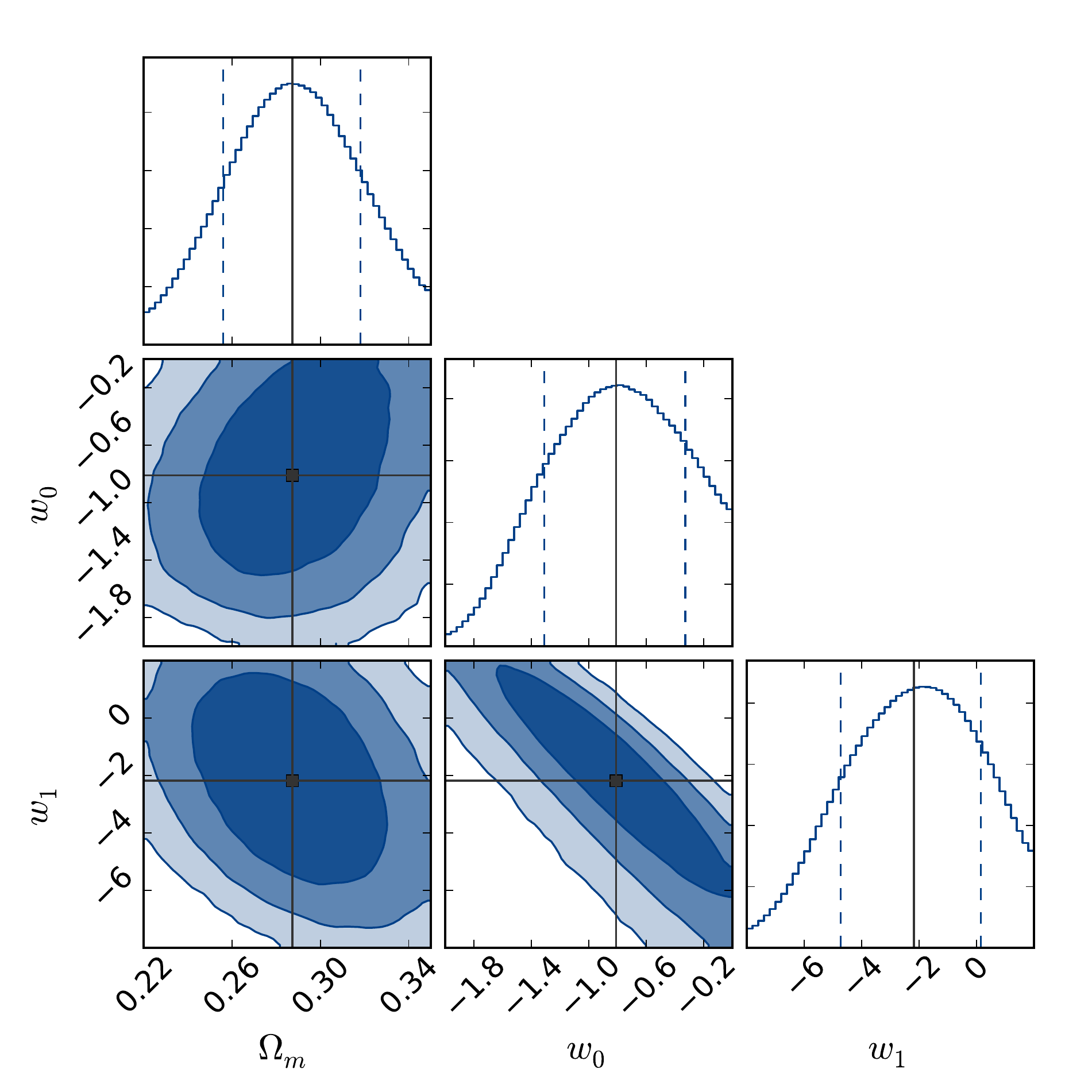}} & 
\subfloat[$\fgas^{\mathrm{Allen}}+H(z)$+BAO+WMAP]{\includegraphics[width=0.35\textwidth]{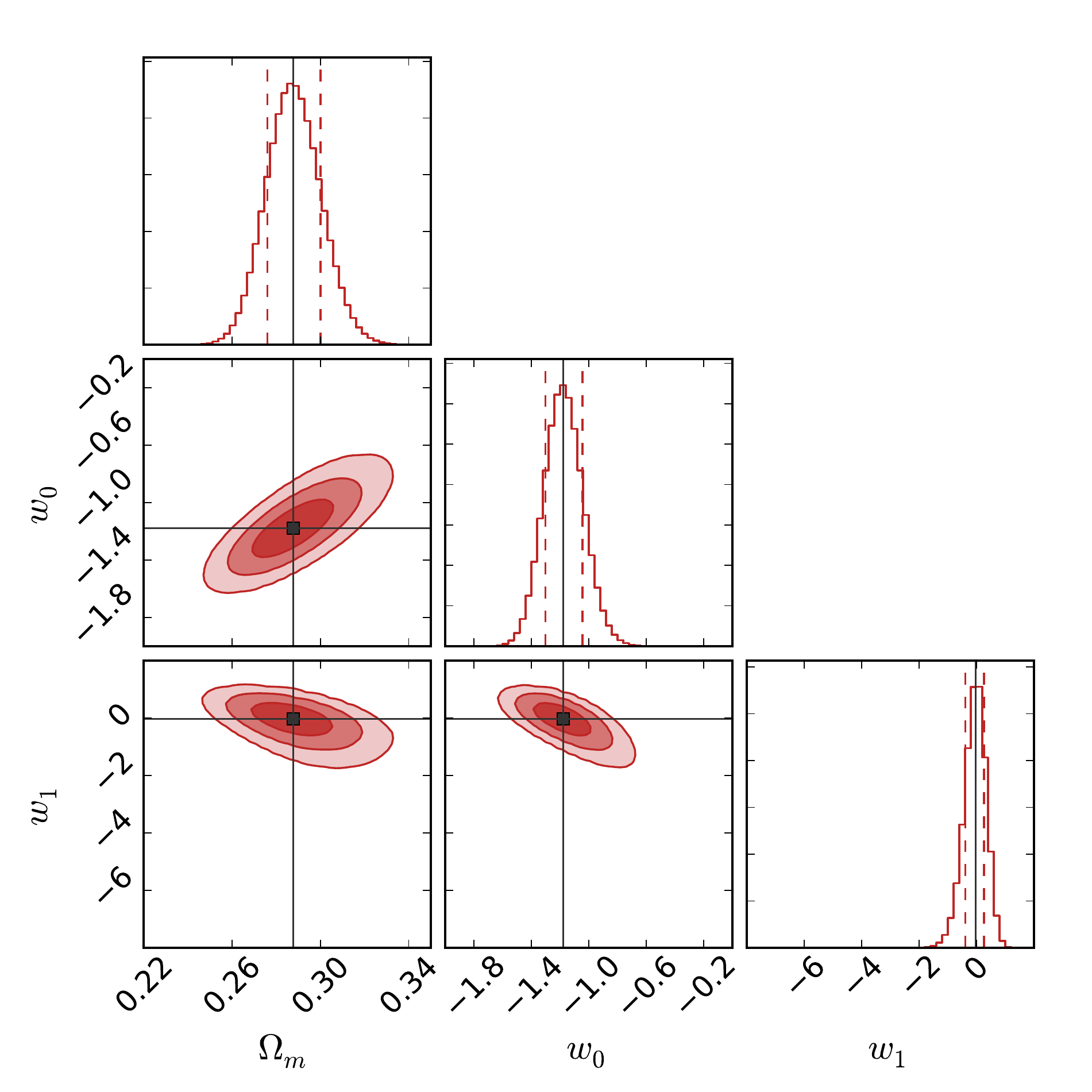}} &
\subfloat[$\fgas^{\mathrm{Allen}}+H(z)$+BAO+Planck]{\includegraphics[width=0.35\textwidth]{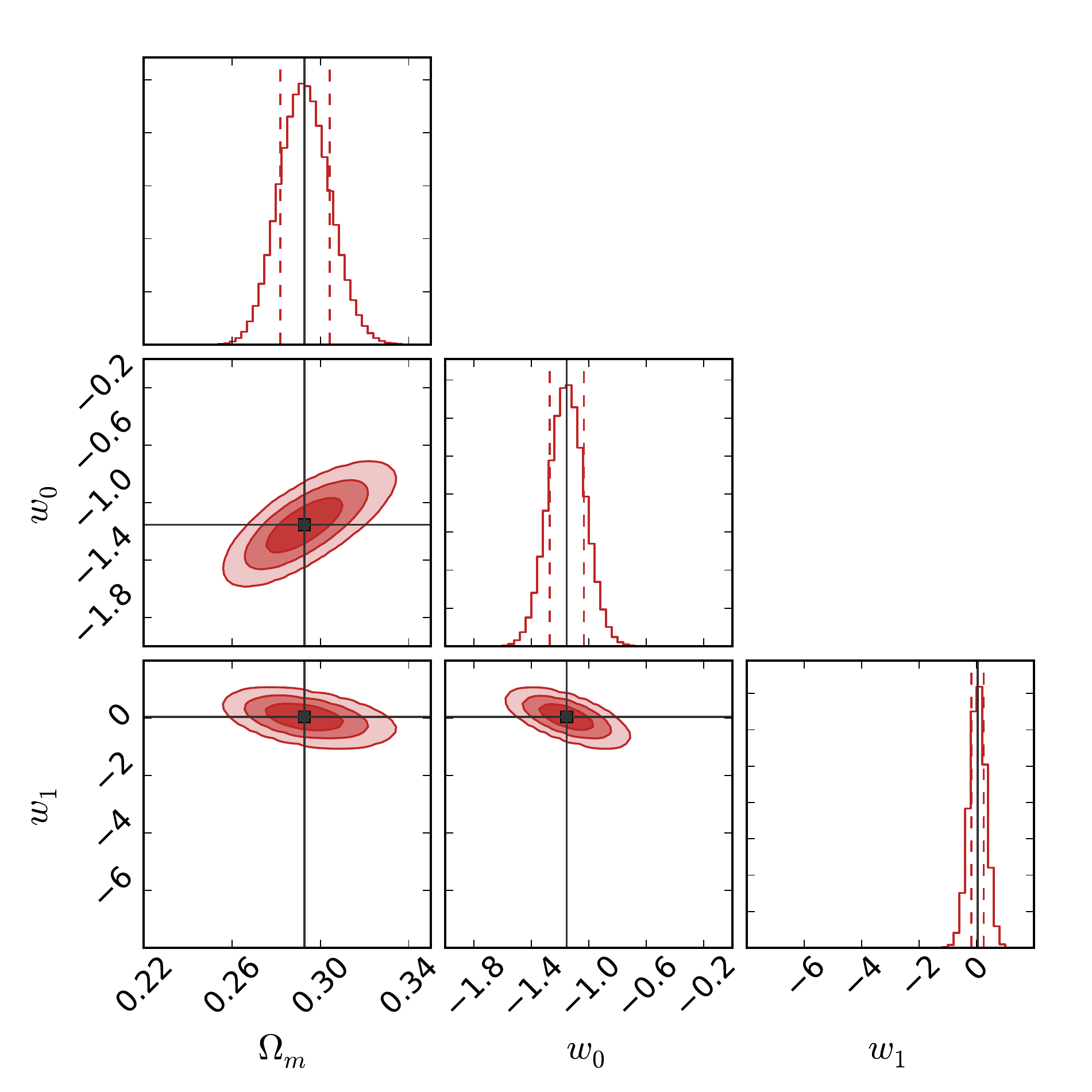}}\\
\subfloat[$\fgas^{\mathrm{ACT}}$]{\includegraphics[width=0.35\textwidth]{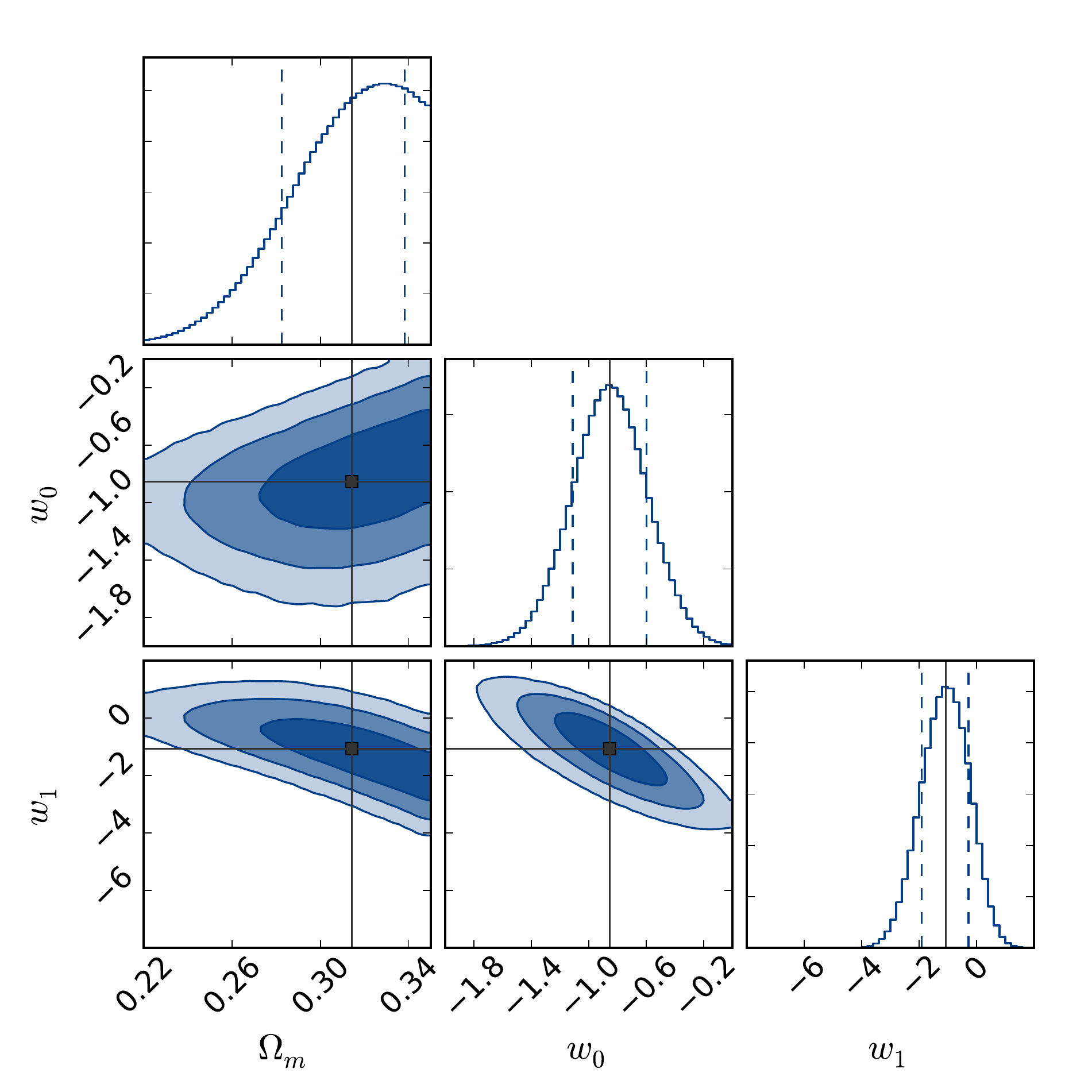}} & 
\subfloat[$\fgas^{\mathrm{ACT}}+H(z)$+BAO+WMAP]{\includegraphics[width=0.35\textwidth]{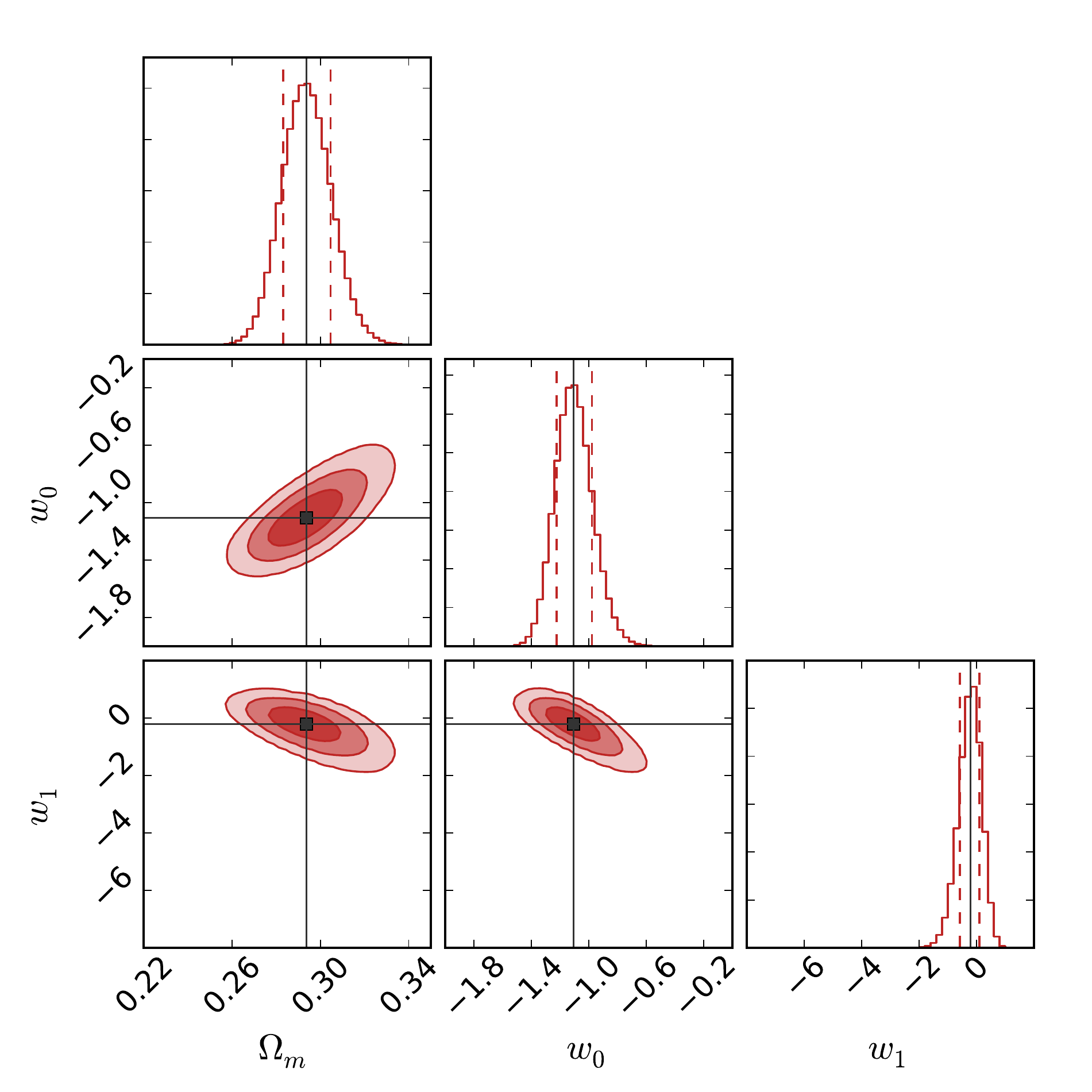}} &
\subfloat[$\fgas^{\mathrm{ACT}}+H(z)$+BAO+Planck]{\includegraphics[width=0.35\textwidth]{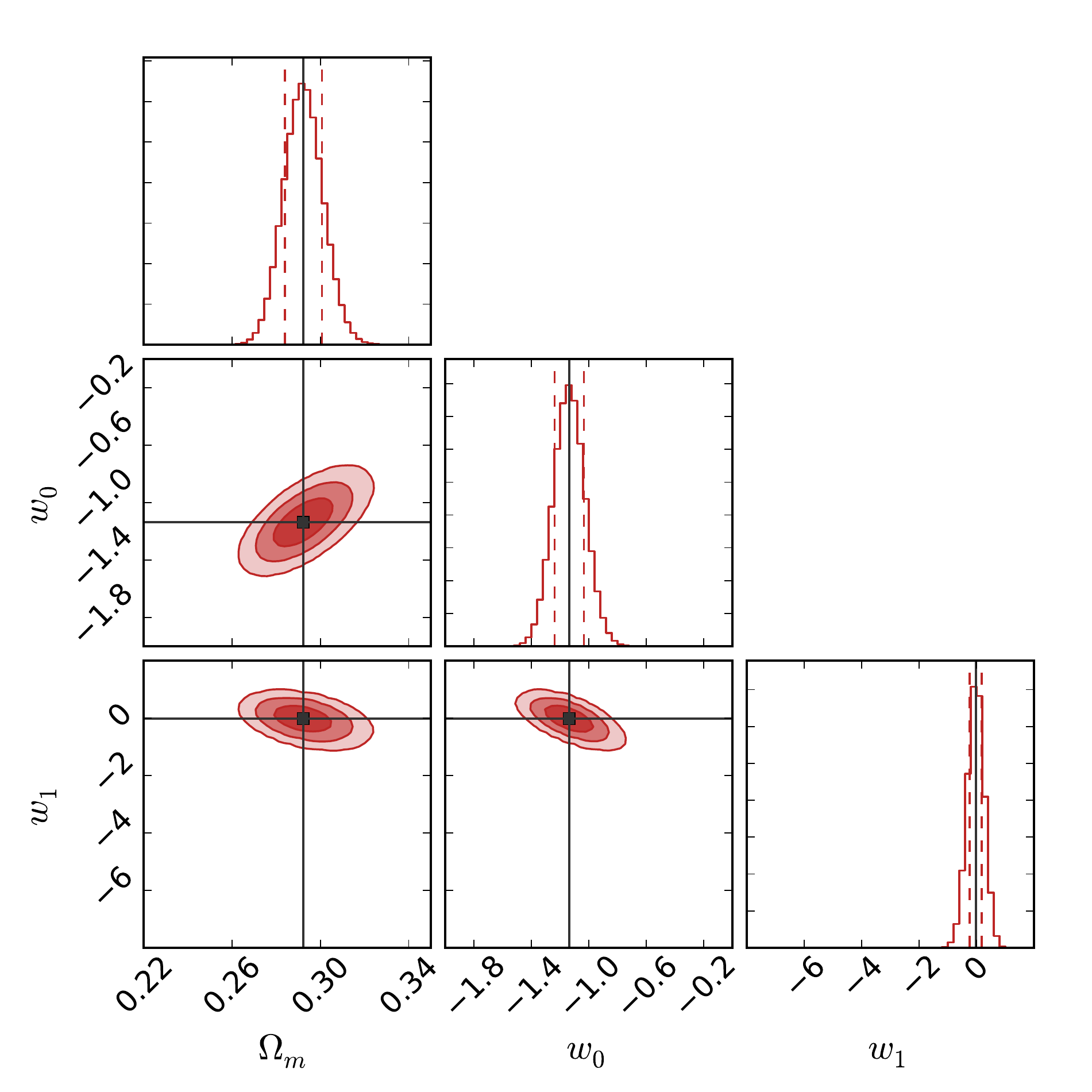}}\\

\end{tabular}
\caption{The same as Fig. C1 for the BA parameterization.}
\label{fig:BA}
\end{figure*}

\begin{figure*}
\centering
FSLLI parameterization\par\smallskip
\begin{tabular}{ccc}
\subfloat[$\fgas^{\mathrm{Allen}}$]{\includegraphics[width=0.35\textwidth]{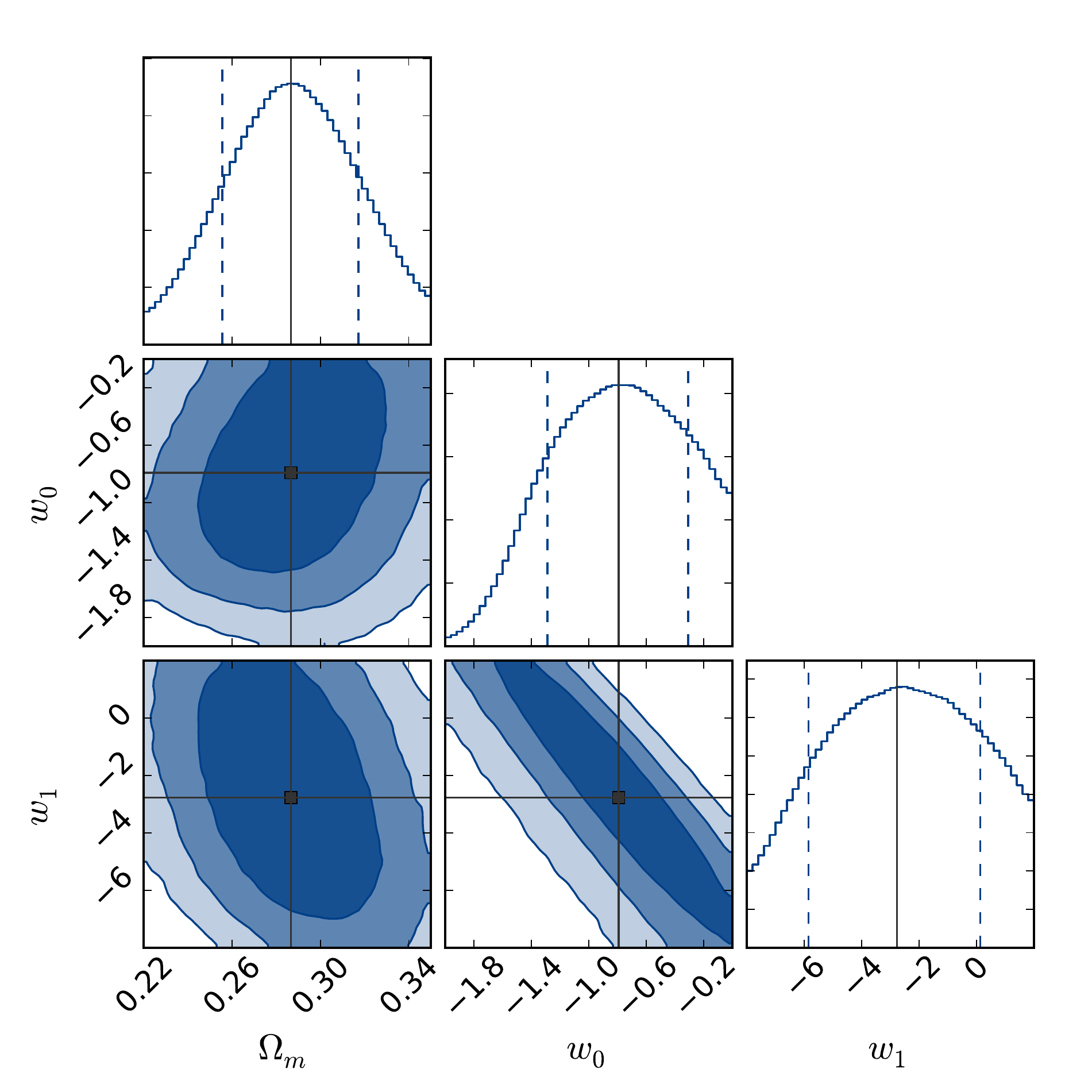}} & 
\subfloat[$\fgas^{\mathrm{Allen}}+H(z)$+BAO+WMAP]{\includegraphics[width=0.35\textwidth]{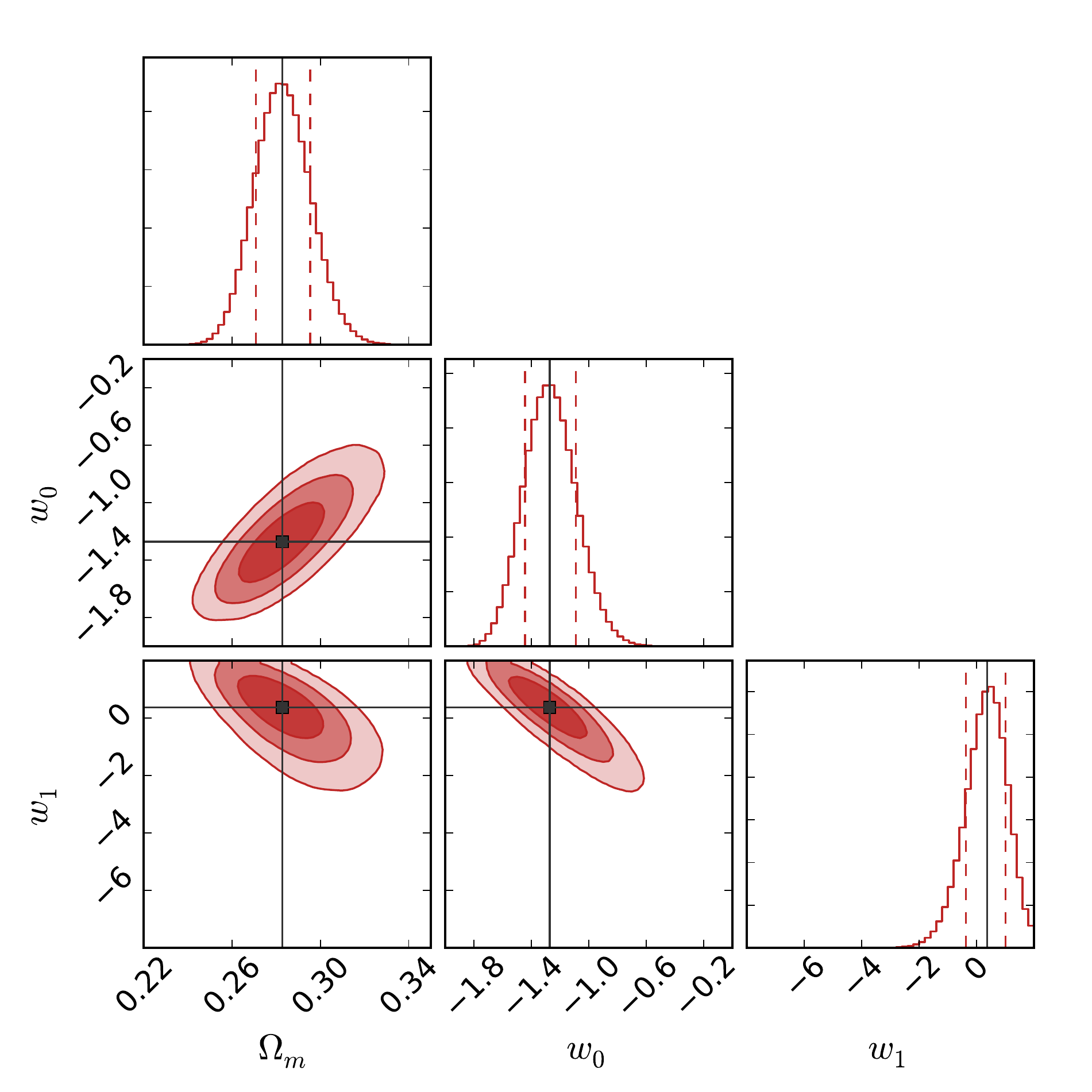}} &
\subfloat[$\fgas^{\mathrm{Allen}}+H(z)$+BAO+Planck]{\includegraphics[width=0.35\textwidth]{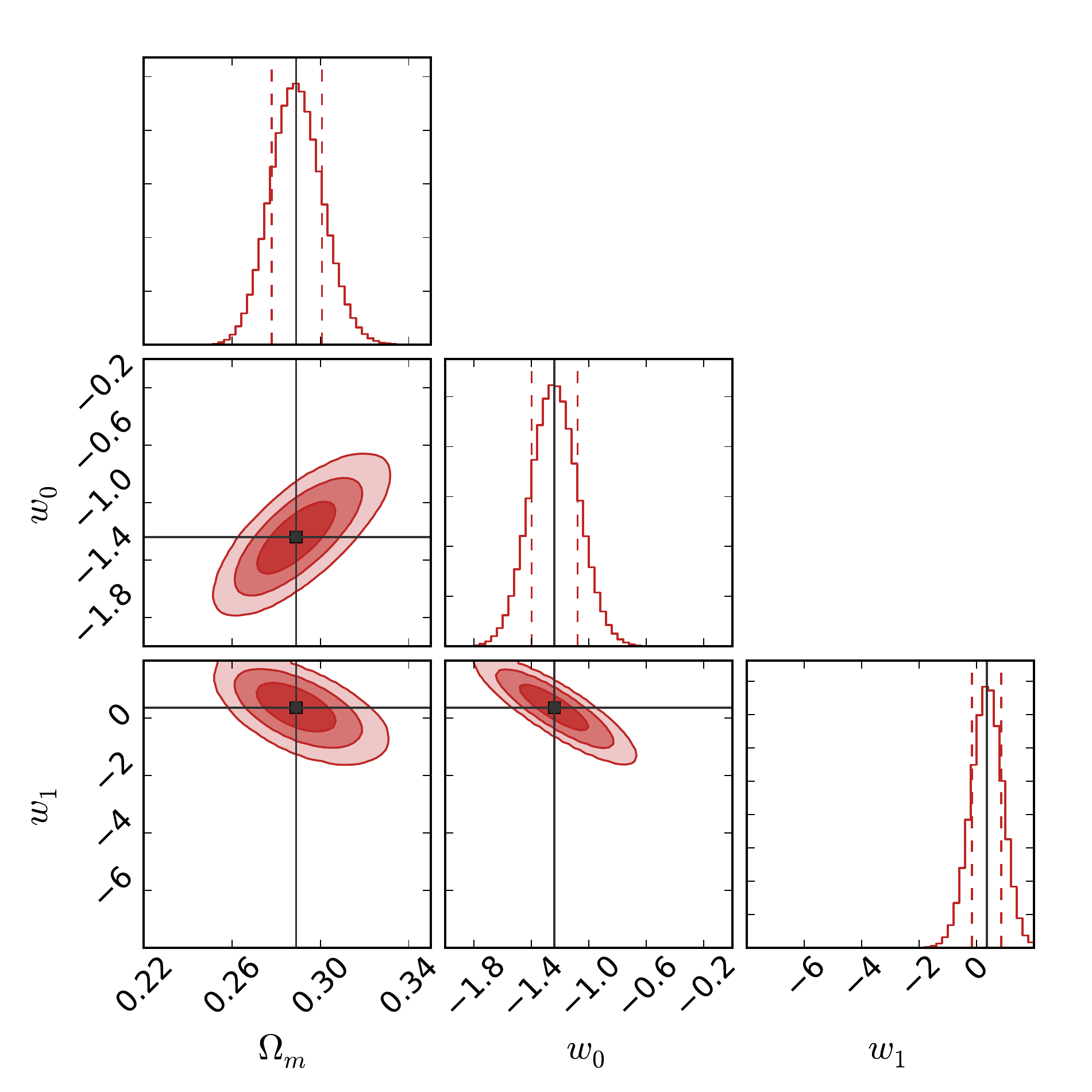}}\\
\subfloat[$\fgas^{\mathrm{ACT}}$]{\includegraphics[width=0.35\textwidth]{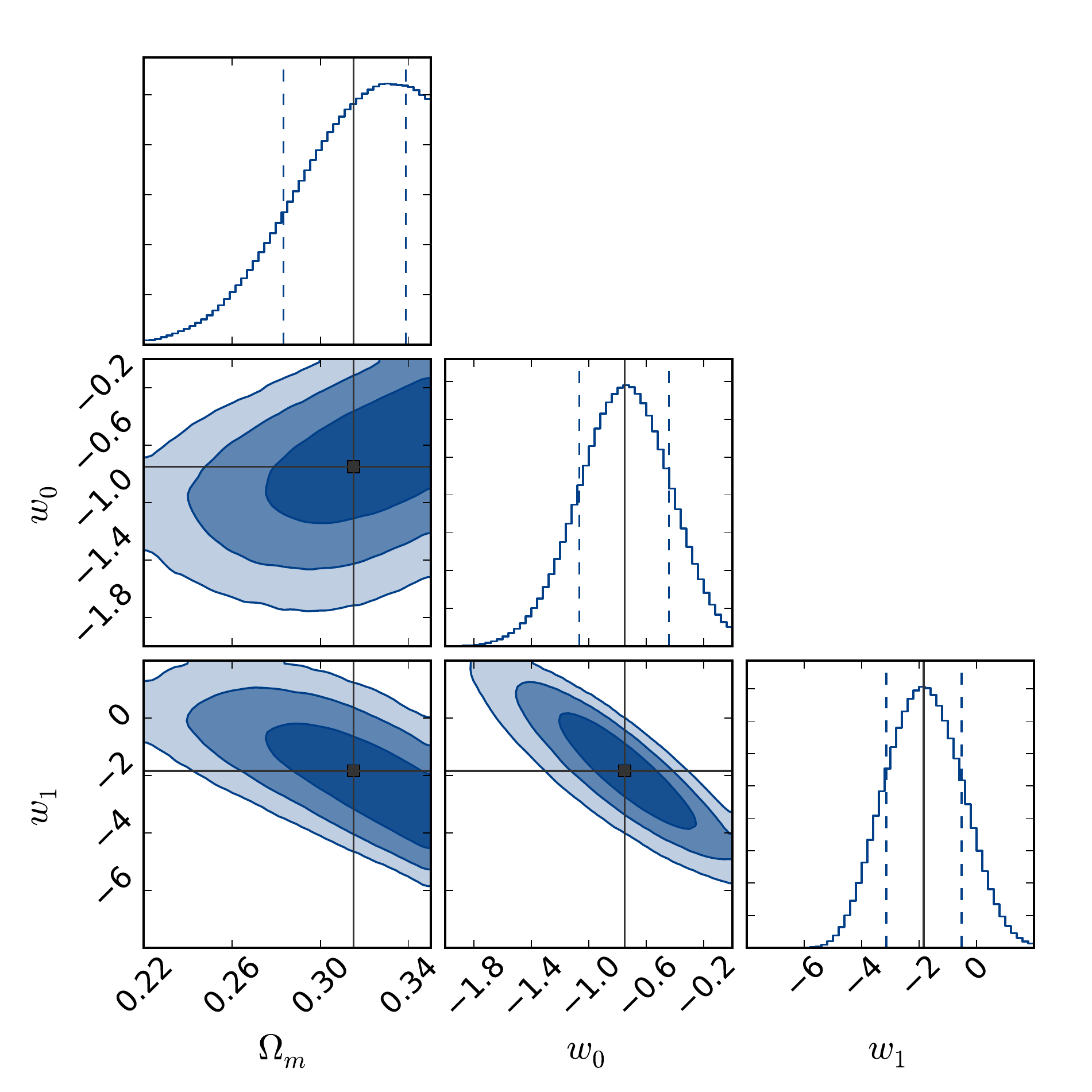}} & 
\subfloat[$\fgas^{\mathrm{ACT}}+H(z)$+BAO+WMAP]{\includegraphics[width=0.35\textwidth]{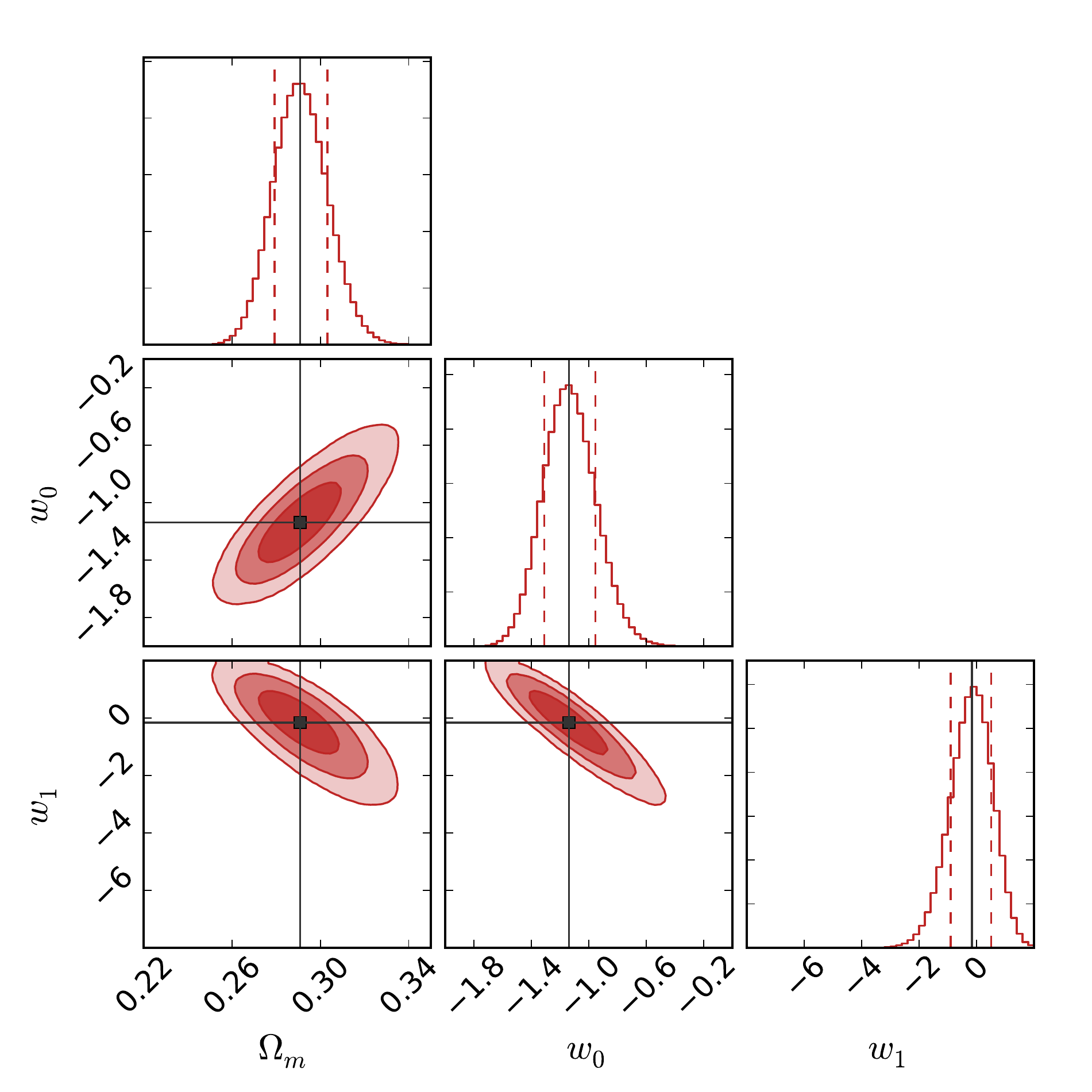}} &
\subfloat[$\fgas^{\mathrm{ACT}}+H(z)$+BAO+Planck]{\includegraphics[width=0.35\textwidth]{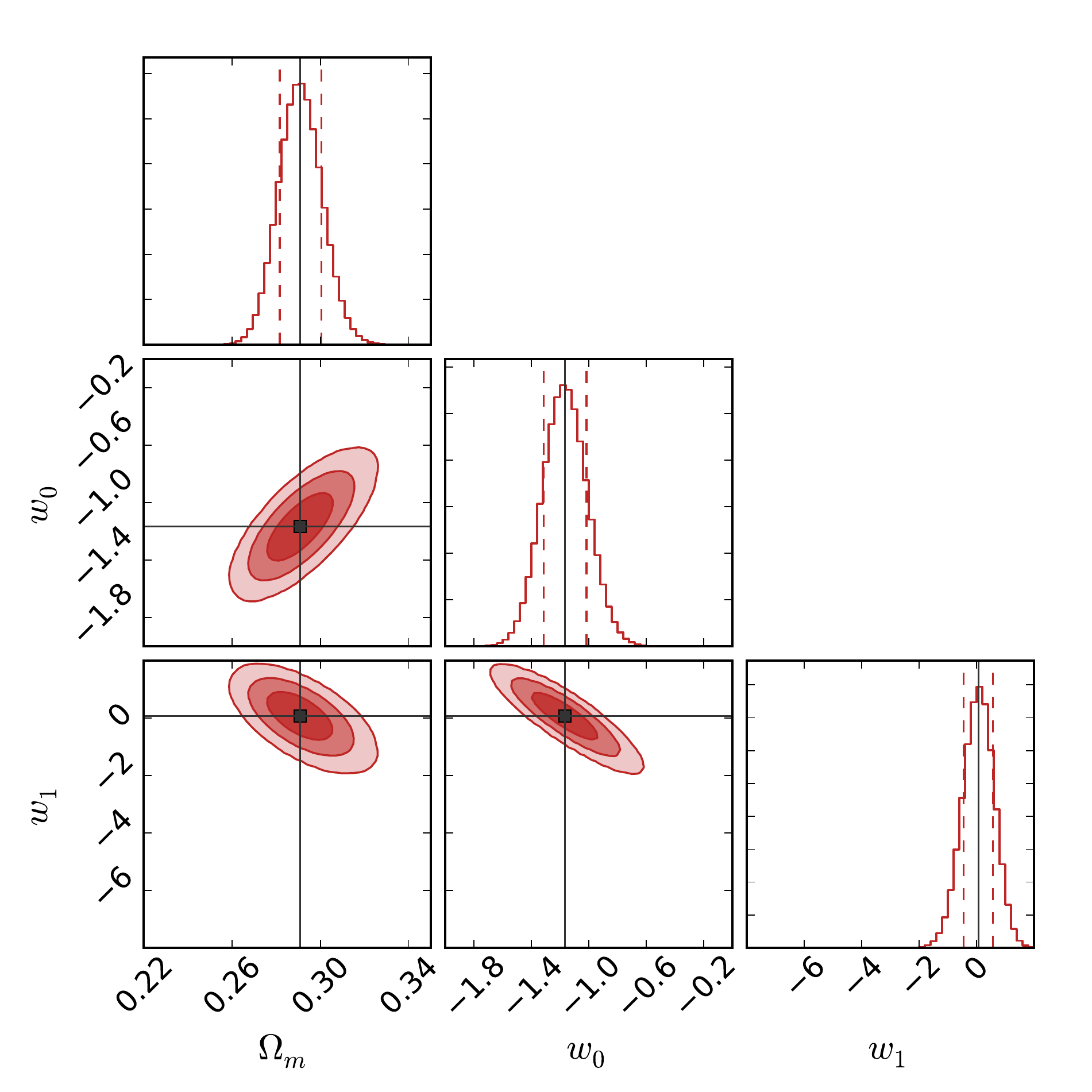}}\\

\end{tabular}
\caption{The same as Fig. C1 for the FSLLI parameterization.}
\label{fig:FSLLI}
\end{figure*}

\begin{figure*}
\centering
FSLLII parameterization\par\smallskip
\begin{tabular}{ccc}
\subfloat[$\fgas^{\mathrm{Allen}}$]{\includegraphics[width=0.35\textwidth]{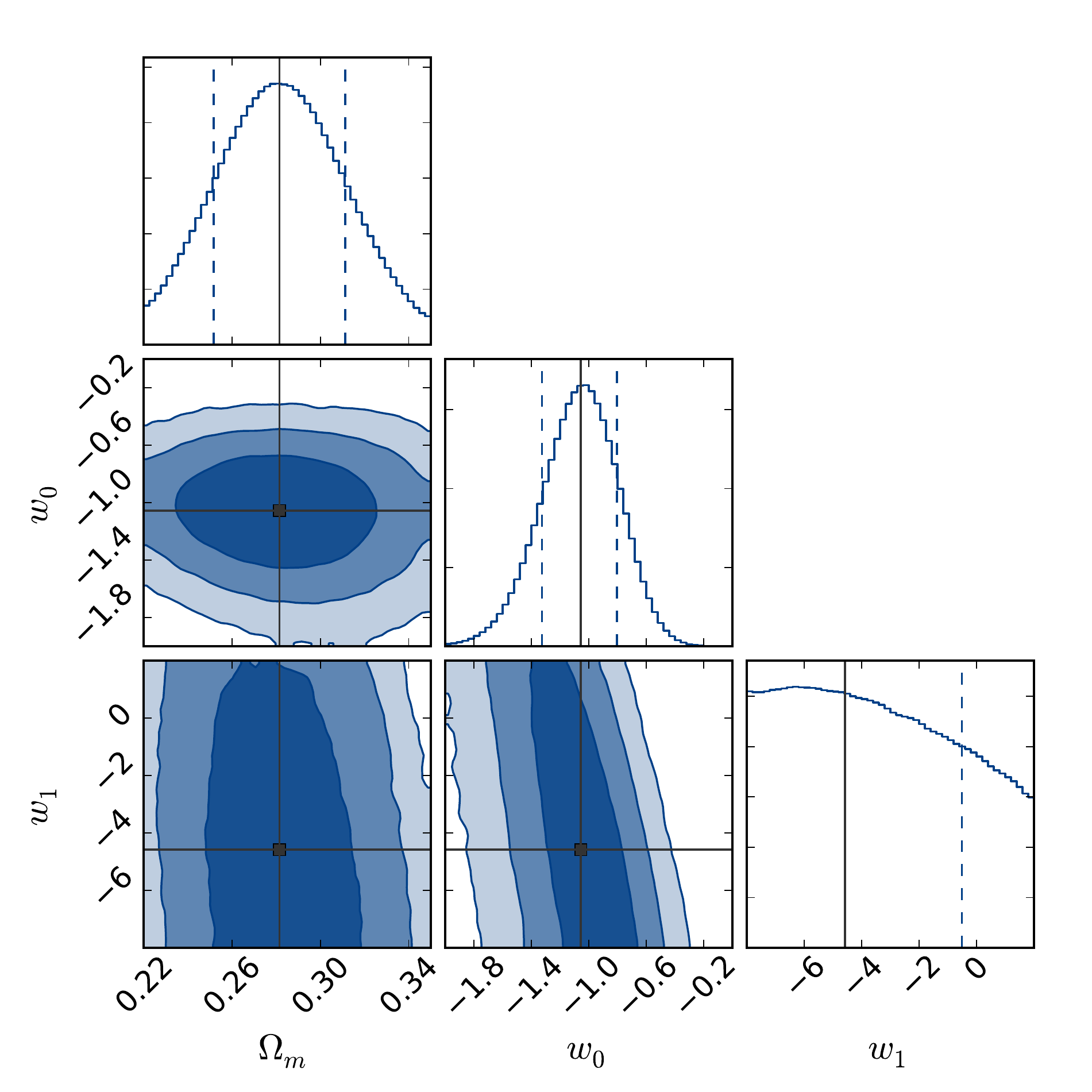}} & 
\subfloat[$\fgas^{\mathrm{Allen}}+H(z)$+BAO+WMAP]{\includegraphics[width=0.35\textwidth]{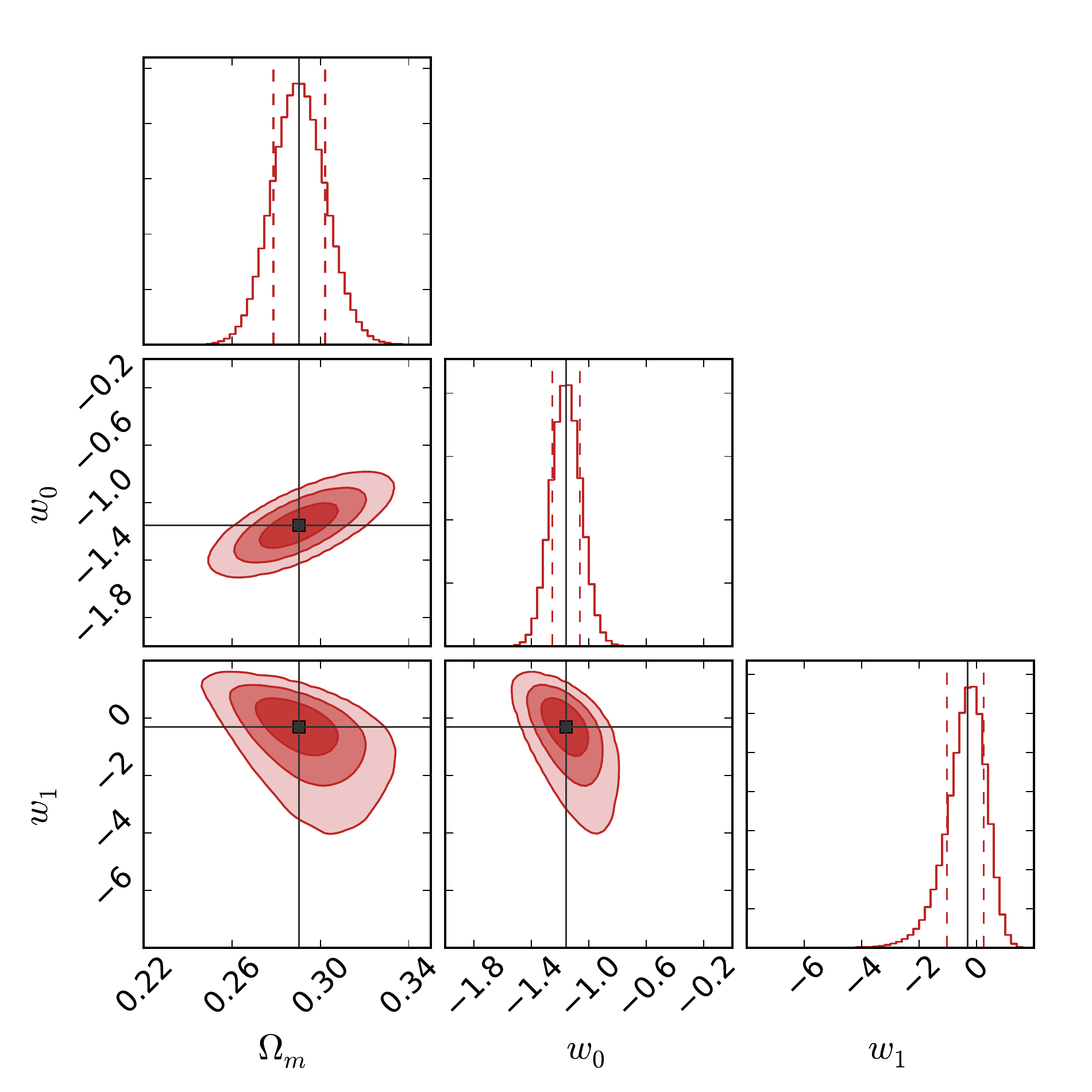}} &
\subfloat[$\fgas^{\mathrm{Allen}}+H(z)$+BAO+Planck]{\includegraphics[width=0.35\textwidth]{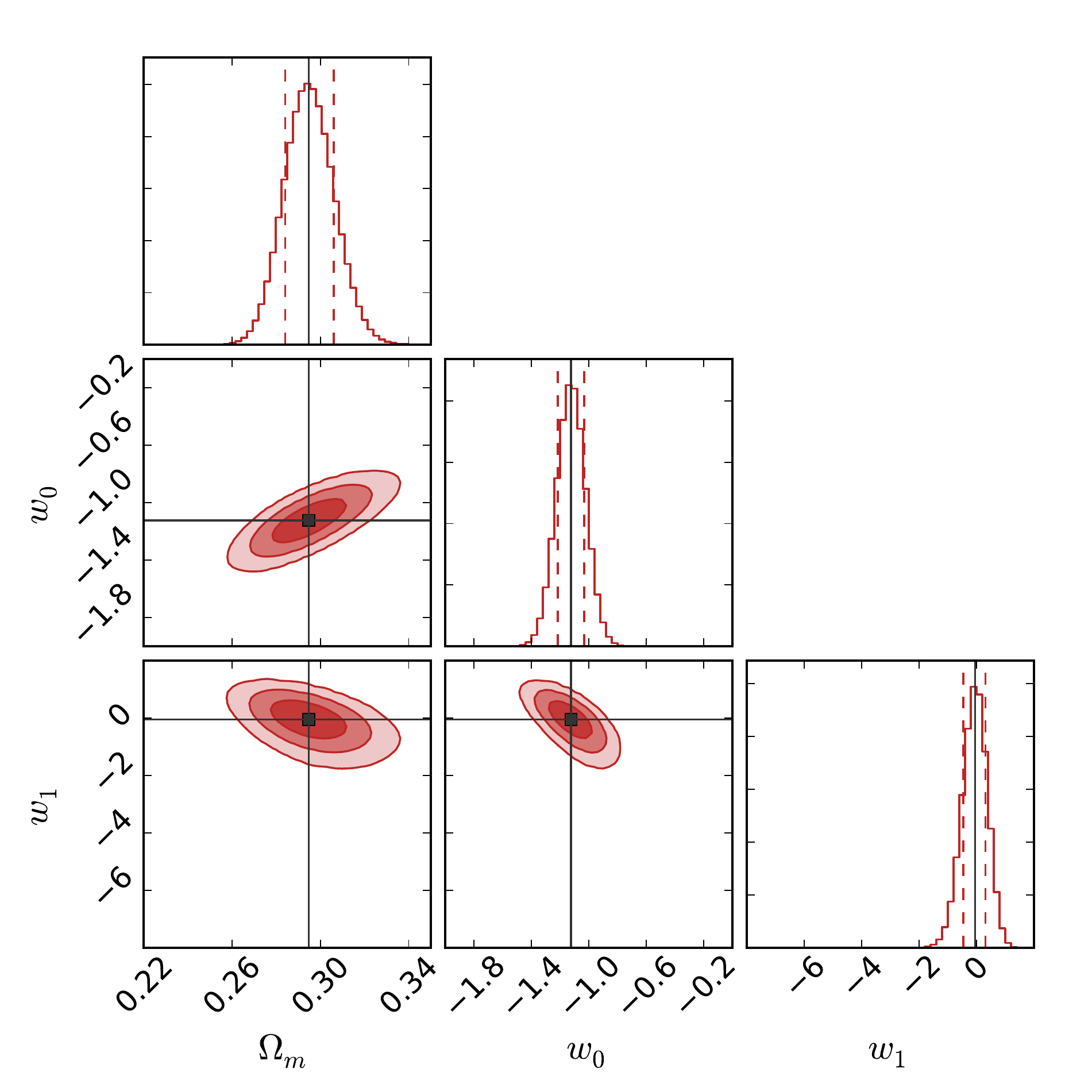}}\\
\subfloat[$\fgas^{\mathrm{ACT}}$]{\includegraphics[width=0.35\textwidth]{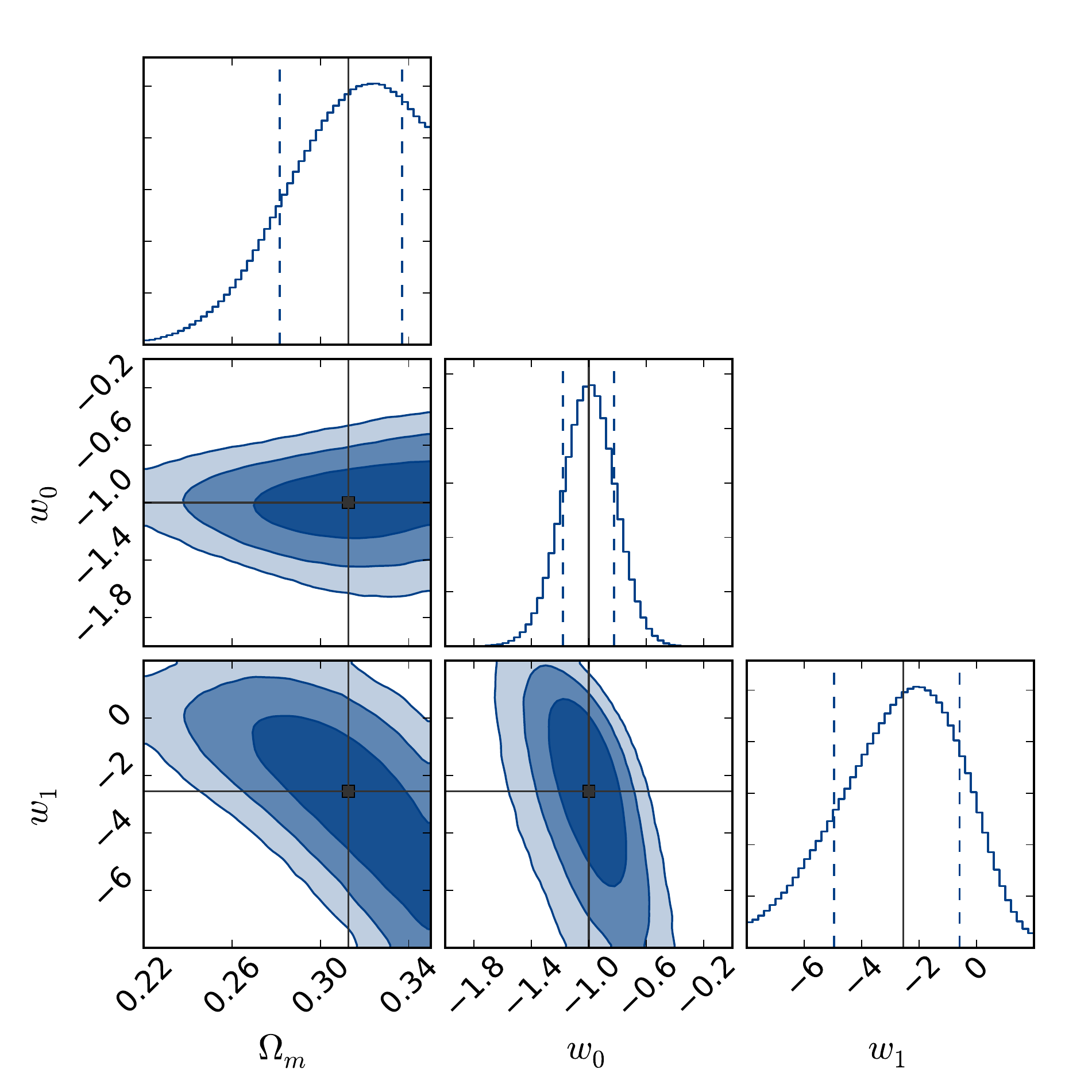}} & 
\subfloat[$\fgas^{\mathrm{ACT}}+H(z)$+BAO+WMAP]{\includegraphics[width=0.35\textwidth]{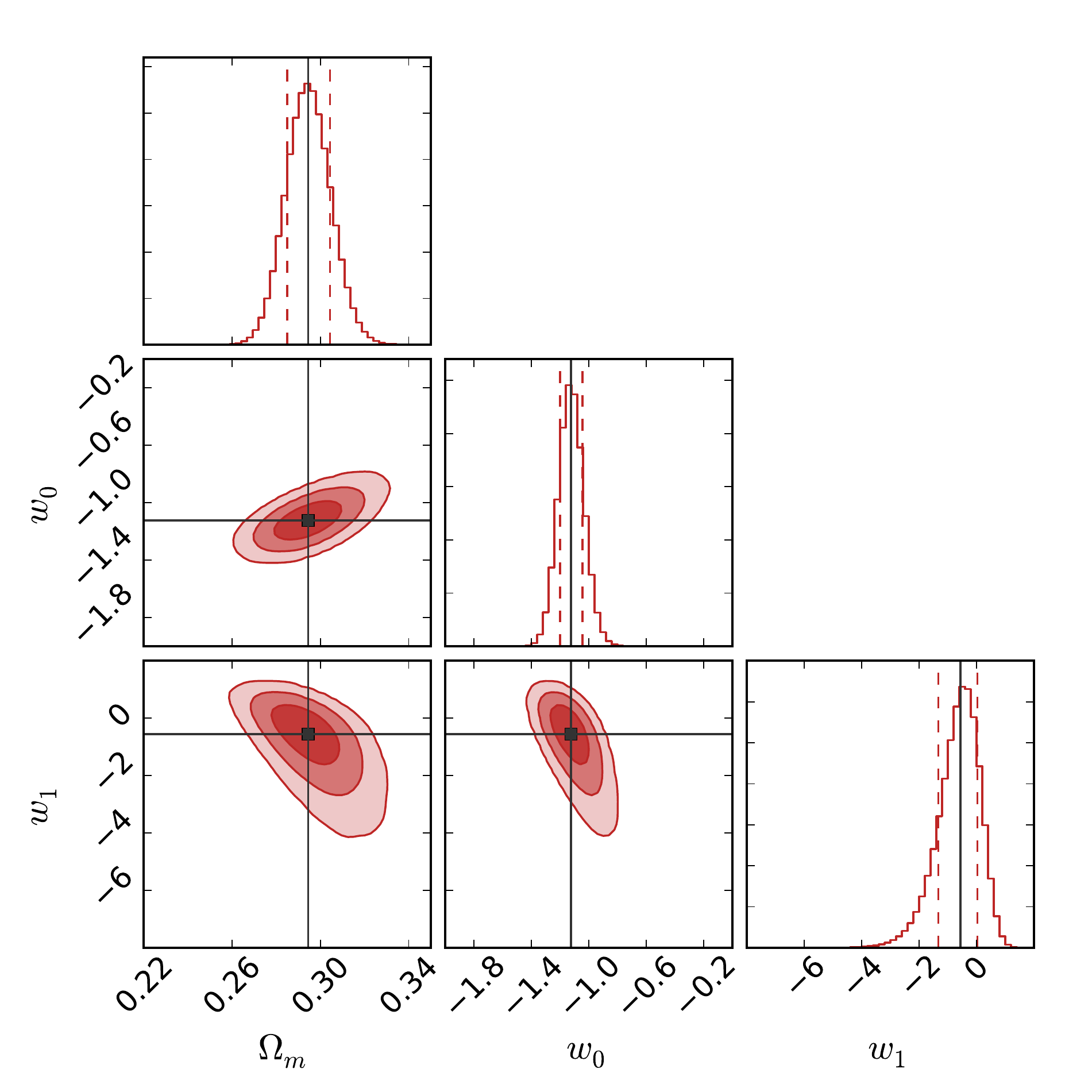}} &
\subfloat[$\fgas^{\mathrm{ACT}}+H(z)$+BAO+Planck]{\includegraphics[width=0.35\textwidth]{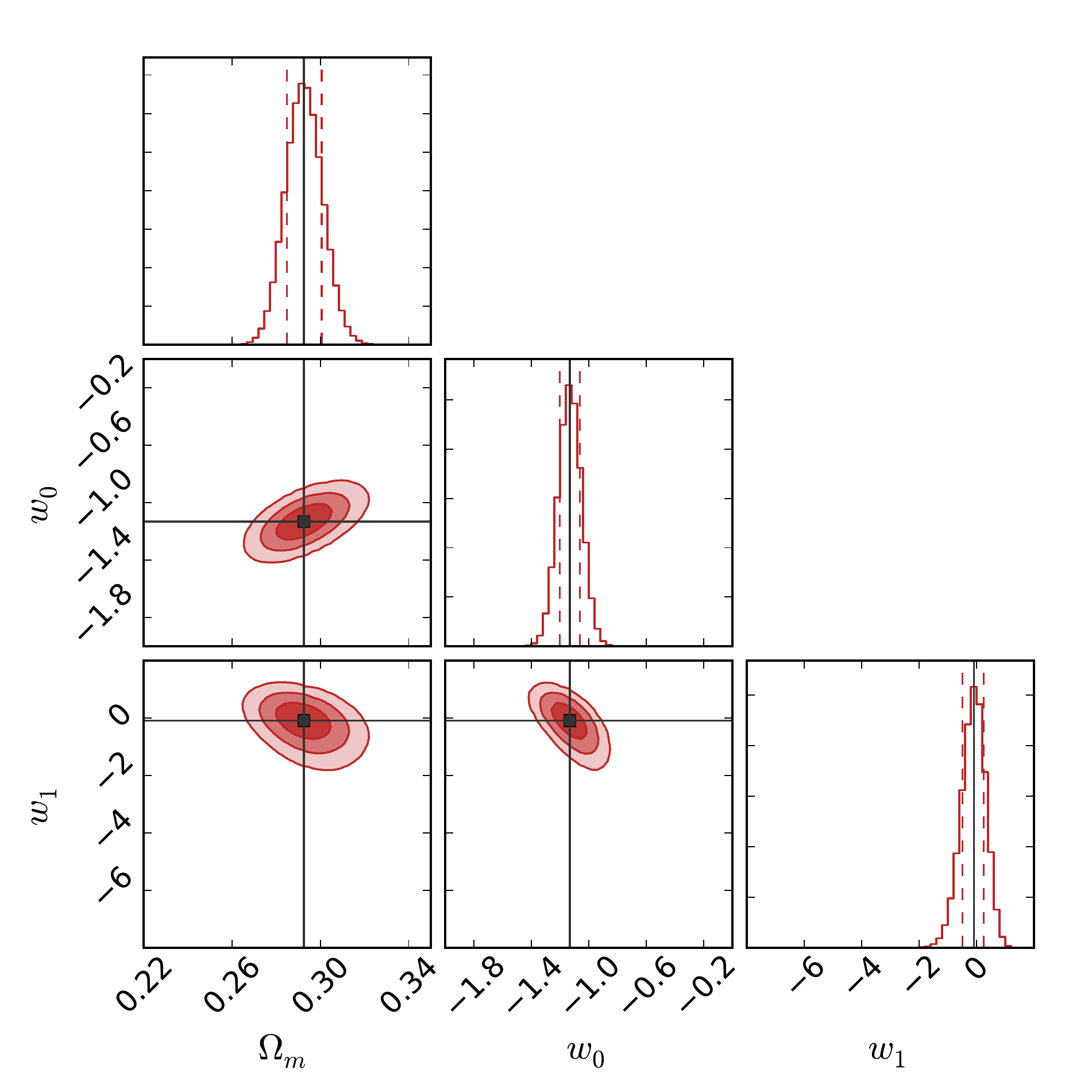}}\\

\end{tabular}
\caption{The same as Fig. C1 for the FSLLII parameterization.}
\label{fig:FSLLII}
\end{figure*}

\begin{figure*}
\centering
SL parameterization\par\smallskip
\begin{tabular}{ccc}
\subfloat[$\fgas^{\mathrm{Allen}}$]{\includegraphics[width=0.35\textwidth]{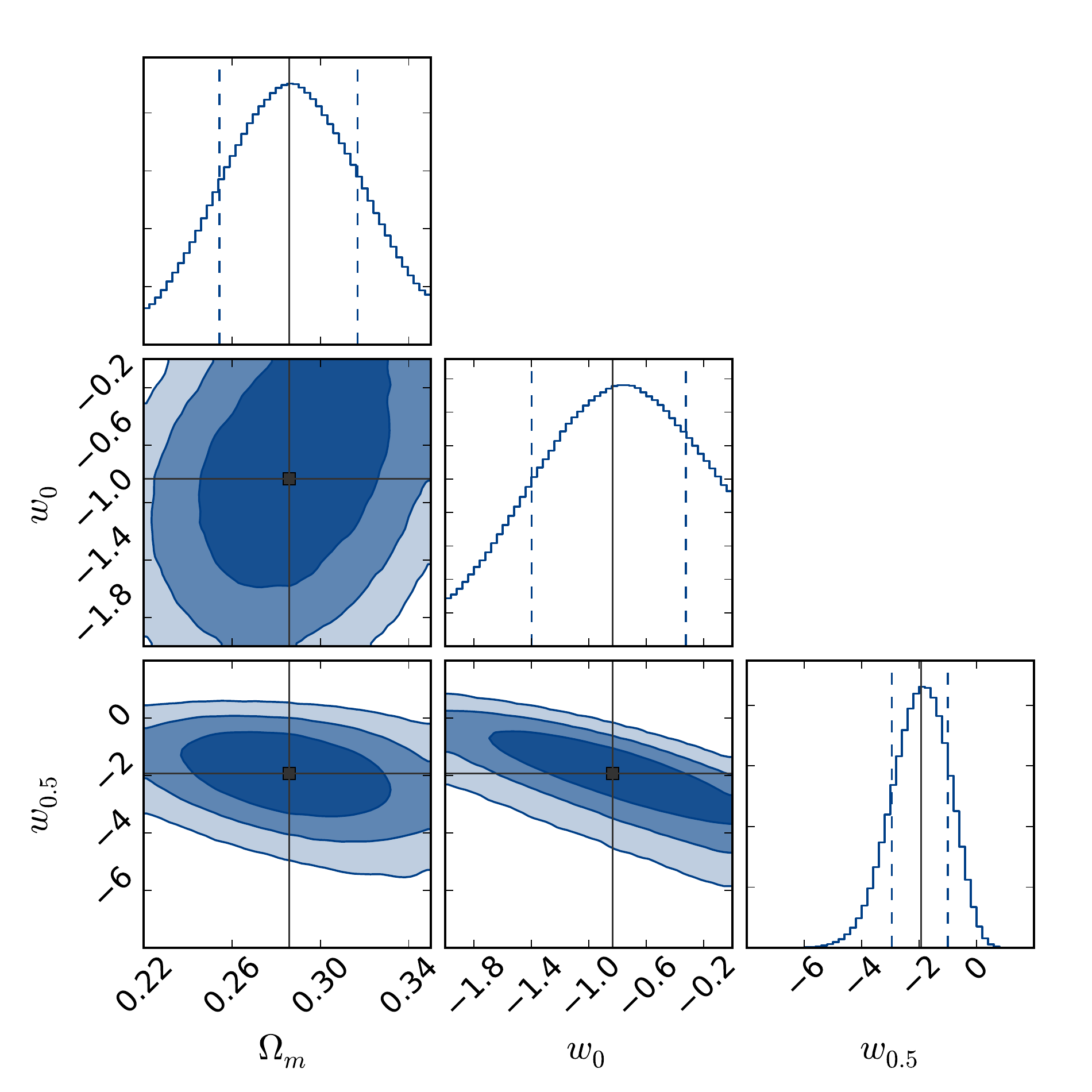}} & 
\subfloat[$\fgas^{\mathrm{Allen}}+H(z)$+BAO+WMAP]{\includegraphics[width=0.35\textwidth]{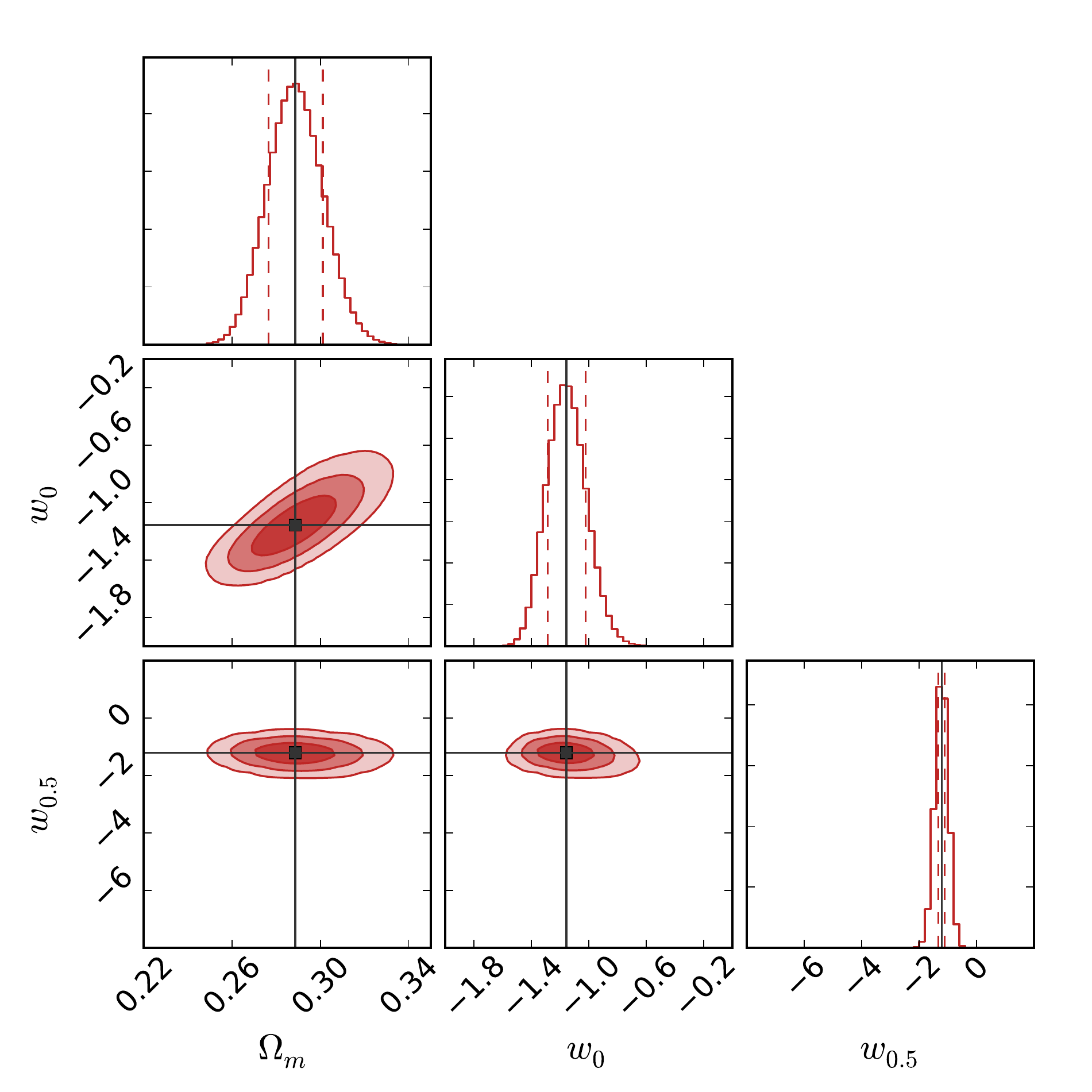}} &
\subfloat[$\fgas^{\mathrm{Allen}}+H(z)$+BAO+Planck]{\includegraphics[width=0.35\textwidth]{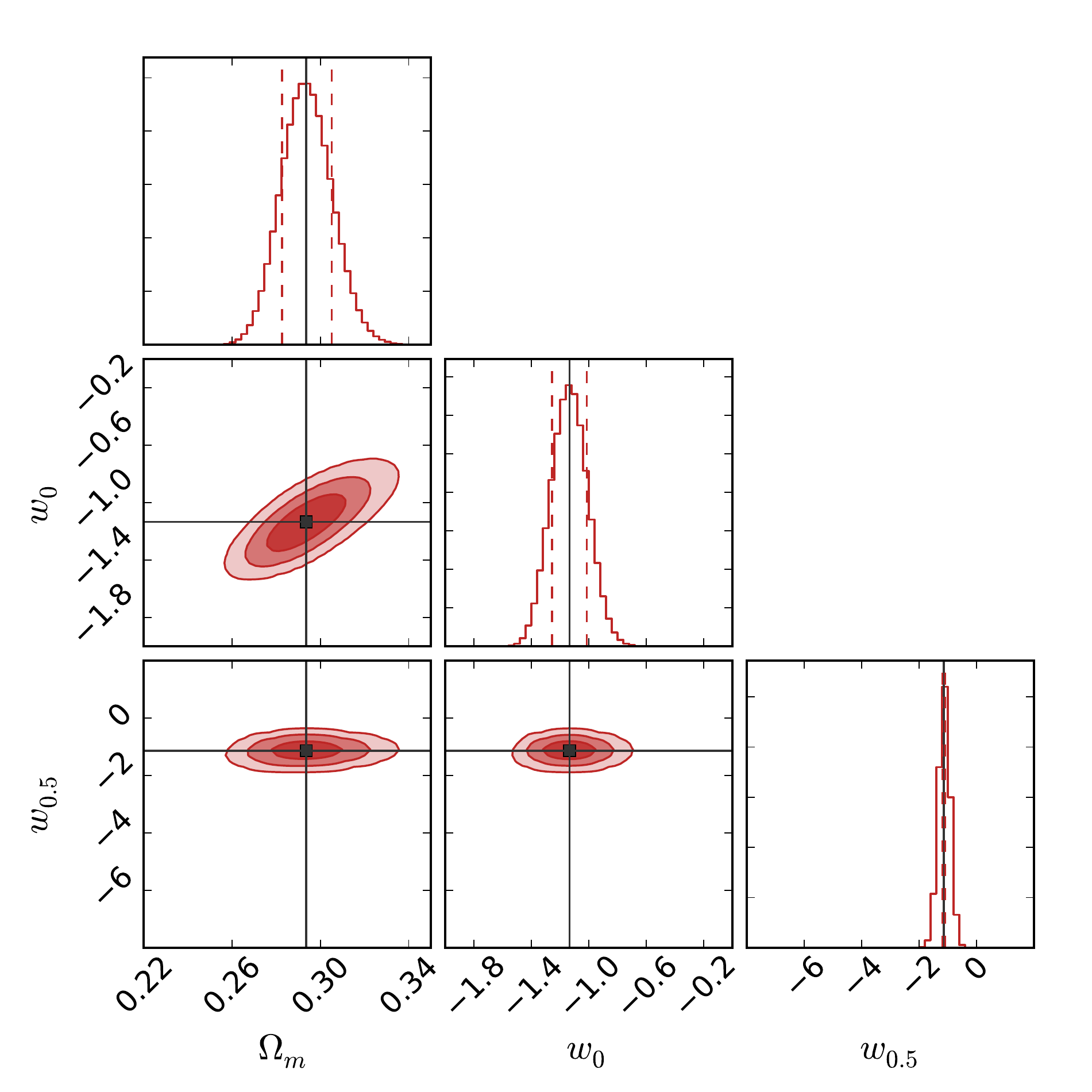}}\\
\subfloat[$\fgas^{\mathrm{ACT}}$]{\includegraphics[width=0.35\textwidth]{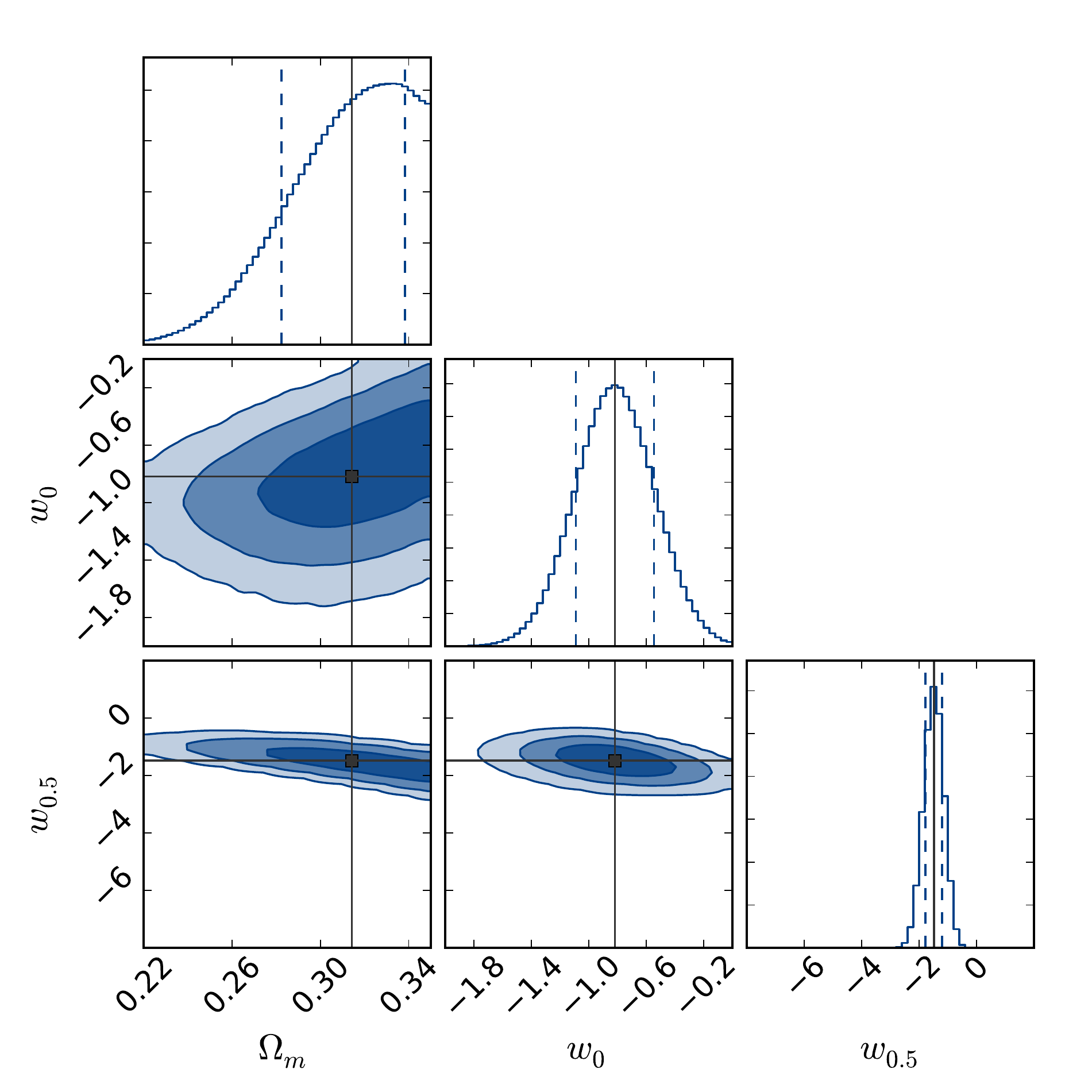}} & 
\subfloat[$\fgas^{\mathrm{ACT}}+H(z)$+BAO+WMAP]{\includegraphics[width=0.35\textwidth]{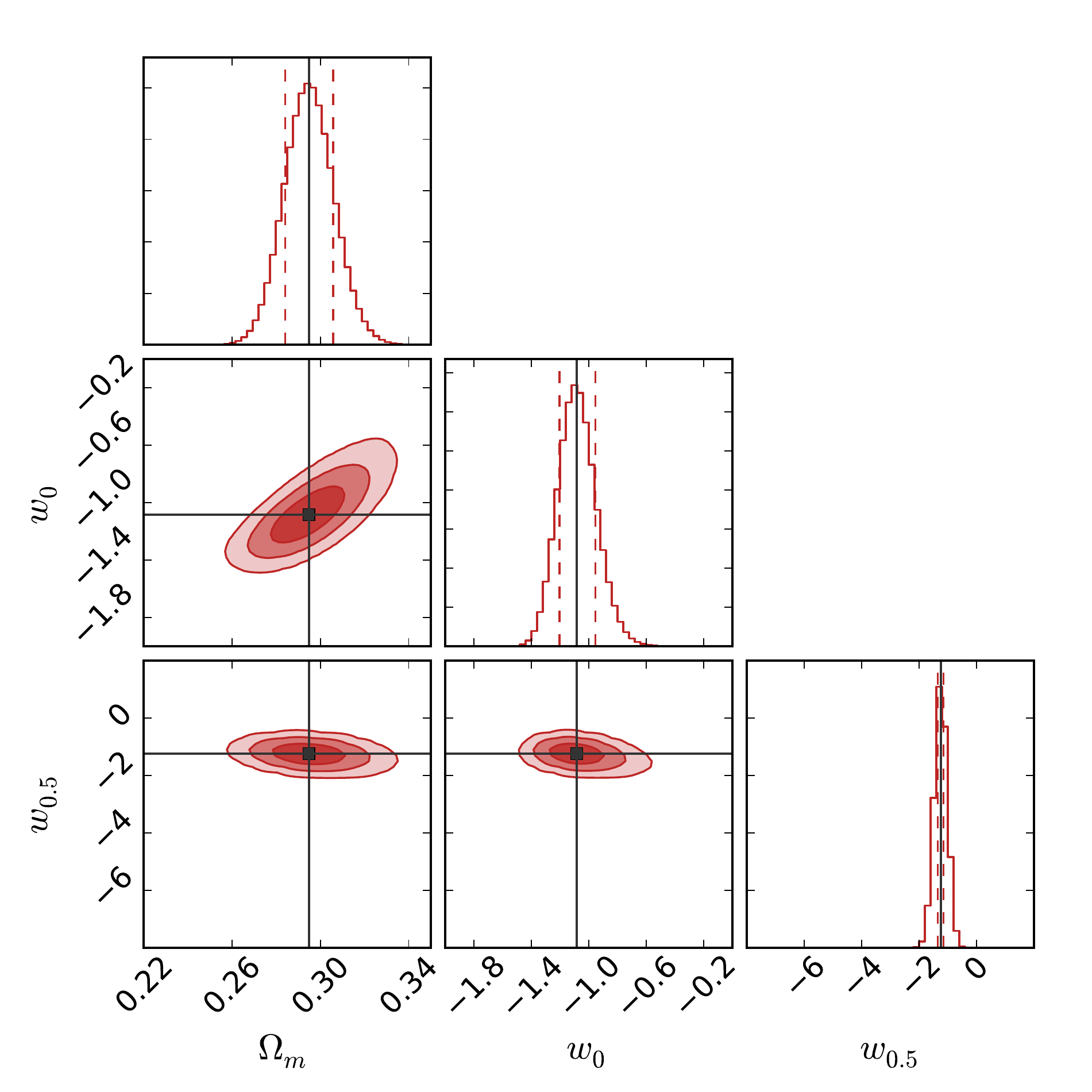}} &
\subfloat[$\fgas^{\mathrm{ACT}}+H(z)$+BAO+Planck]{\includegraphics[width=0.35\textwidth]{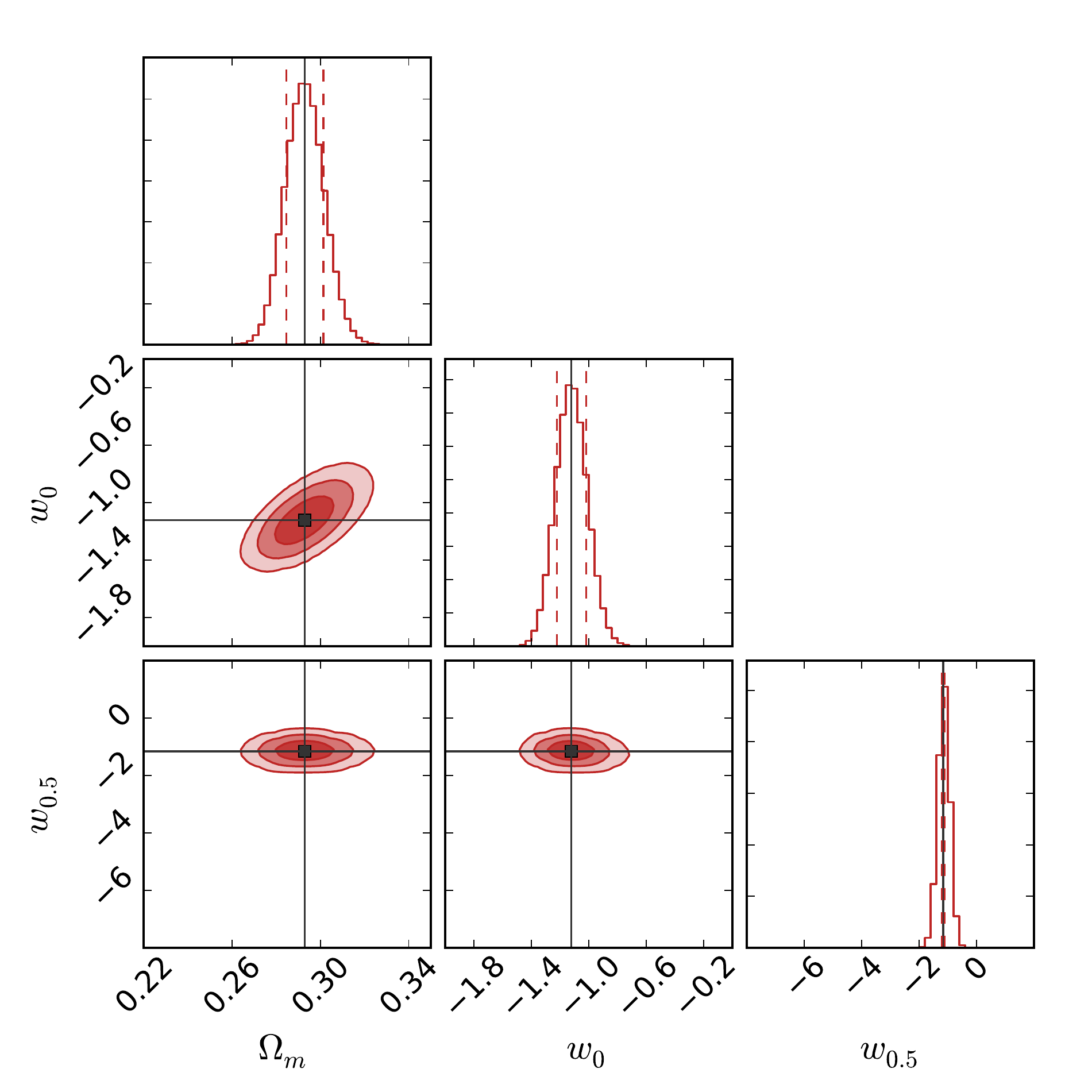}}\\
\end{tabular}
\caption{The same as Fig. C1 for the SL parameterization.}
\label{fig:SL}
\end{figure*}
\section{The H(z) data set} \label{Ap:Hz}
\begin{table*}
\begin{tabular}{|lllll|}
\hline
$z$ & $H(z)$ & $\sigma_{H}$ & Reference & Method \\
\hline
0.07 & 69 & 19.6 & \cite{Zhang2012} & DA\\
0.1 & 69 & 12 & \cite{Stern2010} & DA\\
0.12 & 68.6 & 26.2 & \cite{Zhang2012} & DA\\
0.17 & 83 & 8 & \cite{Stern2010} & DA \\
0.179 & 75 & 4 & \cite{Moresco2012} & DA\\
0.199& 75 & 5 & \cite{Moresco2012} & DA\\
0.2& 72.9 & 29.6 & \cite{Zhang2012} & DA\\
0.24 & 79.69 & 2.65 & \cite{Gaztanaga2009} & Clustering \\
0.27 & 77 & 14 & \cite{Stern2010} & DA\\
0.28 & 88.8 & 36.6 & \cite{Zhang2012} & DA\\
0.3 & 81.7 & 6.22 & \cite{Oka:2014}& Clustering \\
0.35 & 82.7 & 8.4 & \cite{Chuang2013} & Clustering\\
0.352 & 83 & 14 & \cite{Moresco2012} & DA\\
0.4 & 95 & 17 & \cite{Stern2010} & DA\\
0.43 & 86.45 & 3.68 & \cite{Gaztanaga2009} & Clustering\\
0.44 & 82.6 & 7.8 & \cite{Blake2012} & Clustering\\
0.48 & 97 & 62 & \cite{Stern2010} & DA\\
0.57 & 92.9 & 7.8 & \cite{Anderson2014} & Clustering\\
0.593& 104 & 13 & \cite{Moresco2012} & DA\\
0.600& 87.9 & 6.1& \cite{Blake2012} & Clustering \\
0.68 & 92 & 8 & \cite{Moresco2012} & DA\\
0.73 & 97.3 & 7 & \cite{Blake2012} & Clustering\\
0.781 & 105 & 12 & \cite{Moresco2012} & DA\\
0.875 & 125 & 17 & \cite{Moresco2012} & DA\\
0.88 & 90 & 40 & \cite{Stern2010} & DA\\
0.9 & 117 & 23 & \cite{Stern2010} & DA\\
1.037 & 154 & 20 & \cite{Moresco2012} & DA\\
1.3 & 168 & 17 & \cite{Stern2010} & DA\\
1.43 & 177 & 18 & \cite{Stern2010} & DA\\
1.53 & 140 & 14 & \cite{Stern2010} & DA\\
1.75 & 202 & 40 & \cite{Stern2010} & DA\\
2.3  & 224 & 8 & \cite{Busca2013} & Clustering\\
2.34 & 222 & 7 & \cite{Delubac2014}& Clustering\\
2.36 & 226 & 8 & \cite{Font-Ribera:2014}& Clustering\\
\hline
\end{tabular} 
\caption{Hubble parameter measurements $H(z)$ (in km s$^{-1}$Mpc$^{-1}$) and its error, $\sigma_{H}$, 
at redshift $z$. The method refers to as $H(z)$ was obtained: DA stands for diameter angular distance.}
\label{tab:Hz}
\end{table*}

\end{document}